\documentclass[11pt,a4paper]{article}
\usepackage{jheppub}
\usepackage{amsmath,amsfonts,epsfig,color}
\usepackage{amssymb}
\usepackage[english, USenglish]{babel}
\usepackage{epsfig,psfrag}

\usepackage{hyperref}

\csname @addtoreset\endcsname{equation}{section}

\makeatletter
\def\timenow{\@tempcnta\time
  \@tempcntb\@tempcnta
  \divide\@tempcntb60
  \ifnum10>\@tempcntb0\fi\number\@tempcntb
  \multiply\@tempcntb60
  \advance\@tempcnta-\@tempcntb
  :\ifnum10>\@tempcnta0\fi\number\@tempcnta}
\makeatother

\def\oonoo#1#2#3{\vbox{\ialign{##\crcr
	\hfil\hfil\hfil{$#3{#1}$}\hfil\crcr\noalign{\kern1pt\nointerlineskip}
	$#3{#2}$\crcr}}}
\def\oon#1#2{\mathchoice{\oonoo{#1}{#2}{\displaystyle}}
	{\oonoo{#1}{#2}{\textstyle}}{\oonoo{#1}{#2}{\scriptstyle}}
	{\oonoo{#1}{#2}{\scriptscriptstyle}}}

\def\dt#1{\oon{\hbox{\bf .}}{#1}}  
\def\ddt#1{\oon{\hbox{\bf .\kern-1pt.}}#1}
\def\slap#1#2{\setbox0=\hbox{$#1{#2}$}
	#2\kern-\wd0{\hfuzz=1pt\hbox to\wd0{\hfil$#1{/}$\hfil}}}

\title{$\cN=4$ super-Yang-Mills in LHC superspace \\ Part I: Classical and quantum theory}

\author[a]{Dmitry Chicherin}
\author[a,b]{and Emery Sokatchev}

\affiliation[a]{LAPTH $^{*}$\note{Laboratoire d'Annecy-le-Vieux de Physique Th\'{e}orique, UMR 5108}, Universit\'{e} de Savoie, CNRS, B.P. 110,  F-74941 Annecy-le-Vieux, France}
\affiliation[b]{Theoretical Physics Department, CERN, CH -1211, Geneva 23, Switzerland}

\abstract{We present a formulation of the maximally supersymmetric $\cN = 4$ gauge theory  in Lorentz harmonic chiral (LHC) superspace. 
It is closely related to the twistor formulation of the theory but employs the simpler notion of Lorentz harmonic variables. They parametrize a two-sphere and allow us to handle efficiently infinite towers of higher-spin auxiliary fields defined on ordinary space-time. In this approach the chiral half of  $\cN=4$ supersymmetry is manifest. The other half is realized non-linearly and the algebra closes on shell. We give a straightforward derivation of  the Feynman rules in coordinate space.  We show that the LHC formulation of the $\cN = 4$ super-Yang-Mills theory is remarkably similar to  the harmonic superspace formulation of the  $\cN=2$ gauge and hypermultiplet matter theories. In the twin paper arXiv:1601.06804 we apply the LHC formalism to the study of the non-chiral multipoint correlation functions of the $\cN = 4$ stress-tensor supermultiplet.}

\emailAdd{chicherin@uni-mainz.de}
\emailAdd{emeri.sokatchev@cern.ch}


\arxivnumber{1601.06803}

\dedicated{Dedicated to the memory of Boris Zupnik,\\ a colleague and a friend}

\begin{document}

\renewcommand{\thefootnote}{\fnsymbol{footnote}}


\newcommand{\norm}[1]{{\protect\normalsize{#1}}}
\newcommand{\p}[1]{(\ref{#1})}
\newcommand{\half}{{\ts \frac{1}{2}}}
\newcommand \vev [1] {\langle{#1}\rangle}
\newcommand \ket [1] {|{#1}\rangle}
\newcommand \bra [1] {\langle {#1}|}

\newcommand{\cM}{{\cal M}} 
\newcommand{\cR}{{\cal R}} 
\newcommand{\cS}{{\cal S}} 
\newcommand{\cK}{{\cal K}}
\newcommand{\cL}{{\cal L}} 
\newcommand{\cF}{{\cal F}}
\newcommand{\cN}{{\cal N}}
\newcommand{\cA}{{\cal A}}
\newcommand{\cB}{{\cal B}}
\newcommand{\cG}{{\cal G}}
\newcommand{\cO}{{\cal O}}
\newcommand{\cY}{{\cal Y}}
\newcommand{\cX}{{\cal X}}
\newcommand{\cT}{{\cal T}}
\newcommand{\cW}{{\cal W}}
\newcommand{\cP}{{\cal P}}
\newcommand{\nt}{\notag\\} 
\newcommand{\pa}{\partial}
\newcommand{\ep}{\epsilon}
\newcommand{\om}{\omega}
\newcommand{\bep}{\bar\epsilon}
\renewcommand{\a}{\alpha}
\renewcommand{\b}{\beta}
\newcommand{\g}{\gamma}
\newcommand{\s}{\sigma}
\newcommand{\la}{\lambda}
\newcommand{\da}{{\dot\alpha}}
\newcommand{\db}{{\dot\beta}}
\newcommand{\dg}{{\dot\gamma}}
\newcommand{\dd}{{\dot\delta}}
\newcommand{\q}{\theta}
\newcommand{\bq}{\bar\theta}
\newcommand{\bQ}{\bar Q}
\newcommand{\tx}{\tilde{x}}
\newcommand{\tr}{\mbox{tr}}
\newcommand{\+}{{\dt+}}
\renewcommand{\-}{{\dt-}}

\maketitle
\flushbottom

\setcounter{page}{1}\setcounter{footnote}{0}
\renewcommand{\thefootnote}{\arabic{footnote}}

\section{Introduction}

A very old problem in supersymmetric field theory is how to formulate the maximally supersymmetric  $\cN=4$ super-Yang-Mills (SYM) theory off shell with full manifest  supersymmetry. A counting argument \cite{Siegel:1981dx} claims that there exist no \emph{finite} sets of fermionic and bosonic auxiliary fields which could remove the mismatch of the off-shell degrees of freedom of the physical fields. The same counting argument \cite{Stelle:1985jh} in the case of the $\cN=2$ hypermultiplet has been successfully circumvented by introducing \emph{infinite} sets of auxiliary fields, in the framework of harmonic superspace \cite{Galperin:1984av}.\footnote{For a review see \cite{Galperin:2001uw}. See also \cite{Roslyi:1989ya,Roslyi:1986yg,Karlhede:1984vr,Roslyi:1985hn} for alternative approaches. } A similar approach exists to $\cN=3$ SYM \cite{Galperin:1984bu}, a theory with the same physical degrees of freedom as $\cN=4$ SYM but with less manifest supersymmetry. However, the $\cN=4$ barrier still remains a formidable challenge.

Here we do not propose a solution to this old problem but perhaps a step towards it. We present a formulation of  $\cN=4$ SYM in Lorentz harmonic chiral (LHC) superspace.  In it the chiral half of supersymmetry is manifest and realized off shell. The other half is realized non-linearly and the algebra closes modulo equations of motion and gauge transformation. An important advantage of this `semi-off-shell' formulation is that we are able to maintain the full  R-symmetry $SU(4)$ combined with manifest chiral supersymmetry, unlike the alternative formulations of $\cN=4$ SYM in terms of $\cN=1,2,3$ superfields.

One of our aims is the simple and clear derivation of the Feynman rules which preserve the symmetries. We apply these rules in the twin paper \cite{twin} to construct the non-chiral sector of the Born-level correlation functions of the $\cN=4$ stress-tensor multiplet. Our main result is surprisingly simple: the full non-chiral correlator is obtained from its chiral truncation studied in \cite{Chicherin:2014uca} by a Grassmann shift of the space-time coordinates.

 We use Lorentz harmonics (LHs) as a convenient tool for introducing infinite sets of auxiliary fields and pure gauges. Our formulation  closely follows, in many aspects,  that of $\cN=2$ SYM and hypermultiplet matter  in harmonic superspace \cite{Galperin:1984av, Zupnik:1987vm}. The main difference is that here the auxiliary and pure gauge fields have arbitrarily high Lorentz spins instead of the $SU(2)$ isospins in the harmonic formulation of $\cN=2$ SYM and matter.

The Lorentz harmonics are auxiliary variables, in addition to the Euclidean space-time $R^4$ with  coordinates $x^{\da\a} = (\sigma_\mu)^{\da\a} x^\mu$ and Lorentz group $SO(4) \sim SU(2)_L\times SU(2)_R$. The harmonics  describe the two-sphere  $S^2 \sim SU(2)_L/U(1)$. They are defined as matrices $u^{\pm}_{\a} \in SU(2)_L$ carrying an $SU(2)_L$ index $\a=1,2$ and a $U(1)$ charge $\pm$.  The harmonic fields $f(x,u)$ are defined on $R^4\times S^2$ by their harmonic expansion as functions on $SU(2)_L$ homogeneous under the action of the subgroup $U(1)$. For example, a harmonic field of charge $+1$ has the infinite expansion
\begin{align}\notag
f^+(x,u) = f^\a (x)u^+_\a + f^{(\a\b\gamma)}(x) u^+_\a u^+_\b u^-_\gamma + \ldots\ .
\end{align}
The terms in this expansion carry  $SU(2)_L$ spins $1/2, 3/2, \ldots\ $. Thus, a harmonic field is a collection of infinitely many higher-spin fields $f^{\a\b\ldots}(x)$, most of which are pure gauges or auxiliary fields. Fixing the appropriate gauge and eliminating the auxiliary fields via their kinematical field equations, we obtain a description of the theory in terms of a finite number of ordinary propagating fields. As an example, the correlators of the $\cN=4$ stress-tensor multiplet {do not depend} on the LHs (they are integrated out). To put it differently, at the end of the calculation we are able to eliminate the infinite sets of auxiliary and pure gauge fields. This clearly shows the auxiliary character of the LHs.

The idea to use LH variables has a long history. Soon after the development of the $\cN=2$ harmonic superspace method, the close connection between harmonics and twistors was observed in \cite{Galperin:1987wc}. Harmonic variables on $S^2$ were used to solve the self-dual Yang-Mills equations,   as an alternative to the well-known Ward construction of instantons \cite{Ward:1977ta}. This approach to self-dual (super)-Yang-Mills was further developed in \cite{Devchand:1992st,Devchand:1993ba,Devchand:1995gp}. Some aspects of the correspondence harmonics/twistors were discussed in \cite{Evans:1992az}.

Later on, LHs were used by one of the authors of the present paper to reformulate Siegel's self-dual $\cN=4$ SYM theory \cite{Siegel:1992xp} in chiral superspace \cite{Sokatchev:1995nj}. The theory is described by two dynamical superfields $A^+_\da(x, \q^+,u)$ and $A^{++}(x, \q^+,u)$. Here $\q^{+A}=u^+_\a \q^{\a A}$ is the LH projection of the chiral odd variable $\q^{\a A}$ ($A=1,\ldots,4$ is an $SU(4)$ R-symmetry index). These superfields play the role of gauge connections for the harmonic-projected space-time derivative $\pa^+_\da= u^{+\a} \pa/\pa x^{\da\a}$ and the harmonic derivative $\pa^{++}=u^{+\a} \pa/\pa u^{-\a}$, respectively. The vanishing of the corresponding curvatures  is equivalent to the field equations of self-dual $\cN=4$ SYM. These equations are derived from a superspace \emph{Chern-Simons} action.

Our formulation is closely related to the twistor one developed in  \cite{Witten:2003nn,Mason:2005zm,Boels:2006ir,Boels:2007gv,Boels:2007qn,Adamo:2011cb}. The main conceptual difference is how one treats the harmonic variables and their cousins twistors. In the twistor approach one introduces the twistor projective variable $Z^M=(\la_\a, \mu^\da)$ to describe the space   $CP^3$. Contact with (complexified) space-time is made via the incidence relation $\mu^\da = x^{\da\a}\la_\a$. The ordinary  fields $f(x)$ are obtained from the twistor fields $F(Z)$ by an integral (Penrose) transform. Thus, the concept of space-time reappears at the end, after performing this very non-trivial transform. In contrast, in the LH approach space-time is present from the very beginning, and the integral Penrose  transform is replaced by the simpler spherical harmonic expansion on $S^2$. In the twistor approach the accent is put on the conformal symmetry (the twistor $Z^M$ is a linear representation of the conformal group $GL(4)$). In the harmonic approach one insists on the manifest Lorentz symmetry of the ordinary fields $f^{\a\b\ldots}(x)$, viewed as harmonic modes on $S^2$.

The development of the twistor approach to $\cN=4$ SYM started with the work of Witten \cite{Witten:2003nn}. He found the twistor analog of the Chern-Simons action of  \cite{Sokatchev:1995nj}. Siegel's self-dual action  contains a Lagrange multiplier $G^{\a\b}$ for the self-dual YM tensor $F_{\a\b}$ in the form $\int d^4x\; G^{\a\b} F_{\a\b}$. Witten proposed to add another superspace gauge invariant to the Chern-Simons action, which contains the square of the Lagrange multiplier, $\int d^4x\; G^{\a\b} G_{\a\b}$. The sum of the two  terms is equivalent to the standard YM action $\int d^4x F^{\a\b} F_{\a\b}$, up to a topological term.  In a more explicit form the twistor version of the second action term was given by Mason  \cite{Mason:2005zm}. The detailed  supersymmetric construction was elaborated in \cite{Boels:2006ir}. 

If the action term $\int d^4x\; G^{\a\b} F_{\a\b}$ can be upgraded to a supersymmetric Chern-Simons term in LHC \cite{Sokatchev:1995nj} or twistor \cite{Witten:2003nn} superspace, the generalization of the term $\int d^4x\; G^{\a\b} G_{\a\b}$ takes a completely different form. Its Lagrangian density is given by $\log \det \left(\bar\pa^{-1}(\bar\pa + \cA)  \right)$, where $\bar\pa$ is a holomorphic derivative on twistor space and $\cA$ is its (super)connection. At close look (especially at the most explicit form shown in \cite{Boels:2006ir}) we realize that this peculiar action term has been known for almost 30 years, in the context of $\cN=2$ SYM formulated in harmonic superspace. Switching from twistors to LHs, we can identify the operator $\bar\pa$ with the harmonic derivative $\pa^{++}$ and $\cA$ with the gauge connection $A^{++}(x, \q^+,u)$ mentioned above. Then the Lagrangian term $\log \det \left(\bar\pa^{-1}(\bar\pa + \cA)  \right)$ becomes practically identical with the $\cN=2$ SYM harmonic superspace Lagrangian proposed by Zupnik  in  \cite{Zupnik:1987vm}. 

This remarkable similarity between the LHC formulation of $\cN=4$ SYM  and the $SU(2)$ harmonic formulation of the $\cN=2$ supersymmetric theories  goes even farther. In Sect.~\ref{n2}  we show that the $\cN=4$ Chern-Simons term has its analog as the $\cN=2$ hypermultiplet action in harmonic superspace. The common point between the two theories is the use of harmonics on $S^2$, either as LHs or as R-symmetry harmonics.  We also propose an eight-dimensional SYM theory  whose dimensional reduction gives rise either to the LHC  formulation of this paper or to the non-chiral $\cN=2$ formulation of $\cN=4$ SYM.

We present the LHC formulation of  $\cN=4$ SYM in Sects.~\ref{s2} and \ref{s3}. After recalling some basic definitions and properties of Euclidean LHC superspace, we introduce the two dynamical superfields $A^+_\da(x, \q^+,u)$ and $A^{++}(x, \q^+,u)$ as gauge connections. We then construct the supercurvatures and try to match them with the field content of  $\cN=4$ SYM. In this way we derive the two defining supercurvature constraints of the theory. As usual in  $\cN=4$ SYM, these are also equations of motion. The next step is to find an action from which the constraints follow. As mentioned above, the action consists of the Chern-Simons term first proposed in \cite{Sokatchev:1995nj} and of Zupnik's $\log \det$ term adapted to the LHC setup. 

An important aspect of the LHC/twistor formulation of  $\cN=4$ SYM is that only the chiral half ($Q$) of the $\cN=4$ supersymmetry is manifest. Indeed, the dynamical superfields  involve only chiral odd variables. As a consequence, the anti-chiral half ($\bQ$) remains on shell. This phenomenon was already observed in \cite{Sokatchev:1995nj} where the $\bQ$ transformation rules for $A^+_\da$ and $A^{++}$ were found. The addition of the $\log \det$ term modifies  the $\bQ$ transformations and the way  their algebra closes. We present a detailed discussion of the supersymmetry of the LHC  $\cN=4$ SYM action in Sect.~\ref{s4}. We use these results in the twin paper \cite{twin} for constructing the full (non-chiral) supermultiplet of the $\cN=4$ stress-energy tensor. 

The quantization of the theory is discussed in Sect.~\ref{s6}. We adopt the light-cone gauge (also called CSW or axial gauge) first proposed in \cite{Cachazo:2004kj} and also used in \cite{Boels:2006ir}. The gauge-fixing parameter  is an anti-chiral commuting spinor $\xi^\da$. This gauge has the nice feature that all the propagators are just delta functions in $x,u,\q$ space. This is very helpful in application to the Born-level correlators studied in \cite{twin} (and earlier in \cite{Chicherin:2014uca}). As shown there, the integrals at the vertices in the Feynman graphs are lifted and we obtain directly the expected rational expressions for the Born-level correlators. We give some simple examples of LHC supergraphs, postponing the main application of the Feynman rules to Ref.~\cite{twin}.

Appendix A summarizes  some properties of the harmonic/space-time distributions that we need for the propagators. 
In Appendix B we transform the propagators to momentum space.
Appendix C gives a dictionary between the twistor and harmonic languages, for the sake of the reader with a twistor background.

\section{Euclidean Lorentz  harmonic chiral superspace}\label{s2}

The Euclidean four-dimensional space with Lorentz group $SO(4) \sim
SU(2)_L\times SU(2)_R$ is parametrized by coordinates
$x^{\da\a}$, where $\a$  and $\da$  are spinor indices of
$SU(2)_L$ and $SU(2)_R$, respectively. The $\cN$-extended
superspace has coordinates
\begin{equation}\label{1}
x^{\da\a}\,,  \  \q^{\a A}\,, \ \bq^\da_A \,,
\end{equation}
where $A=1,\ldots,\cN$ are co- or contravariant indices of the automorphism group
$U(\cN)$ (or $SU(4)$ in the case $\cN=4$).  In it one can realize $\cN$-extended
supersymmetry in the following way\footnote{To simplify the formulae we skip the imaginary unit  in the supersymmetry variation. Consequently, our translation generator $P=\pa_x$ is anti-hermitian. If needed, the imaginary unit can easily be reinserted.  }
\begin{equation}\label{1'}
\delta x^{\da\a} = -{1\over 2}(\bep^\da_A\q^{\a A}
+ \ep^{\a A}\bq^\da_A)\,,
\ \ \delta\bq^\da_A = \bep^\da_A\,,  \ \ \delta\q^{\a A}
= \ep^{\a A} \,.   \end{equation}
The corresponding algebra of supercovariant derivatives is
\begin{equation}\label{0}
\{D_{\a A},  D_{\b B}\}=0\,,
\quad \{\bar D^A_\da,  \bar D^B_\db\} =0\,,  \quad \{\bar D^A_\da,  D_{\b B}\} = \delta^A_B \pa_{\b\da} \,,
\end{equation}
where 
\begin{align}\label{DDbar}
D_{\a A} = \pa_{\a A} + \frac1{2} \bq^\da_A \pa_{\a\da}\,, \qquad \bar D^A_\da = \bar \pa^A_\da + \frac1{2} \q^{\a A} \pa_{\a\da}
\end{align}
and $\pa_{\a A}= \pa/\pa \q^{\a A}\,, \ \bar \pa^A_\da= \pa/\pa\bq^\da_A\,, \  \pa_{\a\da}= \pa/\pa x^{\da\a}$.

We choose to `harmonize' half of the
Lorentz group,  e.g.,  the factor $SU(2)_L$.\footnote{LHs of this type were first used in \cite{Devchand:1992st,Devchand:1993ba,Devchand:1995gp} for solving the self-dual SYM constraints.}  We introduce 
harmonic variables $u^{\pm \a}$ defined as two $SU(2)_L$ spinors forming
an $SU(2)_L$ matrix:
\begin{equation}\label{2}
u^{\pm \a} \in SU(2)_L \ : \qquad \left\{ \begin{array}{ll} u^{+\a} u^-_{\a} \equiv  u^{+\a} \ep_{\a\b} u^{-\b}= 1 \\ \\ (u^{+\a})^* = u^-_\a\,, \quad (u^+_\a)^* = - u^{-\a} \end{array} \right.  
\end{equation}
(the $SU(2)_L$ spinor indices are raised  and lowered with the
Levi-Civita tensor, $\ep_{12}=-\ep^{12}=1$).
The index $\pm$ refers to the charge of these variables with respect to
 $U(1)_L \subset SU(2)_L$.  Thus, the harmonic
variables defined in this way describe the compact coset
$S^2 \sim SU(2)_L/U(1)_L$.  We are going to apply to them the 
rules of harmonic calculus \cite{Galperin:1984av}.  Here we give a short summary (see Appendix \ref{Hd} for more detail).  

Harmonic functions are defined by their harmonic expansion on $S^2$
\begin{equation}\label{3}
f^{(q)}(u) = \sum^\infty_{n=0} f^{\a_1 \ldots \a_{2n+q}}
u^+_{(\a_1} \ldots u^+_{\a_{n+q}} u^-_{\a_{n+q+1}}
\ldots u^-_{\a_{2n+q})} \,.
\end{equation}
By definition, they are homogeneous under the action of $U(1)_L$, i.e.,
they carry a certain charge $q$ (in (\ref{3}) $q\geq 0$).   From (\ref{3})
it is clear that the harmonic functions are collections of infinitely many finite-dimensional 
irreducible representations of $SU(2)_L$ (multispinors). 

The differential operators compatible with the defining constraint
(\ref{2}) are the covariant harmonic derivatives
\begin{align}\label{4}
& \pa^{++}= u^{+\a}{\pa\over\pa u^{-\a}} \ : \ \ \
\pa^{++}u^{+\a} =0\,,\  \ \pa^{++} u^{-\a} = u^{+\a}\nt
&\pa^{--}= u^{-\a}{\pa\over\pa u^{+\a}} \ : \ \ \
\pa^{--}u^{+\a} =u^{-\a}\,,\  \ \pa^{--} u^{-\a} = 0\,.
\end{align}
Together   with the charge operator $\pa^0$ ($\pa^0 u^{\pm\a}= \pm u^{\pm\a}$) the harmonic derivatives  form the algebra of $SU(2)$ realized on the indices $\pm$ of the harmonics,
\begin{align}\label{2.7}
[\pa^{++}, \pa^{--}] = \pa^0\,.
\end{align} 
The derivatives $\pa^{++}$ and $\pa^{--}$ have the meaning of the raising and lowering operators of $SU(2)_L$,  while $\pa^0$ counts the $U(1)_L$ charge of the harmonic functions, $\pa^0f^{(q)}(u) = q f^{(q)}(u)$. The restriction to functions on $SU(2)$ with definite charge gives a particular realization of the harmonic coset $SU(2)_L/U(1)_L$.

A direct consequence of the above definitions is the following lemma:
\begin{equation}\label{5}\pa^{++}f^{(q)}(u) = 0 \ \ \Rightarrow \
\ \left\{ \begin{array}{ll} f^{(q)}(u) = 0\,,  \ \ q<0 \\ \\ f^{(q)}(u) =
f^{\a_1 \ldots \a_{q}} u^+_{\a_1} \ldots u^+_{\a_{q}}
\,, \ \ q\geq 0 \end{array} \right.  \,. \end{equation}
This condition can be interpreted as defining the highest weight of a finite-dimensional unitary irreducible representation of $SU(2)$ of spin $q/2$. This irrep is carried by the totally symmetric tensor (multispinor) coefficient $f^{\a_1 \ldots \a_{q}}$.

Finally,  harmonic integration amounts to projecting out the singlet part of
a chargeless integrand,  according to the  rule
\begin{equation}\label{6}
\int du\; f^{(q)}(u) =  \left\{ \begin{array}{ll}
0,  \ q \neq 0 \\ f_{\rm singlet} ,  \ q= 0 \end{array} \right.
\,.
\end{equation}
This integration rule is designed to give an $SU(2)$ invariant result. It
is compatible with integration by parts for the harmonic
derivative $\pa^{++}$,
\begin{align}\label{2.10}
\int du\; \pa^{++} f^{(-2)}(u) =  0\,,
\end{align}
due to the absence  of a singlet term in the harmonic expansion of the function $ f^{(-2)}(u) $. 

Now, let us come back to the superspace \p{1}. The structure of the algebra \p{0} admits realizations of the $\cN$-extended supersymmetry algebra in subspaces of the
 superspace involving only part of the Grassmann variables
$\q$. One of them is the chiral superspace which does not contain the
variables $\bq^\da_A$. It is characterized by the coordinate shift
\begin{equation}\label{CB}
\mbox{Chiral basis:} \ \ \ x^{\da\a}_L \ = \ x^{\da\a}
-{1\over 2}
\bq^\da_A \q^{\a A} \,.
\end{equation}
In the chiral superspace the supersymmetry transformations are
\begin{equation}\label{11''}
\delta x^{\a \da}_L = - \bep^\da_A \q^{\a A} \,, \quad
\delta \q^{\a A} = \ep^{\a A}\,, \quad
\delta \bq^{\da}_A =\bep^{\da}_A  \,.
\end{equation}
In this basis the antichiral covariant derivative becomes $\bar D^A_\da =  \bar \pa^A_\da$, therefore   the \emph{chiral} superfields
defined by the constraint
\begin{equation}\label{chiral}
\bar D^A_\da \Phi = 0  \ \ \ \Rightarrow \ \ \
\Phi=\Phi(x^{\da\a}_L,\q^{\a A}, u)
\end{equation}
do not depend on $\bq^\da_A$.

The LHs $u^\pm$ allow us to make a step further and define \emph{chiral-analytic}  `semi-superfields' which depend only on one quarter of the Grassmann variables, 
\begin{align}\label{chan}
\Phi(x^{\da\a}_L,\q^{+ A}, u)\,, \qquad {\rm where} \ \q^{+ A} = u^+_{\a} \q^{\a A}\,.
\end{align}
We call such objects   `semi-superfields' because such an extreme shortening of a superfield is not possible without putting it on shell, if we insist on maintaining the full supersymmetry manifest. Indeed, 
in addition to the chirality constraint \p{chiral} it should satisfy the \emph{Lorentz analyticity} shortening condition\footnote{The name comes from the analogy with the Grassmann analytic superfields introduced in \cite{Galperin:1984av} for the off-shell formulation of theories with $\cN=2$ supersymmetry. There one uses $SU(2)$ harmonics $u^\pm_A$ to projects the R symmetry indices of $\q^{\a A}$ and $\bq^{\da A}$. The analytic superfields depend only on $\q^{\a+},\bq^{\da +}$ (see Sect.~\ref{n2} for more detail). } 
\begin{align}\label{analytic}
D^+_A \Phi=0\,, \qquad {\rm where} \ D^+_A = u^+_\a D^\a_A = \pa^+_A + \bq_{\da A} \pa^{+\da}\,.   \end{align}
Here and it what follows we use the harmonic projected spinor and space-time  derivatives
\begin{align}\notag
\pa^\pm_A = u^{\pm \a} \frac{\pa}{\pa \q^{\a A}}\,, \ \pa^{\pm}_\da = u^{\pm \a} \frac{\pa}{\pa x^{\da\a}_L}\,.
\end{align}
The two constraints and the algebra \p{0} imply $\{\bar D^A_\da, D^+_B\} \Phi = -\delta^A_B \partial^+_\da \Phi = 0 $, which puts the superfield on shell, since $\Box \Phi = 4\partial^{+\da}\partial^-_\da
\Phi = 0 $. What we really mean by the term `semi-superfield' is that we have set $\bq=0$ by hand, thus dropping the space-time derivative term in  \p{analytic}. Of course, this brakes  the manifest $\bQ$  supersymmetry. As mentioned above, this half of supersymmetry will be realized in a non-linear manner on the semi-superfields \p{chan} (see Sect.~\ref{s4}). However, our formulation will maintain the $Q$ half manifest. In the absence of $\bq$  we cannot tell the difference between the chiral basis \p{CB} and the original one, so in the sequel we will drop the subscript of $x_L$.

\section{Lorentz  harmonic chiral formulation of $\cN=4$ SYM }\label{s3}

In this Section we develop the classical $\cN=4$ SYM theory in LHC superspace. We introduce the two dynamical semi-superfields as the gauge connections for the derivatives $\pa^+_\da$ and $\pa^{++}$. From  them we construct the supercurvatures and try to match them with the physical fields. In this way we derive the two defining constraints of the theory, equivalent to the field equations. We then find a superspace action from which these equations follow. The action consists of a Chern-Simons term for the self-dual part of the equations and a log det (Zupnik) term for the deviation from self-duality. We display the ordinary component field content of the theory in the non-supersymmetric Wess-Zumino gauge.

\subsection{Dynamical superfields as gauge super-connections}\label{s3.1}

We start by introducing the main objects of the theory, two chiral-analytic  `semi-superfields' 
depending  only on a quarter of the Grassmann variables:
\begin{equation}\label{09}
A^+_\da = A^+_\da (x,\q^+,u) \,, \qquad A^{++} = A^{++}(x,\q^+,u) \,.
\end{equation}
They are defined as the gauge connections for two covariant derivates, 
\begin{align}\label{3.3}
\nabla^+_\da =  \pa^+_\da + A^+_\da \,, \qquad \nabla^{++} = \pa^{++} + A^{++}\,,
\end{align}
with respect to a gauge group with  a {chiral-analytic parameter} 
\begin{equation}\label{21} \delta A^+_{\da} = \pa^+_\da \Lambda
+ [A^+_{\da}, \Lambda]\,,  \ \ \ \delta A^{++} = \pa^{++} \Lambda
+ [A^{++}, \Lambda]\,, \qquad \Lambda=\Lambda(x,\q^+,u) \,.
\end{equation}
The commutators of these gauge covariant derivatives  generate super-curvatures:
\begin{align}\label{3.5}
&[\nabla^+_\da,\nabla^{++}]=\pa^+_\da A^{++} - \pa^{++} A^+_\da + [A^+_\da,
A^{++}] =: W^{+3}_\da\nt
&[\nabla^+_\da,\nabla^+_\db]=\pa^+_\da A^+_\db - \pa^+_\db A^+_\da + [A^+_\da,
A^+_\db]  =: \ep_{\da\db} W^{++}\,.
\end{align}
Following the standard logic in supersymmetric gauge theories, we should try to identify the physical components of the $\cN=4$ SYM multiplet with the super-curvatures evaluated at $\q=0$. Thus, we immediately see that $W^{+3}_\da = W^{\a\b\gamma}_\da u^+_\a u^+_\b u^+_\gamma + \ldots$ is a bosonic curvature of dimension 1 (the dimension of the space-time derivative $\pa_{\a\da}$; the harmonic variables and derivatives are dimensionless), carrying Lorentz spin $(3/2,1/2)$. Such a field has no match in the $\cN=4$ SYM multiplet, so it should be set to zero. We arrive at
\begin{align}\label{3.6}
{\rm Constraint\ I:} \qquad W^{+3}_\da =\pa^+_\da A^{++} - \pa^{++} A^+_\da + [A^+_\da,
A^{++}]= 0\,.
\end{align} 
This is an example of a curvature constraint typical for all supersymmetric gauge theories. 

The other super-curvature in \p{3.5}, $W^{++} = F^{\a\b}(x) u^+_\a u^+_\b + \ldots$,  can be identified with the self-dual half of the Yang-Mills curvature, $F^{\a\b} = \sigma^{\a\b}_{\mu\nu} F^{\mu\nu}$. Indeed, it has the right dimension 2 and Lorentz spin $(1,0)$, and is made of the gauge superfield $A^+_\da = \cA_{\a\da} (x) u^{+\a} + \ldots$, whose first component plays the role of the usual gauge field (gluon). 

The next question is where to find the six scalars of the $\cN=4$ SYM multiplet. In the standard approach (see, e.g., \cite{Sohnius:1978wk}) to $\cN=4$ SYM they are identified with  the super-curvature $W_{AB}=-W_{BA}$, which appears in the anticommutator of two chiral spinor covariant derivatives,
\begin{align}\label{3.7}
\{\nabla^\a_A , \nabla^\b_B\} = \ep^{\a\b} W_{AB}\,.
\end{align}
The absence of a curvature symmetric in the Lorentz indices $\a,\b$ on the right-hand side is the defining constraint of $\cN=4$ SYM in this standard formulation.\footnote{In a space-time with Minkowski signature another constraint is needed, the reality condition $W_{AB} = \frac1{2} \ep_{ABCD} \bar W^{CD}$. This point is discussed in Sect.~\ref{reality}.} In our LHC approach we can construct the curvature $W_{AB}$ directly in terms of the harmonic gauge connection $A^{++}$. The construction goes through several steps. First of all, we notice that, with the  gauge parameter from \p{21} satisfying the analyticity condition $\pa^+_A \Lambda=0$, the harmonic projected spinor derivative $\pa^+_A$ needs no gauge connection. This suggests to project relation \p{3.7} with the harmonics $u^-_\a u^+_\b$:
\begin{align}\label{3.8}
\{\nabla^-_A , \pa^+_B\} = \pa^+_B A^-_A=  W_{AB}\,,
\end{align}
where we have used the defining property \p{2}. Notice that the right-hand side in this relation is antisymmetric in $AB$ while the left-hand side is not. This implies the constraint $\pa^+_{(B} A^-_{A)}=0$, which we solve explicitly by constructing the gauge super-connection $A^-_A$. This is done by covariantizing the obvious commutation relation
\begin{align}\label{3.9}
[\pa^{--}, \pa^+_A] = \pa^-_A \ \Rightarrow \ [\nabla^{--}, \pa^+_A] = \nabla^-_A \ \Rightarrow \ A^-_A = - \pa^+_A A^{--}\,.
\end{align}
In this way we have introduced yet another gauge super-connection, for the second harmonic derivative (the lowering operator of $SU(2)_L$):
\begin{align}\label{a--}
\nabla^{--} = \pa^{--} + A^{--}\,, \qquad \delta A^{--} = \pa^{--}\Lambda + [A^{--},\Lambda]\,. 
\end{align} 
To determine it we covariantize the $SU(2)$ algebraic relation \p{2.7},
\begin{align}\label{3.10}
[\nabla^{++}, \nabla^{--}] = \pa^0\,.
\end{align}
Notice that the charge operator $\pa^0$ needs no gauge connection since the gauge parameter is a harmonic function of charge zero, $\pa^0 \Lambda=0$. The commutation relation \p{3.10} implies a harmonic differential equation for the unknown harmonic connection $A^{--}$ in terms of the given $A^{++}$:
\begin{align}\label{3.11}
\pa^{++} A^{--} - \pa^{--} A^{++} + [A^{++},A^{--}]=0\,.
\end{align}
Having in mind our definition \p{3} of the harmonic functions as expansions in terms of finite-dimensional irreps of $SU(2)$, we can easily convince ourselves that this equation has a unique solution for $A^{--}$. Indeed, the differential operator $\pa^{++} $ is invertible on a harmonic function with negative charge due to the first property on the right-hand side of \p{5}. The details of the solution are  given below in Sect.~\ref{s3.3}. Here we only remark that while the gauge connection $A^{++}(x,\q^+,u)$ is chiral-analytic by definition, the new one depends on all chiral Grassmann variables, $A^{--}(x, \q^\a, u)$. 

To summarize the above procedure, we start with the chiral-analytic super-connection $A^{++}(x,\q^+,u)$, from it we determine $A^{--}(x, \q^\a, u)$, which we then substitute in \p{3.9} and \p{3.8}. The result is the following manifestly antisymmetric expression for the super-curvature
\begin{align}\label{3.12}
W_{AB} = \pa^+_A \pa^+_B A^{--}\,,
\end{align}
which contains the six scalars of the $\cN=4$ SYM multiplet, $W_{AB} = \phi_{AB}(x) + \ldots\;$.  Let us check that this curvature is indeed gauge covariant:
\begin{align}\label{313}
\delta W_{AB} = \pa^+_A \pa^+_B \delta A^{--} = \pa^+_A \pa^+_B (\pa^{--}\Lambda + [A^{--},\Lambda]) = [W_{AB},\Lambda]
\end{align}
because $\pa^+_A \pa^+_B \pa^{--}\Lambda=0$ due to the analyticity $\pa^+_A\Lambda=0$ and the (anti)commutation relations $[\pa^{--}, \pa^+_A] = \pa^-_A$ and $\{\pa^+_A, \pa^-_B\}=0$. Further, it is (covariantly) harmonic independent, 
\begin{align}\label{314}
\nabla^{++} W_{AB}  = \pa^+_A \pa^+_B \nabla^{++} A^{--}= \pa^+_A \pa^+_B \pa^{--} A^{++}=0\,,  
\end{align} 
where we have used \p{3.11} and  the analyticity of $A^{++}$. 

From the curvature \p{3.12} with the gauge transformation \p{313} we can construct other covariant curvatures by applying one or two spinor derivatives $\pa^+_A$. In particular,  
\begin{align}\label{3.13}
{\mathbb W}^{++} \equiv \frac1{4!}\ep^{ABCD} \pa^+_A \pa^+_B W_{CD} = (\pa^+_A)^4  A^{--} = G^{\a\b}(x) u^+_\a u^+_\b + \ldots \ .
\end{align} 
The component field $G^{\a\b}$ has the same characteristics (dimension 2, Lorentz spin $(1,0)$, R-symmetry singlet)  as the self-dual part of the gauge curvature $F^{\a\b}$. The latter, as we pointed out earlier, is the lowest component of the super-curvature $W^{++}$ from \p{3.5}. Notice that the new curvature ${\mathbb W}^{++}$ is built entirely from the gauge super-connection $A^{++}$ (via $A^{--}$ and the differential equation \p{3.11}, see Sect.~\ref{s3.3} for the detail), while the old one $W^{++}$ is made from the connection $A^+_\da$.  We should not let two similar components  coexist, otherwise we would double the $\cN=4$ SYM multiplet. So,  we impose the identification
\begin{align}\label{3.14}
{\rm Constraint\ II:} \qquad W^{++}= \om {\mathbb W}^{++}\,,
\end{align}
or in terms of the gauge connections,
\begin{align}\label{3.14'}
\pa^{+\da} A^+_\da + A^{+\da} A^+_\da = \om(\pa^+)^4 A^{--}\,.
\end{align}
Notice that this is a non-trivial relation between the connections $A^+_\da$ and $A^{++}$ appearing on the left-hand and right-hand side, respectively. 

We have left the proportionality constant $\om$ on the right-hand side of \p{3.14} arbitrary. In the twistor literature (Ref.~\cite{Witten:2003nn} and thereafter) this constant is treated as a perturbative parameter (coupling). We prefer to introduce the gauge coupling $g$ in the traditional way by rescaling the gauge connections $A \to g A$ and then dividing the action by $g^2$ (see the end of Sect.~\ref{action}). Nevertheless, it is helpful to keep $\om$ as a parameter which measures the `deviation from self-duality'. Indeed, if we set $\om=0$ we obtain a stronger version of the constraint \p{3.14},
\begin{align}\label{sd}
W^{++}=0\,,
\end{align}
which yields the self-dual YM equation $F_{\a\b}=0$, see \cite{Sokatchev:1995nj} for details. The relaxed form of the constraint  \p{3.14} simply identifies the components $F_{\a\b}= \om G_{\a\b}$, which is part of the first-order (Lagrange multiplier) formulation of the YM equations (see eq.~\p{3.42} below). 

The fermions (gluinos) from the $\cN=4$ SYM multiplet can be identified with spinor derivatives of the above curvatures. Hitting \p{3.12} with another derivative $\pa^+_C$ we obtain the fermionic curvature
\begin{align}\notag
W^{+A} = \frac1{3} \ep^{ABCD}   \pa^+_B \pa^+_C \pa^+_D A^{--} = \psi^{A}_\a (x) u^{+\a} + \ldots\ .
\end{align}
It starts with the chiral gluino field $\psi^{\a A}$. The anti-chiral gluino can be found in the curvature $[\nabla^+_\da, \nabla^-_A] = \bar\psi_{\da A}(x) + \ldots$ (see also the discussion of the components in the Wess-Zumino gauge in Sect.~\ref{s3.4}). 

In conclusion, we have given sufficient evidence that the $\cN=4$ SYM multiplet is indeed described by the set of chiral-analytic gauge super-connections \p{09}, provided that we impose two constraints on the super-curvatures, eqs.~\p{3.6} and \p{3.14}. Being necessary for the correct identification of the super-curvatures with the components of the $\cN=4$ SYM multiplet, these constraints turn out to also impose the equations of motion for  $A^+_\da$ and $A^{++}$. This will become clear in Sect.~\ref{action} where we discuss the $\cN=4$ SYM action. 

\subsubsection{Reality properties}\label{reality}

One of the constraints  defining $\cN=4$ SYM is the reality condition on the six scalars, $\phi_{AB} = \frac1{2} \ep_{ABCD} \bar\phi^{CD}$. Without restricting the super-curvature in \p{3.7} appropriately the content of the multiplet will be doubled (complexified).  In a superspace with Minkowski signature $(1,3)$ the Grassmann variables $\q^\a$ transform under the fundamental representation of $SL(2,C)$ and their conjugates $\bq_{\da}$ under the inequivalent anti-fundamental. So, imposing a reality condition on the super-curvature \p{3.12} would require turning on $\bq$, something we do not wish to do in our chiral formulation. An easy way out  is to change the signature to $(2,2)$. There the Lorentz groups is $SL(2,R)\times SL(2,R)$ and the spinors $\q^\a$ as well as the super-connections can be real. This choice was made in \cite{Siegel:1992xp} and \cite{Sokatchev:1995nj}, however, it requires using LHs on the non-compact coset $SL(2,R)/R$. Then the harmonic expansions and integration become problematic and have to be treated formally. Here we prefer to keep the Euclidean signature $(4,0)$, in order to have a well-defined harmonic analysis on $S^2 \sim SU(2)/U(1)$. A reality condition on our chiral superfields can be imposed if we restrict the R-symmetry group $SU(4)$ of $\cN=4$ SYM to its subgroup $Sp(4)$. It has an invariant symplectic tensor $\Omega_{AB}$ and one can impose the pseudo-Majorana condition (see, e.g., \cite{Kugo:1982bn})  $\bar\q_{\a A} = \ep_{\a\b} \q^{\b B} \Omega_{BA}$. This is the option that we adopt in the present paper. 

In practice, as we show in Sect.~\ref{s3.4}, the component field content of our super-connections exactly matches that of the $\cN=4$ multiplet (a gluon, 4 chiral and 4 antichiral gluinos and 6 scalars), with the above reality condition implied.

\subsection{Action for ${\cal N}=4$ SYM in Lorentz harmonic chiral superspace} \label{action}

Our next task is to derive the two constraints \p{3.6} and \p{3.14} from an action principle. Let us start with  \p{3.6}. We remark that the super-curvature $W^{+3}_\da(x,\q^+,u)$ is chiral-analytic by construction, just like the gauge super-connections.  It seems logical to use $A^+_\da$ as a Lagrange multiplier for the constraint  \p{3.6} in an action of the type
\begin{align}\label{3.15}
\int d^4x du d^4\q^+\;  {\rm Tr} \left( A^{+\da} W^{+3}_\da\right)\,.
\end{align}
Notice that the integrand carries Lorentz charge $+4$ which cancels that of the Grassmann measure. This is necessary  for the harmonic integral to give a non-vanishing result. But there is an obvious problem with this action term  -- it is not gauge invariant. 
This can be repaired by noticing that  \p{3.15} involves three non-Abelian gauge super-connections, $A^{++}$ and $A^+_\da$. So, we might try  a gauge invariant action of the Chern-Simons type\footnote{The action \p{3.18} was proposed for the first time in  \cite{Sokatchev:1995nj} as a superspace formulation of $\cN=4$  self-dual  SYM, following the component field version of Siegel \cite{Siegel:1992xp}. An equivalent twistor reformulation of this Chern-Simons self-dual action appeared more recently in \cite{Witten:2003nn} and \cite{Boels:2006ir}.}
 \begin{equation}\label{3.18}
S_{\rm CS} = \int d^4x du d^4\q^+\; L_{\rm CS}(x,\q^+,u)\,,
\end{equation}
where
\begin{align}\label{CS}
L_{\rm CS}(x,\q^+,u) =  \tr\; (A^{++}\pa^{+\da}A^+_\da-
{1\over 2} A^{+\da}\pa^{++} A^+_\da + A^{++} A^{+\da} A^+_\da) \,.
\end{align}
This action term is gauge invariant up to total derivatives, including the harmonic derivative $\pa^{++}$. The property \p{2.10} of the harmonic integral makes it possible. 
Now we have a gauge invariant action, but we have created a new problem: the variation with respect to $A^{++}$ yields the self-dual field equation \p{sd}. So, we have to add yet another term to the action which will give rise to the weaker constraint \p{3.14}. The new term should be built from the super-connection $A^{++}$, it should be gauge invariant and its variation should supply the right-hand side of eq.~\p{3.14}. How to construct such an invariant?

The answer to this question has been known for almost 30 years, but in a different context. The formulation of $\cN=2$ SYM in $\cN=2$ harmonic superspace first proposed in \cite{Galperin:1984av} makes use of harmonics on the R-symmetry group $SU(2)$ (and not on the Lorentz group as we do now). The dynamical superfield  of $\cN=2$ SYM is the gauge prepotential $V^{++}$ which is the super-connection for the harmonic derivative $\pa^{++}$, exactly as in \p{3.3}. The gauge invariant action is non-polynomial in $V^{++}$. The first few terms in its expansion were found in \cite{Galperin:1985bj}, but the complete expansion, as well as a very compact form of the action were proposed by Zupnik in  \cite{Zupnik:1987vm}. In our  present context it is straightforward to adapt Zupnik's construction of the invariant. Unlike the action term \p{3.18}, which is given by an integral over the chiral-analytic superspace, the new term is an integral over the full chiral superspace:
\begin{align}\label{sint}
S_{\rm Z}  = \int d^4x d^8\q\; L_{\rm Z}(x, \q)\, .
\end{align}
This part of the action involves non-liner (interaction) terms in $A^{++}$.\footnote{The cubic term in the Chern-Simons action  \p{3.18} is an interaction between $A^{++}$ and $A^+_\da$ which can be eliminated in the light-cone gauge, see  Sect.~\ref{s6.1}. Thus, the true interaction is only in \p{sint}.} Its explicit form is
\begin{align}\label{lint}
L_{\rm Z}    = \om\, \tr\sum^\infty_{n=2}{(-1)^n\over n} \int du_1\ldots du_n\; {A^{++}(x, \q^+_1,u_1) \ldots A^{++}(x, \q^+_n,u_n) \over
(u^+_1u^+_2) \ldots (u^+_nu^+_1)}\,,
\end{align} 
where $\q^{+A}_i = \q^{\a A} (u_i)^+_{\a}$ with $i=1,\ldots,n$ and $(u^+_i u^+_j) \equiv u^{+\a}_i u^+_{j \a}$.  This action term is local in $(x,\q)$ space but non-local in the harmonic space (each copy of $A^{++}$ depends on its own harmonic variable). The properties of the action term \p{sint} are discussed in detail in Sect.~\ref{s3.3}. 

The Lagrangian density \p{lint} can be rewritten  in a compact form  by introducing an integral harmonic operator with the following kernel 
\begin{align}\notag
\left( \frac1{\pa^{++}} \nabla^{++}\right)_{ab}(u_1,u_2) = \delta_{ab} \delta(u_1,u_2) + \frac{f_{abc} A^{++}_c(x, \q^+_2,u_2)}{(u^+_1 u^+_2)}\,, 
\end{align}
where $f_{abc}$ is the structure constant of the gauge group. 
Then \p{lint} takes the form \cite{Zupnik:1987vm} \footnote{The type of gauge invariant action \p{sint}, \p{logdet} is close in spirit to Witten's formulation of the Wess-Zumino model \cite{Witten:1983ar}, although the context is quite different. An on-shell version of the invariant was considered in \cite{Abe:2004ep} for the purpose of reproducing the MHV gluon scattering amplitudes. }
\begin{align}\label{logdet}
 L_{\rm Z}  = \om\,  \log\det \left( \frac1{\pa^{++}} \nabla^{++}\right) .
\end{align}

The main claim we are making  now is that the sum of the two action terms \p{3.18} and \p{sint} is equivalent to the full $\cN=4$ SYM action,
\begin{align}\label{N4}
S_{\cN=4} = S_{\rm CS} + S_{\rm Z}  \,.
\end{align}
In Sect.~\ref{s3.4} we examine the component field content of the theory and show that the action \p{N4} is indeed equivalent to the $\cN=4$ SYM action in a first-order formulation for the gauge field. 

\subsubsection{Coupling constant}

In the above we have not displayed the gauge coupling constant. To see it, we need to rescale both gauge super-connections, $A \to g A$ and divide the action by $g^2$:
\begin{align}\label{3.23}
\frac1{g^2}   S_{\cN=4} (gA) &=  \tr\; \int d^4x du d^4\q^+\;  (A^{++}\pa^{+\da}A^+_\da-
{1\over 2} A^{+\da}\pa^{++} A^+_\da)\nt
& -  \frac{\om}{2}\,   \tr\; \int d^4x d^8\q du_1 du_2\; {A^{++}(x, \q^+_1,u_1) A^{++}(x, \q^+_2,u_2) \over
(u^+_1u^+_2)^2} + O(g)\,.  
\end{align}
This definition associates the gauge coupling $g$ with the non-Abelian color structure. Every color commutator is accompanied by a factor of $g$.  The bilinear terms shown in \p{3.23}  exist also in the Abelian (or free, or $g=0$) theory. The interaction terms include the cubic term from the Chern-Simons Lagrangian \p{CS} and all the non-linear terms with $n>2$ from \p{lint}. We can set the gauge coupling constant $g=0$ and still have both terms in \p{N4}. This is in contrast with the philosophy of the twistor formulation of \cite{Witten:2003nn}, \cite{Mason:2005zm}, \cite{Boels:2006ir}, where the full SYM action is treated as a `perturbation around a self-dual background', $S_{\cN=4} = S_{\rm CS} +  \om S_{\rm Z}$, with the second term proportional to $\om$ (recall \p{lint}). In other words, there the `free theory' ($\om=0$) is self-dual SYM, not the usual Abelian gauge theory. When quantizing the theory, following \cite{Boels:2006ir} we prefer to treat the bilinear term in the second line in \p{3.23} as an `interaction term' giving rise to a `bivalent vertex', while the propagators will be determined from the terms in the first line.  The advantage of this treatment is the simpler form of the propagators  (see Sect.~\ref{s6} for detail).  

This is one  of the reasons why we keep the constant $\om$ in \p{lint} arbitrary. Another reason is that it helps us keep track of the non-polynomial modification terms in the $\bQ$ supersymmetry transformations (see Sect.~\ref{s4.2}). This will be important in \cite{twin} where we use $\bQ$ to reconstruct the full non-chiral stress-tensor multiplet. There we show that the terms proportional to $\om$ do not contribute to the correlation functions at Born level. If needed, $\om$ can be fixed at some conventional (non-zero) value by computing a correlator and comparing the result to another calculation based on standard Feynman rules.

\subsection{Gauge invariance of the $\cN=4$ SYM action}\label{s3.3}

As mentioned earlier, the CS action term \p{3.18} is gauge invariant in the standard way, up to total derivatives. The invariance of the interaction term \p{sint} is less obvious. Here we adapt the argument originally given by Zupnik in \cite{Zupnik:1987vm} (its analog can also be found in the twistor literature).

Let us first compute the variation of $L_{\rm Z}$ with respect to the superfield $A^{++}$:
\begin{align}\label{444}
\delta L_{\rm Z} &=  \om\, \tr\sum^\infty_{n=2} (-1)^n \int du_1\ldots du_n\; {\delta A^{++}(x, \q^+_1,u_1) \ldots A^{++}(x, \q^+_n,u_n) \over
(u^+_1u^+_2) \ldots (u^+_nu^+_1)}\nt
  &= -\om\, \tr \int  du \; \delta A^{++}(x,\q^+,u) A^{--}(x, \q,u)\,.
\end{align}
Here
\begin{align}\label{448}
A^{--}(x,\q,u) =  \sum^\infty_{n=1} (-1)^n\int du_1\ldots du_n\; { 
A^{++}(1) \ldots A^{++}(n) \over (u^+u^+_1)(u^+_1u^+_2) \ldots
(u^+_nu^+)}
\end{align} 
{with $A^{++}(k) = A^{++}(x,\theta\cdot u^+_k,u_k)$,}
is the gauge super-connection for the harmonic derivative $\pa^{--}$ defined in \p{a--}. Unlike the connection $A^{++}$, this one is not chiral-analytic but only chiral, i.e. it depends on the full chiral $\q^\a$. Let us check that it satisfies the defining differential equation \p{3.11}.

For $n > 1$ the derivative $\pa^{++}$ acting on the $n$-th term in
(\ref{448}) gives (see (\ref{4.9.1}))
\begin{eqnarray}
\pa^{++}A^{--}_{(n)} &\equiv& (-1)^n \pa^{++}\int du_1\ldots du_n\; {  A^{++}(1)
\ldots A^{++}(n) \over (u^+u^+_1) \ldots (u^+_nu^+)} \nonumber
\\ &=& (-1)^n\int du_1\ldots du_n\;   A^{++}(1) \ldots
A^{++}(n)\nonumber \\ &&\ \times \left[  {\delta(u,u_1)\over (u^+_1u^+_2)
\ldots (u^+_nu^+)}- {\delta(u,u_n)\over (u^+u^+_1) \ldots
(u^+_{n-1}u^+_n)}\right] \nonumber \\ &=& -A^{++}A^{--}_{(n-1)} + A^{--}_{(n-1)}A^{++}\,, \qquad n > 1\,. \notag
\end{eqnarray}
For $n=1$ we {differentiate the harmonic distribution using \p{4.10.2} and \p{4.8.6b}, afterwards we  integrate the harmonic derivative by parts}
\begin{eqnarray}
\pa^{++}A^{--}_{(1)} &\equiv& \pa^{++}\int du_1\; {A^{++}(u_1)\over (u^+u^+_1)^2} =
\int du_1\; \pa^{--}\delta^{(2,-2)} (u,u_1) A^{++}(u_1) \nonumber
\\ &=& \pa^{--}A^{++}\,.  \notag
\end{eqnarray}
Putting all of this together we see that \p{3.11} is indeed satisfied.

Now, consider a variation $\delta A^{++}$ corresponding to the gauge transformation \p{21},
\begin{align}\notag
\delta A^{++} = \pa^{++}\Lambda + [A^{++},\Lambda] = \nabla^{++}\Lambda\,.
\end{align}  
Under the trace in \p{444} we can integrate the covariant harmonic derivatives by parts and then use \p{3.11} in the form $\nabla^{++} A^{--}  = \pa^{--} A^{++}$ to obtain
\begin{align}\label{450}
\delta_\Lambda L_{\rm Z} =\om\, \tr \int  du \; \Lambda\, \pa^{--} A^{++} \,.
\end{align}
This variation does not vanish by itself. To make it vanish we need to act upon $\pa^{--} A^{++}$ with at least two  spinor derivatives $\pa^+_A$. The mechanism is the same as in \p{313}.  In the action \p{sint} we have four such derivatives. Indeed, we can rewrite the integration measure in the form 
\begin{align}\notag
\int du d^4x d^8\q = \int du d^4x (\pa_\q)^8 = \int du d^4x (\pa^-)^4 (\pa^+)^4\,.
\end{align}
The four derivatives $(\pa^+)^4$ are more than sufficient to kill $\pa^{--} A^{++}$. 
In summary, the gauge invariance of $S_{\rm Z}$ is based on the following property of the gauge variation of $L_{\rm Z}$:
\begin{align}\label{3.32}
\delta_{\rm gauge} L_{\rm Z} =  \int du \Delta L\,, \qquad \pa^+_A \pa^+_B   \Delta L =0\,,
\end{align}
i.e. $ \Delta L$ is at most a linear function of $\q^{-A}$. 

Finally, let us perform a general variation of the action term \p{sint} with respect to the gauge super-connection $A^{++}$. Using \p{444}, we find 
\begin{align}\notag
\delta S_{\rm Z} &=-\om\, \tr \int du d^4x (\pa^-)^4 (\pa^+)^4 \; \left[ \delta A^{++}(x,\q^+,u) A^{--}(x, \q,u)\right]\nt
& =-\om\, \tr \int du d^4x d^4\q^+ \; \delta A^{++} (\pa^+)^4A^{--}\,,  \notag
\end{align}
hence (see \p{3.12} and \p{3.13})
\begin{align}\notag
\frac{\delta S_{\rm Z}}{\delta A^{++}} =-\om(\pa^+)^4A^{--} =-\om\, {\mathbb W}^{++}\,.
\end{align}
Recalling that the variation of the Chern-Simons term \p{3.18} produces the super-curvature $W^{++}$, we see that the variation of the full action \p{N4} yields the desired constraint \p{3.14}. In principle, one can perform a thorough analysis of the equations of motion \p{3.6} and \p{3.14} and obtain all the component field equations, but it is simpler to reveal the component content of the action itself.

\subsection{Component field content}\label{s3.4}

Let us explore the component content of the super-connections 
$A^+_\da(x,\q^+,u)$, $A^{++}(x,\q^+,u)$. We first consider the simplified example 
of ${\cal N}=1$ supersymmetry. The harmonic
connection has a very short Grassmann expansion:
\begin{equation}\label{-22}
A^{++} = a^{++}(x, u) + \q^{+}\sigma^+(x, u) \,.
\end{equation}
The fields in (\ref{-22}) are harmonic, i.e., they contain infinitely many
ordinary fields (recall (\ref{3})). However, we still have the gauge
transformations
(\ref{21}) with parameter
\begin{equation}\label{-22'}
\Lambda = \la(x, u) + \q^{+}\rho^-(x, u)\,.  \end{equation}
Let us compare the harmonic expansions (\ref{3}) in
(\ref{-22}) and (\ref{-22'}) (bosons only):
\begin{eqnarray}
&&a^{++}(x, u) = u^+_{(\a}u^+_{\b)} a^{\a\b}(x) +
u^+_{(\a}u^+_{\b} u^+_{\g}u^-_{\delta)}
a^{\a\b\g\delta}(x) + \ldots \,, \nonumber \\
&&\la(x, u) = \la(x) + u^+_{(\a}u^-_{\b)}
\la^{\a\b}(x) +
u^+_{(\a}u^+_{\b} u^-_{\g}u^-_{\delta)}
\la^{\a\b\g\delta}(x) + \ldots \,. \nonumber
\end{eqnarray}
Clearly, the parameter $\la(x, u)$ contains enough components to completely
gauge away the harmonic field $a^{++}(x, u)$ (note that the singlet part
$\la(x)$ in $\la(x, u)$ is not used in the process; it remains
non-fixed and plays the role of the ordinary gauge parameter). Similarly, the
parameter $\rho^-(x, u)$ can gauge away the entire field $\sigma^+(x, u)$.
Thus, we arrive at the following\footnote{We call the  gauges \p{-23} and \p{23} ``Wess-Zumino gauges" by analogy with the standard $\cN=1$ WZ gauge. The latter uses the entire freedom in the chiral superfield gauge parameter $\Lambda(x,\q)$  (and its conjugate) to gauge away as much as possible from the real gauge superpotential $V(x,\q,\bq)$. Only  the ordinary gauge parameter $\la(x)$ is left intact as it serves the usual component gauge field $A_\mu(x)$.  In the twistor approach of \cite{Mason:2005zm}, \cite{Boels:2006ir} this gauge is called `space-time' because it reveals the space-time field content of the theory, as opposed to the fields on  twistor space. In the Lorentz harmonic approach the notion of space-time is always present. The harmonic expansion on $S^2$ gives rise to infinitely many space-time auxiliary and gauge degrees of freedom.  The WZ gauges \p{-23} and \p{23}  eliminate the (infinite sets of) gauge degrees of freedom, leaving us with propagating and auxiliary fields only.}
\begin{equation}\label{-23}
\mbox{${\cal N}=1$ Wess-Zumino gauge:}\qquad A^{++}= 0\,.
\end{equation}

The other $\cN=1$ gauge connection has the expansion
\begin{equation}\label{-25}
A^{+}_\da = {\cal A}^{+}_\da(x, u) + \q^{+}
\bar\psi_\da(x, u) \,.
\end{equation}
The harmonic dependence in it can be eliminated with the help of the
constraint (\ref{3.6}).  Substituting (\ref{-23}) and (\ref{-25})
into \p{3.6}, we obtain $\pa^{++}A^+_\da = 0$, from where follow the
harmonic equations
\begin{eqnarray}
&&\pa^{++}{\cal A}^+_\da(x, u) = 0 \ \ \Rightarrow
{\cal A}^+_\da(x, u) = u^{+\a}{\cal A}_{\a\da}(x) \,, \nonumber
\\
&&\pa^{++}\bar\psi_{\da }(x, u) = 0 \ \ \Rightarrow
\bar\psi_\da(x, u) = \bar\psi_\da(x) \,. \label{hars}
\end{eqnarray}
These component fields can be identified with the gluon ${\cal A}_{\a\da}(x)$ and the anti-chiral gluino $\bar\psi_\da(x)$. The  $\cN=1$ SYM multiplet includes also a chiral gluino $\psi_\a(x)$, but there is no room for it in our LHC formulation. The reason is the use of chiral gauge connections. In such a superfield the span of Lorentz helicities is limited by the maximal power of the chiral odd variable $\q^{+A}$ with $A=1,\ldots,\cN$.  Only in the maximally supersymmetric case $\cN=4$ we can accommodate the entire self-conjugate SYM multiplet with helicities from $-1$ to $+1$. As we show below, half of them live in the gauge connection $A^{++}$, the other half in $A^+_\da$. 

The maximal case ${\cal N}=4$ follows the same pattern. Comparing the harmonic dependence of the components of $A^{++}$ with that of the gauge parameter $\Lambda$, we can fix the
\begin{equation}\label{23}
\mbox{${\cal N}=4$ WZ gauge:}\qquad A^{++}= (\q^+)^{2\, AB}\phi_{AB}(x) +
(\q^+)^3_A
u^{-\a}\psi^A_{\a}(x) + 3(\q^+)^4 u^{-\a}u^{-\b}
G_{\a\b}(x)\,,
\end{equation}
where we use the shorthand notation 
$$
(\q^+)^{2\, AB} = {1\over 2!}\q^{+A}\q^{+B},  \ (\q^+)^3_A  =
{1\over 3!} \ep_{ABCD}\q^{+B}\q^{+C}\q^{+D},  \
(\q^+)^4 = {1\over 4!}
\ep_{ABCD}\q^{+A}\q^{+B}\q^{+C}\q^{+D}
\,.  $$
By this we have exhausted the entire freedom in the gauge super-parameter but for its first component, $\Lambda(x,\q^+,u) = \la(x) + \ldots$, which plays the role of the ordinary  gauge parameter. 

The other gauge connection $A^{+}_\da$ has the expansion
\begin{equation}\label{25}
A^{+}_\da = {\cal A}^{+}_\da(x, u) + \q^{+A}
\bar\psi_{\da A}(x, u) +
(\q^+)^{2\, AB}B^-_{\da AB}(x, u) + (\q^+)^3_A \tau^{--A}_\da(x, u)
+ (\q^+)^4 C^{-3}_\da(x, u)\,.
\end{equation}
We recall that it appears only in the Chern-Simons action term \p{3.18}, the interaction term \p{sint} is made  of  $A^{++}$. So, let us first examine the component content of \p{3.18}.  Inserting the Wess-Zumino gauge (\ref{23}) for $A^{++}$ and the expansion (\ref{25}) of $A^+_{\da}$ into it and doing the Grassmann integral,  we obtain
\begin{eqnarray}
S_{\rm CS} &=& \int d^4x du \; \tr\;\left\{ {1\over 2} \phi^{AB} \nabla^{+\da}
B^-_{\da AB}
+ \psi^{-A} \nabla^{+\da} \bar\psi_{\da A} + 3 G^{--}{F}^{++}  - C^{-3\da}
\pa^{++}{\cal A}^+_\da \right. \nonumber \\
&&\left. \ \ \ \ \ - \tau^{--\da A} \pa^{++} \bar\psi_{\da A}
-{1\over 4} B^{-\da AB}\pa^{++}B^-_{\da AB} - \phi^{AB}
\bar\psi^\da_A
\bar\psi_{\da B}\right\} \,,  \label{32}
 \end{eqnarray}
where $\phi^{AB}=1/2\ep^{ABCD}\phi_{CD}$, $\nabla^{+\da}  =
\pa^{+\da}  + [{\cal A}^{+\da},  \cdot\ ]$, ${F}^{++} =
\pa^{+\da}
{\cal A}^+_\da + {\cal A}^{+\da}{\cal A}^+_\da$.
The fields $B,\, C, \,\tau$ are auxiliary and their variation gives rise to the
harmonic equations
$$
\pa^{++}{\cal A}^+_\da(x,u) = 0\,,  \ \ \pa^{++}
\bar\psi_{\da A}(x,u) =0\,,  \ \ \pa^{++}B^{-\da AB}(x,u)  =\nabla^{+\da} \phi^{AB}(x) \,,
$$
which allow us to eliminate the harmonic dependence of ${\cal A}^+_\da$
and $\bar\psi_{\da A}$ {(see eq.~\p{hars})} and to express $B$ in terms of $\phi$. The variation of \p{32} {with respect to} the physical fields { in $A^+_{\da}$} gives further harmonic equations, from which we can express the auxiliary  fields $\tau$ and $C$ in terms of the physical ones. The result is
\begin{eqnarray}
&& {\cal A}^+_\da(x, u) = u^{+\a}{\cal A}_{\a\da}(x)\,,
\qquad
\bar\psi_{\da A}(x, u) = \bar\psi_{\da A}(x)\,,  \nonumber \\
&& B^-_{\da AB}(x, u) = u^{-\a}\nabla_{\a\da}\phi_{AB}(x)\,,
\qquad \tau^{--A}_\da(x, u) = {1\over 2}u^{-\a}u^{-\b}
\nabla_{\a\da}\psi^A_{\b}(x)\,, \nonumber \\
&& C^{-3}_\da(x, u) = u^{-\a}u^{-\b}u^{-\g}
\nabla_{\a\da} G_{\b\g}(x) \,.  \label{59}
\end{eqnarray}
Putting this back in \p{32} and doing the harmonic integral (it just picks the Lorentz singlet part, see \p{6}), we arrive at the component action 
\begin{equation}
S_{\rm CS} = \int d^4x \; \tr\;\left\{ -{1\over 8} \nabla^{\da\a}\phi^{AB}
\nabla_{\a\da} \phi_{AB} +{1\over 2} \psi^A_{\a}
\nabla^{\da\a}
\bar\psi_{\da A} +  G^{\a\b} F_{\a\b}
 - \phi^{AB}\bar \psi^\da_A \bar\psi_{\da B}\right\} \,.  \label{33}
 \end{equation}

If we restrict ourselves to the Chern-Simons action term \p{3.18}, we recover the action for $\cN=4$  self-dual SYM, first given in \cite{Siegel:1992xp,Siegel:1992wd,Ketov:1992ix}.  The ${\cal N}=4$ self-dual multiplet
contains all the helicities from $+1$ (described by the self-dual field
${\cal A}$) down to $-1$ (the field $G$).  The latter serves as a  Lagrange multiplier for the
self-duality condition $F_{\a\b}=0$ on ${\cal A}_{\a\da}$. Similarly, the spinor
fields $\psi^A_{\a}$ and $\bar\psi_{\da A}$ form a Lagrangian pair. 

However, here we are not interested in the self-dual but  in the full $\cN=4$ SYM theory. The key step is to relax the self-duality constraint $F_{\a\b}=0$ following from \p{33} by adding a quadratic term in the Lagrange multiplier $G$,\footnote{In the twistor literature  the action \p{3.42}  is referred to as the Chalmers-Siegel action, often quoting \cite{Chalmers:1996rq}, although this paper deals only with the self-dual action of Siegel \cite{Siegel:1992xp}. This first-order form of the YM action appears in the paper \cite{Chalmers:1997sg},   but the authors quote an earlier source \cite{Ashtekar:1986yd}.}
\begin{align}\label{3.42}
\int d^4x \;  G^{\a\b} F_{\a\b} \ \to \ \int d^4x \;  [ G^{\a\b} F_{\a\b} -{\frac{\om}{2} }G^{\a\b} G_{\a\b}]\,. 
\end{align} 
This is the usual \emph{first-order form} of the YM action. 
Now the variation with respect to $G$ simply identifies $\om G =  F$. Eliminating $G$ from \p{3.42}, we obtain the second-order action $\int d^4x \;  F^{\a\b} F_{\a\b} \sim \int d^4x \;  F^{\mu\nu} F_{\mu\nu}$ up to a topological term. 

We can now clearly see the LHC analog of the above transition from self-dual to full SYM. The self-duality equation is contained in the super-curvature constraint $W^{++}=0$ and its relaxed form is the constraint \p{3.14}. As explained above, the right-hand side of \p{3.14} follows from the variation of  $S_{\rm Z}$ \p{sint}. We should then expect that $S_{\rm Z}$ contains the quadratic Lagrange multiplier term $GG$ from \p{3.42}. Indeed, this is straightforward to show in the WZ gauge \p{23}. The bilinear term in \p{sint}, \p{lint} has the form
\begin{equation}\label{76}
- {\om \over 2} \tr\int
d^4x d^8\q \, du_1 du_2\; {A^{++}(x, \q^+_1,u_1)   A^{++}(x, \q^+_2,u_2) \over
(u^+_1u^+_2)^2}\; .
\end{equation}
The Grassmann integral requires $(\q)^8$ and this can only come from the last term in \p{23}, $\int d^8\q\;  (\q^+_1)^4 (\q^+_2)^4 = (u^+_1 u^+_2)^4$. This cancels the harmonic denominator in \p{76}, after which the harmonic integration picks the singlet term  and we obtain the quadratic Lagrange multiplier term $G_{\a\b} G^{\a\b}$  from \p{3.42}. Further, the non-linear part of  \p{sint}, \p{lint} is truncated at $n=4$ due to the nilpotent character of $A^{++}$ in the WZ gauge \p{23}. The cubic term $(\q^+_1)^2 \phi (\q^+_2)^3 \psi(\q^+_3)^3 \psi$ supplies the chiral Yukawa interaction $\phi\psi\psi$ (the complement of the anti-chiral Yukawa term in \p{33}). Note that $(\q^+_1)^2 \phi (\q^+_2)^2 \phi (\q^+_3)^4 G$ does not contribute due to the harmonic integration (absence of a Lorentz singlet in the product $\phi\phi G$). Finally, the quartic term $(\q^+_1)^2 \phi (\q^+_2)^2 \phi (\q^+_3)^2 \phi (\q^+_4)^2 \phi $ supplies the $(\phi)^4$ interaction. 
So, we get\footnote{This coincides with eq. (3.27) in  \cite{Boels:2006ir}. Likewise, their eq. (3.16) coincides with our \p{33}. The latter first appeared in \cite{Siegel:1992xp,Siegel:1992wd,Ketov:1992ix} and was recast in superspace form in  \cite{Sokatchev:1995nj}.}
\begin{equation}\label{76'}
S_{\rm Z}  = \om\int
d^4x\,  \tr\left(- \frac{1}{2}  G_{\a\b} G^{\a\b} - \frac{1}{4}  \phi_{AB} \psi^{\a A} \psi^B_{\a} 
+  \frac{1}{32}  [\phi^{AB}, \phi^{CD}] [\phi_{AB},  \phi_{CD}]  \right)\; .
\end{equation}
Together, the action terms \p{33} and \p{76'} form the complete $\cN=4$ SYM action with a first-order (Lagrange multiplier) formulation of the YM Lagrangian.

\subsection{Yang-Mills theory in Lorentz harmonic space}

The formulation of $\cN=4$ SYM described above can be truncated to the ordinary (non-supersymmetric) Yang-Mills theory. In doing so we recover the result of \cite{Lovelace:2010ev}. We refer the reader to that paper for the detailed comparison with the original twistor version.

The truncation is done by dropping all the components of the semi-superfields $A^{++}$ and $A^+_\da$ which carry $SU(4)$ indices:
\begin{align}\label{3.50}
A^{++} = \cA^{++}(x,u) + 3(\q^+)^4 \cG^{--}(x,u)\,, \qquad A^+_\da = \cA^+_\da(x,u) + (\q^+)^4 C^{-3}_\da(x,u)\,. 
\end{align}
Notice that unlike the WZ gauge \p{23}, here the connection $A^{++}$ still has a component at level $(\q)^0$. Together with $\cA^+_\da$ they play the role of the gauge connections for the covariant harmonic and projected space-time derivatives (cf. \p{3.3}):
\begin{align}\label{3.51}
\nabla^{++} = \pa^{++} + \cA^{++}   \,, \qquad  \nabla^+_\da =  \pa^+_\da + \cA^+_\da   \,.
\end{align}
They undergo gauge transformations like in \p{21} but with a bosonic parameter $\la(x,u)$. The $(\q^+)^4$ components in \p{3.50} are Lagrange multipliers, as we show below.  

Let us now substitute the truncated expansions \p{3.50}  into the action \p{N4} and integrate out the Grassmann variables. The Chern-Simons part \p{3.18} gives 
\begin{align}\label{3.52}
S_{\rm CS} = \tr \int d^4x du\left[ 3\cG^{--} F^{++} - C^{-3 \da}(\pa^{++} \cA^+_\da - \pa^+_\da \cA^{++} +[\cA^{++},\cA^+_\da])  \right]\,
\end{align}
{with $F^{++} = \frac{1}{2} [ \nabla^{+\da},\nabla^{+}_{\da}] = \pa^{+\da} A^+_\da + A^{+\da} A^+_\da$.}
The Lagrange multiplier $ C^{-3}$ imposes the zero-curvature condition 
\begin{align}\label{3.53}
0=\pa^{++} \cA^+_\da - \pa^+_\da \cA^{++} +[\cA^{++},\cA^+_\da]\ \Leftrightarrow \ [\nabla^{++}, \nabla^+_\da]=0\,.
\end{align}
This harmonic differential equation for the connection $\cA^+_\da$ allows us to express it in terms of the other connection $\cA^{++}$, except for the first term in the harmonic expansion $\cA^+_\da(x,u) = u^{+\a} \cA_{\a\da}(x) + \ldots$ (cf. \p{59}). This component is identified with the  gluon field. 
The other Lagrange multiplier $\cG^{--}$, if left alone, would impose the self-duality constraint $F^{++} =0$. However, $\cG^{--}$ appears also in the interaction term \p{sint}, \p{lint}:
\begin{align}\label{3.54}
&S_{\rm Z} = 9\om\, \tr \sum^\infty_{n=2}{(-1)^n\over n}\sum_{k\neq j} \int du_1\ldots du_n\; (u^+_k u^+_j)^4   \ \times \nt
& {\cA^{++}(1) \ldots \cA^{++}(k-1) \over
(u^+_1u^+_2) \ldots (u^+_{k-1} u^+_k)} {\cG^{--}(k)\over (u^+_k u^+_{k+1})} {\cA^{++}(k+1) \ldots \cA^{++}(j-1) \over
(u^+_{k+1} u^+_{k+2}) \ldots (u^+_{j-1} u^+_j)}{\cG^{--}(j)\over (u^+_j u^+_{j+1})} {\cA^{++}(j+1) \ldots \cA^{++}(n) \over
(u^+_{j+1} u^+_{j+2}) \ldots (u^+_{n} u^+_1)}\,,
\end{align}
where the argument of each $\cA^{++}$ or $\cG^{--}$ refers to the harmonic variable at that point {(cf.~\p{lint})}. Eqs.~\p{3.52} and \p{3.54} coincide with eq.~(35) in \cite{Lovelace:2010ev}. As shown there, this YM action in Lorentz harmonic  space is equivalent to the twistor formulation of \cite{Mason:2005zm}, \cite{Boels:2006ir}, \cite{Boels:2007gv}. 

To recover the standard first-order formulation of YM theory in Euclidean space-time, we can profit from the gauge transformation $\delta \cA^{++} = \pa^{++}\la + [\cA^{++},\la]$ to completely gauge away  the harmonic connection, $\cA^{++}=0$. This still leaves the first component in $\la(x,u)=\la(x) + \ldots$ as the usual gauge parameter. In this gauge eq.~\p{3.53} becomes trivial yielding $\cA^+_\da(x,u) = u^{+\a} \cA_{\a\da}(x) $. Substituting this back in \p{3.52} we get
\begin{align}\notag
S_{\rm CS} = \tr\int d^4x\, G_{\a\b} F^{\a\b} \quad {\rm with} \quad  G_{\a\b}(x) = 3\int du\, u^+_\a u^+_\b\, \cG^{--}(x,u)\,. 
\end{align}
Here the second relation is an example of the harmonic version of a `twistor transform'. 
Further, all but the first (bilinear) term in \p{3.54} drop out, thus reducing the `interaction' Lagrangian to the square of the Lagrange multiplier $G$:
\begin{align}\notag
S_{\rm Z} = -{\frac{9\om}{2}}\tr \int d^4x du_1 du_2\, (u^+_1 u^+_2)^2 \cG^{--}(1) \cG^{--}(2) = -{\frac{\om}{2}}\tr \int d^4x \, G^{\a\b} G_{\a\b}\,.
\end{align}


\section{Supersymmetry of the $\cN=4$ SYM action}\label{s4}

In this Section we discuss the realization of $\cN=4$ supersymmetry in the action \p{N4}. We have already shown that in the WZ gauge this action coincides with the full $\cN=4$ SYM component action, which is invariant under $\cN=4$ supersymmetry. The supersymmetry of the component action is not manifest, of course.

As explained at the beginning of Sect.~\ref{s3.1}, the chiral-analytic gauge super-connections (\ref{09}) do not transform as superfields under the full $\cN=4$ supersymmetry, but only under its chiral ($Q$) half. The other half ($\bQ$) is realized non-linearly. In addition, the supersymmetry algebra closes modulo field equations and compensating gauge transformations.  This is typical for an on-shell realization of supersymmetry. We repeat that the purpose of this article is not to break the (so far unsurmountable) $\cN=4$ barrier for a theory with full off-shell supersymmetry. 

{ In Sect.~\ref{s4.1} we recall the supersymmetry transformation in the LHC formulation of the self-dual theory from \cite{Sokatchev:1995nj}. Then in Sect.~\ref{s4.2} we explain how $\bar Q$ supersymmetry is modifies in the full SYM theory. In Sect.~\ref{s43} we verify that the supersymmetry generators form the super-Poincar\'{e} algebra on shell and up to compensating gauge transformations. In Sect.~\ref{s4.4} we demonstrate that the non-linearly realized $\bar Q$ supersymmetry commutes with the gauge transformations. Finally, in Sect.~\ref{s462} we show that  in the WZ gauge the LHC  supersymmetry transformations take the conventional form of the transformation rules for the physical fields. The material of the present Section constitutes our toolbox in \cite{twin} where we reconstruct the $\bar \q$ dependence of the non-chiral stress-tensor multiplet formulated in the  LHC superspace.}

\subsection{Supersymmetry of the Chern-Simons  action} \label{s4.1}

The way $\cN=4$ supersymmetry is realized on the gauge connections $A^{++}$ and $A^+_\da$ was discussed for the first time in \cite{Sokatchev:1995nj}, in the context of the self-dual $\cN=4$ SYM theory. There one imposes the constraint \p{3.6} and the self-dual version \p{sd} of the constraint \p{3.14}. The action is given by the term $S_{\rm CS}$ from \p{3.18}, without the interaction term \p{sint}. In \cite{Sokatchev:1995nj} it was shown that $S_{\rm CS}$ is invariant under the  supersymmetry transformations
\begin{eqnarray}
\delta_{\mathrm{CS}} A^+_\da &=& (\bep^\db_B\q^{+B} \pa^-_\db
-  u^+_\b\ep^{\b B}\pa^-_B)A^+_\da  \,, \label{0241}\\
\delta_{\mathrm{CS}} A^{++} &=& (\bep^\db_B\q^{+B} \pa^-_\db
-  u^+_\b\ep^{\b B}\pa^-_B)A^{++}
 + (\bep^\db_B \q^{+B}) A^+_\db\,. \label{0242}
\end{eqnarray}
These are  transformation laws of `semi-superfields'. We are not transforming the coordinates like in \p{11''} but the superfields themselves. From this point of view the harmonic derivative $\pa^{++}$ does not commute with the translation part of (\ref{0241}), (\ref{0242}),
\begin{equation}\label{025}
[\pa^{++}\,, \ \bep^\db_B\q^{+B} \pa^-_\db
-  u^+_\b\ep^{\b B}\pa^-_B] = \bep^\db_B\q^{+B} \pa^+_\db
-  u^+_\b\ep^{\b B}\pa^+_B \,.
 \end{equation}
Using
(\ref{0241}), (\ref{0242}), (\ref{025}) it is easy to check that
\begin{equation}\label{CStr}
\delta L_{\rm CS}  = (\bep^\db_B\theta^{+B} \partial^-_\db
-  u^+_\b\ep^{\b B}\partial^-_B) L_{\rm CS}  +
{1\over 2} (\bep^\db_B \theta^{+B}) \partial^{+\da}
\tr (A^+_\da A^+_\db) \ ,
\end{equation}
i.e.,  the Lagrangian transforms into a total derivative with respect to the variables
$x^{ \da\a}$ and $\theta^{+A}$. Consequently, the action \p{3.18} is invariant. 

The $\cN=4$ supersymmetry transformations \p{0241}, \p{0242}
form a closed algebra on shell (see Sect.~\ref{s43}), i.e. using the field equation \p{3.6} and the self-dual version  \p{sd} of \p{3.14}:
\begin{align}
\left[ \delta_{\mathrm{CS}}(\kappa) \,,\, \delta_{\mathrm{CS}}(\ep) \right] A^{+}_{\da} 
&= (\ep^{\b B} \bar\kappa^{\db}_B - \kappa^{\b B} \bep^{\db}_B) \,\pa_{\b\db } A^{+}_{\da} +
\delta_{\Lambda}A^{+}_{\da}
\nt
\left[ \delta_{\mathrm{CS}}(\kappa) \,,\, \delta_{\mathrm{CS}}(\ep) \right] A^{++} 
&= (\ep^{\b B} \bar\kappa^{\db}_B - \kappa^{\b B} \bep^{\db}_B) \,\pa_{\b\db } A^{++} +
\delta_{\Lambda}A^{++} \,, 
\notag
\end{align}
up to a gauge transformation \p{21} with the field-dependent parameter
\begin{align}\label{420}
\Lambda_{Q\bQ} = u^-_\b ( \kappa^{\b B}\bep^{\da}_B - \ep^{\b B}\bar\kappa^{\da}_B )\,A^{+}_{\da}\,.  
\end{align}

\subsection{Full action and modified $\bQ$ supersymmetry transformation} \label{s4.2}

After completing the CS action with the `interaction' term $S_{\mathrm{Z}}$ \p{sint}, \p{lint}, we need to  modify the $\bQ-$half of the supersymmetry transformations.\footnote{Such a modification was proposed in \cite{Bullimore:2011kg}, in a component field formulation of the theory. The modification in the form \p{459}  has been known for a long time  in the  context of the formulation of $\cN=4$ SYM in $\cN=2$  harmonic superspace \cite{Galperin:2001uw}, see Sect.~\ref{n2} for the detail. } The basic reason is that the field equation \p{sd} is modified by the interaction term and becomes \p{3.14}. The non-vanishing right-hand side  of \p{3.14} suggests the modification 
\begin{align}\label{4.20}
\delta A^{+\da} = \delta_{\mathrm{CS}}A^{+\da} + \delta_{\mathrm{Z}}A^{+\da}\,,  
\end{align}
where 
\begin{equation}\label{459}
\delta_{\mathrm{Z}} A^{+\da} = \om (\pa^+)^4 [( \bar{\ep}^\da_B \q^{-B}) A^{--}] \,,
\end{equation} 
while $\delta A^{++}$ in \p{0242} remains unchanged,   $\delta A^{++} = \delta_{\rm CS} A^{++}$ and $\delta_Z A^{++} = 0$. 
{We remark that if we switch off the interaction by setting $g=0$, the modification term \p{459} clearly does not vanish. In other words, the bilinear part of the action \p{3.23} is invariant under $\bar Q$ supersymmetry \p{0242}, \p{4.20} including the linearized part of \p{459}.

Let us show that this modification makes the action \p{N4} invariant under the full supersymmetry. 
Evidently, $\delta_{\mathrm{Z}} S_{\mathrm{Z}} = 0$ {since $S_{\rm Z}$ depends only on the  connection $A^{++}$}, however (recall {the variation of $S_{\rm Z}$} \p{444})
\begin{align}
\delta_{\mathrm{CS}} S_{\mathrm{Z}} 
&= -\om\, \tr \int d^4x  du d^8\q \;  \Big[ (\bar{\ep}^\db_B \q^{+B}) A^+_\db  
+  (\bar{\ep}^\db_B\q^{+B} \pa^-_\db
-  u^+_\b\ep^{\b B}\pa^-_B)A^{++}\Big]    A^{--}\,. \label{460}
\end{align}
On the other hand (recall that $\nabla^{++} = \pa^{++} + [A^{++},\cdot\ ]$), 
\begin{align}\label{461}
\delta_{\mathrm{Z}} S_{\mathrm{CS}}  & =  \tr \int d^4x du  d^4 \q^+ \ \delta_{\mathrm{Z}}  A^{+\da} (\pa^+_\da A^{++} - \nabla^{++} A^+_\da)  \nt
& =  \om\, \tr \int d^4x du  d^4 \q^+ \  (\pa^+)^4 [( \bar{\ep}^\da_B \q^{-B}) A^{--}]\  (\pa^+_\da A^{++} - \nabla^{++} A^+_\da)  \nt
& =  \om\, \tr \int d^4x du d^8\q \ ( \bar{\ep}^\da_B \q^{-B}) (\pa^+_\da A^{++} - \nabla^{++} A^+_\da)  A^{--}\,.
\end{align}
{In the second line, when completing the measure $d^4 \q^+ (\pa^+)^4 = d^8 \q$ we used the analyticity of $\pa^+_\da A^{++} - \nabla^{++} A^+_\da$. }
Collecting the terms with $\bep$ and $A^+_\da$ from \p{460} and \p{461}, we get
\begin{align}\notag
& -\om\, \tr \int  \ [  (\bar{\ep}^\da_B\q^{+B}) A^+_\da + ( \bar{\ep}^\da_B \q^{-B}) \nabla^{++} A^+_\da ]  A^{--} = -\om\, \tr \int  \ \nabla^{++} [( \bar{\ep}^\da_B \q^{-B})  A^+_\da]  A^{--}\,.
\end{align}
Then we integrate $\nabla^{++}$ by parts and use the defining relation \p{3.11} {in the form $\nabla^{++} A^{--} = \pa^{--} A^{++}$} to obtain
\begin{align}\label{4.10}
\om \int d^8 \q  \  [( \bar{\ep}^\da_B \q^{-B})  A^+_\da]\ \pa^{--} A^{++} = \om \int  d^4\q^+   \   \bar{\ep}^\da_B A^+_\da (\pa^+)^4 [\q^{-B} \pa^{--} A^{++}] = 0 \,.
\end{align}
{Here we have used the analyticity of $A^+_{\da}$ and $A^{++}$;} one derivative $\pa^+$ eliminates $\q^-$, two others annihilate $\pa^{--} A^{++}$. 

Further, the terms with $A^{++}$ from \p{460} and \p{461} combine into 
\begin{align}\label{463}
\delta_{\mathrm{CS}} S_{\mathrm{Z}}  +   \delta_{\mathrm{Z}} S_{\mathrm{CS}} &= -\om\, \tr \int d^4x du d^8 \q  \ [(\bar{\ep}^\da_B\q^{+B})\pa^-_\da - ( \bar{\ep}^\da_B \q^{-B}) \pa^+_\da  -  u^+_\b\ep^{\b B}\pa^-_B ]A^{++}\cdot  A^{--}\nt
& = -\om\, \tr \int d^4x du d^8 \q  \ [-\bar{\ep}^\da_B\q^{\a B}\pa_{\a\da} +  \ep^{\b B}\pa_{\b B}]A^{++}\cdot  A^{--} \,,
\end{align}
where in the second line we have used the analyticity condition $\pa^+_B A^{++}=0$ to write {the $Q$ supersymmetry transformation \p{0242} in the form} $\delta A^{++} = ( -u^+_\b\ep^{\b B}\pa^-_B + u^-_\b \ep^{\b B}\pa^+_B)A^{++} = {\ep}^{\b B}\pa_{ \b B}A^{++}$. The result is a $\bQ$ shift of $x$ and a $Q$ shift of $\q$ as in \p{11''}. Now, recall the expression \p{lint} for the Lagrangian in \p{sint}. In it each $A^{++}$ depends on the same $(x_{\a\da}, \q^{\a A})$, as in the integration measure in \p{463}, even though in each $A^{++}(i)$ the odd variable  $\q^{\a A}$ appears projected with the  harmonic $(u_i)^+_{\a}$. Thus, the variation \p{463} can be interpreted as a {\sl harmonic independent} shift $\delta x = -\bep \q$, $\delta \q =\ep$  of all $A^{++}$ in \p{lint}. This shift is a total derivative, so $\int d^4 x d^8\q$ annihilates it. {This completes the proof that the action \p{N4} is invariant under the supersymmetry transformations \p{0241},  \p{0242} and \p{4.20}.}

\subsection{Closure of the supersymmetry algebra} \label{s43}

{Here we show that the modified supersymmetry transformations \p{0241}, \p{0242} and \p{4.20} form the super-Poincar\'e algebra on shell, up to field-dependent gauge transformations.
The supersymmetry generators and variations are related by}
\begin{align}
\delta(\epsilon) = \delta_Q(\epsilon) + \delta_{\bQ}(\epsilon)= \ep^{\b B} Q_{\b B} + \bep^{\db}_{B} \bar{Q}^B_{\db}\,.\notag
\end{align}
{The generators of $Q$ supersymmetry are the same for both analytic connections,}
\begin{align}\label{4.27'}
&Q^\b_B = \pa^\b_B \ \rightarrow \ Q^\b_B A(\q^+) = - u^{+\b} \pa^-_B A(\q^+)\,,
\end{align}
{while those of $\bar Q$ supersymmetry are different for $A^{++}$ and $A^{+}_{\da}$ even in the self-dual theory,}
\begin{align}
&\bQ^B_\db A^+_\da  =  \q^{+B} \pa^-_\db A^+_\da + (\bQ_{\rm Z})^B_\db A^+_\da \,, \qquad  (\bQ_{\rm Z})^B_\db A^+_\da  =  \om (\pa^+)^4 (\q^{-B} A^{--})\ep_{\da\db}  \label{4.27}\\
& \bQ^B_\db A^{++} = \q^{+B} \left( \pa^-_\db A^{++} + A^+_\db \right) \,. \label{4.28'}
\end{align}  
The generator    $\bQ^B_\db A^{++}$ can be rewritten  with the help of  the kinematical field equation \p{3.6}, $\pa^+_{\db} A^{++} = \nabla^{++}A^+_{\db}$, in a different form which is not modified by the interaction term:
\begin{align}
\bQ^B_\db A^{++} &= - \q^{\b B} \pa_{\b\db} A^{++} + \q^{-B}\pa^+_\db A^{++} + \q^{+B} A^+_\db = - \q^{\b B} \pa_{\b\db} A^{++} + \q^{-B}\nabla^{++} A^+_\db + \q^{+B} A^+_\db  \nt
 &= - \q^{\b B} \pa_{\b\db} A^{++} + \nabla^{++} (\q^{-B} A^+_\db) \,.  \label{4.28}
\end{align}
From \p{4.28} and the defining relation \p{3.11} we can  find the $\bar{Q}$ transformation {of $A^{--}$}
\begin{align}\label{4.30}
\bQ^B_\db A^{--}   = - \q^{\b B} \pa_{\b\db} A^{--} + \nabla^{--} (\q^{-B} A^+_\db)\quad {\rm with} \quad \nabla^{--} = \pa^{--} + [A^{--},\cdot\ ]\,.
\end{align}
We remark that the terms $\nabla^{\pm\pm} (\q^{-B} A^+_\db)$ in \p{4.28} and \p{4.30} are not gauge transformations \p{21} because the `parameter' $\q^{-B} A^+_\db$ is not LH-analytic.

The $Q$ transformations are just shifts of the odd coordinates, therefore they obviously commute. For the rest of the supersymmetry algebra involving $\bQ$, we expect that it closes up to field equations and compensating gauge transformations.  Let us first check the closure on $A^{++}$. Making two infinitesimal transformations with parameters $\ep$ and $\bar\kappa$ and using  \p{4.27'} and \p{4.28}, we find the commutator {of the $\bar Q$ and $Q$ transformations}
\begin{align}\label{427}
[\delta_{\bQ}(\kappa),\delta_Q(\ep)] 
\, A^{++} = \ep^{\b B} \bar\kappa^{\db}_B \,\pa_{\b\db } A^{++} +
\nabla^{++}\Lambda_{Q\bQ}\,.
\end{align}
{The calculation is the same as in the self-dual case.}
Here the first term on the right-hand side  is the expected translation according to $\{\bQ,Q\}=\pa_x$, the second term is a compensating gauge transformation with the parameter \p{420}.

Further, using \p{4.28'}, \p{4.27} and \p{3.11} we find the commutator {of two $\bar Q$ transformations} on the connection $A^{++}$,
\begin{align}\label{4.35}
[\delta_{\bQ}(\kappa),\delta_{\bQ}(\ep)] 
\, A^{++}  &= 2\om (\pa^{+})^4  (\bep\bar\kappa)_{BC} \q^{+[B}\q^{-C]} A^{--}\,,
\end{align}
{with $(\bep\bar\kappa)_{BC} = \bep_{\db B} \bar\kappa^{\db}_C$\ . Then we use the defining equation for $A^{--}$ \p{3.11} and 
rewrite  \p{4.35} as follows }
\begin{align}
& \om \pa^{++}  (\pa^{+})^4 (\bep\bar\kappa)_{BC} \q^{-B}\q^{-C} A^{--} - \om  (\pa^+)^4 (\bep\bar\kappa)_{BC}\q^{-B}\q^{-C} \pa^{++} A^{--}\nt
 &=\nabla^{++}\Lambda_{\bQ\bQ} - \om  (\pa^+)^4 (\bep\bar\kappa)_{BC}\q^{-B}\q^{-C} \pa^{--} A^{++}\,. \notag
\end{align}
The second term in the last line vanishes due to the analyticity of $A^{++}$ and the four derivatives $\pa^{+}$ 
while the first term is a compensating gauge transformation with parameter
\begin{align}\label{431}
\Lambda_{\bQ\bQ}(\q^+) = \om (\pa^+)^4 \left[(\bep\bar\kappa)_{BC} \q^{-B}\q^{-C} A^{--}\right]\,.  
\end{align}
{We emphasize that $\Lambda_{\bQ\bQ}$ is analytic, as it should be, due to projector $(\pa^+)^4$ in it.}
{Therefore, the supersymmetry algebra closes on the connection $A^{++}$}  
\begin{align}
[\delta(\kappa),\delta(\ep)] \, A^{++} = (\ep^{\b B} \bar\kappa^{\db}_B - \kappa^{\b B} \bep^{\db}_B) \,\pa_{\b\db } A^{++} 
+ \nabla^{++}( \Lambda_{Q\bQ} + \Lambda_{\bQ\bQ})\,. \notag
\end{align}

{Next, let us show the closure of the algebra on the connection $A^{+}_{\da}$.}
Using  \p{4.27'} and \p{4.27} we find the commutator {of the $\bar Q$ and $Q$ transformations} 
\begin{align}\label{432}
[\delta_{\bQ}(\kappa),\delta_Q(\ep)] 
\, A^{+}_{\da} &= - (\ep^{+ B} \bar\kappa^{\db}_B ) \,\pa^{-}_{\db} \,A^{+}_{\da} 
- \om (\pa^+)^4 (\ep^{- B} \bar\kappa_{\da B} ) A^{--} \,.
\end{align}
{Here  $A^{--}$ transforms by shifts of $\theta$ under $Q$ supersymmetry, i.e. $\delta_{Q}(\epsilon) A^{--} = \ep^{\b B} \pa_{\b B} A^{--}$. This agrees with the definition of $A^{--}$ in \p{3.11}.}
In the first term on the right-hand side   we complete the coordinate shift and then use the field equation \p{3.14'} to get
\begin{align}\notag
& - (\ep^{+ B} \bar\kappa^{\db}_B) \,\pa^{-}_{\db} \,A^{+}_{\da}  =  (\ep^{\b B} \bar\kappa^{\db}_B) \,\pa_{\b\db} \,A^{+}_{\da} - (\ep^{- B} \bar\kappa^{\db}_B) \,\pa^{+}_{\db} \,A^{+}_{\da} \nt
& =  (\ep^{\b B} \bar\kappa^{\db}_B) \,\pa_{\b\db} \,A^{+}_{\da} - \nabla^{+}_{\da} \, (\ep^{- B} \bar\kappa^{\db}_B) \,A^{+}_{\db} + \om (\ep^{- B} \bar\kappa_{\da B}) (\pa^+)^4 A^{--} \,.  \notag
\end{align}
The first term in the second line is a translation, the second is a compensating gauge transformation with parameter \p{420} and the third cancels against a similar term in \p{432}. We conclude that
\begin{align}\label{427'}
[\delta_{\bQ}(\kappa),\delta_Q(\ep)] 
\, A^{+}_\da = \ep^{\b B} \bar\kappa^{\db}_B \,\pa_{\b\db } A^{+}_\da +
\nabla^{+}_\da\Lambda_{Q\bQ}\,.
\end{align}

Finally,  using \p{4.27} and  \p{4.30} we evaluate the commutator of two $\bar Q$ transformations  on the connection $A^{+}_{\da}$:
\begin{align}\label{433}
[\delta_{\bQ}(\kappa),\delta_{\bQ}(\ep)] 
\, A^{+}_{\da} &=  -\om(\pa^{+})^4 \left[ (\bep_{\da C}\bar\kappa^{\db}_B )  \q^{+ B}  \q^{-C} \pa^-_\db A^{--} + \bar\kappa_{\da B} \q^{-B} \delta_{\bQ}(\ep) A^{--}\right]  - (\ep \leftrightarrow \kappa)\nt
& =  \om(\pa^{+})^4 \left[(\bep\bar\kappa)_{BC} \q^{-B}\q^{-C} \nabla^+_\da A^{--} -
 \pa^{--}  (\bep\bar\kappa)_{BC} \q^{-B}\q^{-C}A^+_\da \right]\nt
& = \nabla^+_\da  \Lambda_{\bQ\bQ}\,.
\end{align}
This again is a compensating gauge transformation with the parameter \p{431}. {The second term in the second line of \p{433}
drops out due to the projector $(\pa^+)^4$ and the analyticity of $A^+_{\da}$.} 
{So, the supersymmetry algebra closes on the connection $A^{+}_{\da}$ as well,}  
\begin{align}
[\delta(\kappa),\delta(\ep)] \, A^{+}_{\da} = (\ep^{\b B} \bar\kappa^{\db}_B - \kappa^{\b B} \bep^{\db}_B) \,\pa_{\b\db } A^{+}_{\da} 
+ \nabla^{+}_{\da}( \Lambda_{Q\bQ} + \Lambda_{\bQ\bQ})\,.  \notag
\end{align}

\subsubsection{Closure of the $\bQ$ subalgebra off and on shell}

We have checked that the full supersymmetry algebra is realized on the superconnections,  modulo 
gauge transformations and the field equations (the kinematical constraint \p{3.6} and dynamical equation of motion \p{3.14}). The $Q$ supersymmetry transformations \p{4.27'} form an Abelian subalgebra off shell, i.e. the $Q$-half of the supersymmetry is manifest in our formalism. A natural question to ask is if the $\bQ$-half also forms an Abelian subalgebra off shell. This will be important for us in \cite{twin} where we employ $\bQ$ transformations to construct the full stress-tensor supermultiplet starting from its chiral truncation.   We use $\bQ$ in the form \p{4.27} and \p{4.28'}. In Sect.~\ref{s43} we have already shown, 
without using the field equations, that two $\bQ$ anticommute on $A^{++}$ up to a gauge transformation \p{431}\,,
\begin{align}
[\delta_{\bQ}(\kappa),\delta_{\bQ}(\ep)] A^{++} =\nabla^{++} \Lambda_{\bQ \bQ}  \qquad \text{off shell}\,. \notag
\end{align}
However, on $A^{+}_{\da}$ the anticommutation (up to a gauge transformation) is valid only modulo the kinematical constraint \p{3.6}\,.
More precisely, we have  
\begin{align}
[\delta_{\bQ}(\kappa),\delta_{\bQ}(\ep)] A^{+}_{\da} 
& = {\om\ } (\bar\kappa^{\db}_{A} \bep_{\da B} - \bep^{\db}_{A} \bar\kappa_{\da B} )(\pa^+)^4  \q^{-B} \int d v\ 
\Bigl(\q^{-A} W^{+3}_{\db} \frac{\delta}{\delta A^{++}} \Bigr)(v)\ A^{--}
\nt &+ \nabla^{+}_{\da} \Lambda_{\bQ \bQ}   \qquad \text{off shell}\,,  \notag
\end{align}
where the curvature $W^{+3}_{\da}$ is defined in \p{3.5}, the dependence on the dummy harmonic $v$ is indicated explicitly.
We plan to use this anticommutation relation repeatedly on the same superfields. This is possible  because the field equations \p{3.6} and \p{3.14} transform into each other under $\bQ$ supersymmetry,
\begin{align}
&\bQ^{B}_{\db}\ W^{+3}_{\da} = \q^{+B} \pa^{-}_{\db} W^{+3}_{\da} + \ep_{\da\db} \q^{+B}  (W^{++} - \om\,{\mathbb W}^{++})\nt
&\bQ^{B}_{\db}  (W^{++} - \om\,{\mathbb W}^{++}) = \q^{+B} \pa^{-}_{\db}  (W^{++} - \om\,{\mathbb W}^{++}) 
- \om (\pa^+)^4 \int d v\ 
\Bigl(\q^{-B} W^{+3}_{\db} \frac{\delta}{\delta A^{++}} \Bigr)(v)\ A^{--}\,. \notag
\end{align}

\subsection{$\bQ$ supersymmetry and gauge transformations}\label{s4.4}

{The supersymmetry transformations ought to commute with the gauge transformations. This is evident for the $Q-$half (see \p{4.27'}).
The $\bar Q-$half is realized non-linearly and it mixes the two gauge connections, so the commutativity is not obvious. Here we prove that the $\bar Q$ transformations \p{4.27}, \p{4.28'} commute with the gauge transformation \p{21},}
\begin{align} \label{comm}
[ \ \delta_{\bar Q} \ ,\ \delta_{\Lambda}\ ] \ A^{++} \ =\ [ \ \delta_{\bar Q} \ , \ \delta_{\Lambda} \ ] \ A^{+}_{\da} \  = \ 0\,.
\end{align}
{This result will be crucial for us in the twin paper \cite{twin}. There we consider chiral gauge-invariant operators in LHC superspace and use the ${\bQ}$-variations  to construct their non-chiral extensions. The commutativity \p{comm} ensures that the variations are gauge invariant operators as well.} {We emphasize  that the commutativity holds without the field equations.}

We start with  $\delta A^{++}$. Denoting $( \bep^\db\cdot \q^+ ) \equiv \bep_B^\db\q^{+B}$ and recalling \p{21} and \p{4.28'}, we find 
\begin{align}\notag
\delta_\Lambda \, \delta_{\bQ} \, A^{++} &= \delta_\Lambda \, \big((\bep^\db\cdot \q^+)  (\pa^-_\db A^{++} +  A^+_\db) \big) = (\bep^\db\cdot \q^+) (\pa^-_\db  \nabla^{++}\Lambda + \nabla^+_\db  \Lambda)\nt
& = (\bep^\db\cdot \q^+) \bigl( \pa^{++}\pa^-_\db\Lambda + \pa^-_\db[ A^{++}, \Lambda] + [A^+_\db,\Lambda] \bigr)\,. \notag
\end{align}
{The inverse sequence of variations leads to}
\begin{align}\notag
\delta_{\bQ} \, \delta_\Lambda \, A^{++} &= \delta_{\bQ} \, \big(\pa^{++}\Lambda +[A^{++}, \Lambda] \big)\nt
& = \pa^{++} \big((\bep^\db\cdot \q^+)\pa^-_\db\Lambda  \big) + [(\bep^\db\cdot \q^+)  (\pa^-_\db A^{++} +  A^+_\db) , \Lambda] + [A^{++}, (\bep^\db\cdot \q^+) \pa^-_\db\Lambda] \nt
&= \delta_\Lambda \, \delta_{\bQ} \, A^{++} \,.\notag
\end{align}
Here we assume that the gauge parameter transforms {under $\bar Q$ supersymmetry} as a chiral-analytic semi-superfield, $\delta_{\bQ} \Lambda = (\bep^\db\cdot \q^+)\pa^-_\db\Lambda$ {(cf. \p{0241})}. The first part of \p{comm} is proven. 

Now we turn to the  variation of $A^+_\da$ (recall \p{a--}):
\begin{align}\notag
\delta_\Lambda \, \delta_{\bQ} \, A^+_\da &= \delta_\Lambda \, \left((\bep^\db\cdot \q^+)  \pa^-_\db A^+_\da  + \om (\pa^+)^4 (\bep_{\da}\cdot \q^{-}) A^{--}) \right)\nt
& = (\bep^\db\cdot \q^+) \pa^-_\db (\nabla^+_\da\Lambda) + \om (\pa^+)^4 (\bep_{\da} \cdot \q^{-}) (\pa^{--} \Lambda + [A^{--}, \Lambda])\nt
& = (\bep^\db\cdot \q^+) \pa^-_\db \big(\pa^+_\da\Lambda + [A^+_\da, \Lambda] \big) + \om  \left[ (\pa^+)^4 ( (\bep_{\da} \cdot \q^{-}) A^{--}), \Lambda \right]\,, \notag
\end{align}
where we used the fact that $(\pa^{+})^4 (\q^{-B} \pa^{--} \Lambda)=0$ due to the analyticity of the gauge parameter, $\pa^+\Lambda=0$. On the other hand,
\begin{align}\notag
\delta_{\bQ} \, \delta_\Lambda \, A^+_\da &=  \delta_{\bQ} \, \big(\pa^+_\da\Lambda + [A^+_\da, \Lambda] \big) \nt
&=\pa^+_\da (\bep^\db\cdot \q^+)\pa^-_\db\Lambda + \Big[(\bep^\db\cdot \q^+)\pa^-_\db A^+_\da + \om (\pa^+)^4 (\bep_{\da} \cdot \q^{-}) A^{--}, \Lambda\Big] + [A^+_\da, (\bep^\db\cdot \q^+)\pa^-_\db\Lambda]\nt
 & =  \delta_\Lambda \, \delta_{\bQ} \, A^+_\da\,.  \notag
\end{align} 
{ Thus the second relation in \p{comm} is proven.}

\subsection{$\bQ$ supersymmetry in the WZ gauge}\label{s462}

{Here we examine the $\bar Q$ supersymmetry transformations in the WZ gauge \p{23}. We would like to show that the LHC formulae of the previous subsections reproduce the familiar transformation rules for the physical component fields.} For simplicity, we restrict the discussion to the linearized (free) case. {We demonstrate once more that the modification of the supersymmetry transformations \p{459} is necessary even in the free theory.} 
In order  to maintain the gauge we have to make a compensating
gauge transformation (\ref{21}) with parameter $\Lambda =- (\bep^\db_B \q^{+B}){\cal A}^-_\db$
{depending on the gauge field ${\cal A}_{\b\db}$\ . 

Let us start with the $\bar Q$ transformations of $A^+_{\da}$ in the self-dual theory \p{0241}. We use a shorthand notation for the composition of the two transformations $\delta = \delta_{\mathrm{CS} + \Lambda}(\ep)$.} We have
\begin{align}\notag
\delta A^+_\da = \bar{\ep}^\db_B\q^{+B} \pa^-_\db
 A^+_\da  + \pa^+_\da \Lambda \,.
\end{align}
{Inserting the WZ gauge \p{25}, { we deduce the transformation rules for the component fields $\mathcal{A}$ and $\bar\psi$. It is clear that the gauge field $\cA$ does not transform.} At level $(\q^+)^1$ we get
\begin{align}\label{462}
\delta\bar\psi_{\da B} & =  - \bar{\ep}^\db_B(\pa^-_\db
 \cA^+_\da  - \pa^+_\da  {\cal A}^-_\db)  = \bep_{\da B}\ F^{+-} + \bar{\ep}^\db_B \ \tilde F_{\da\db}\,,
\end{align}
where $F^{\a\b} = \pa^{\db(\a }  {\cal A}^{\b)}_\db $ and $\tilde F_{\da\db} =\pa^{\a}_{(\da}  {\cal A}_{\db)\a} $ are the two halves of  the YM curvature. The $\tilde F-$term in the second relation is the expected transformation of the component $\bar\psi_{\da B}$, while the $F-$term is unwanted. In the self-dual theory  we have the field equation $F^{\a\b}  =0$ {(recall \p{sd})} and the unwanted term vanishes. However,  in the full theory the equation  is modified, $F^{\a\b} = \om G^{\a\b} \neq 0$ (recall \p{3.42}). We expect that this term is compensated by the modification \p{459}. Indeed, in the linearized case and in the WZ gauge we have
\begin{align}\notag
\delta_{\rm Z} A^{+\da} = \om (\pa^+)^4 \bep^{\da}_B \q^{-B} A^{--}  = -{\frac{\om}{2}} \bep^{\da}_B\psi^{+B} - \om \q^{+B} \bep^{\da}_B G^{+-}\,.
\end{align}
The fermion term {in this formula} generates the expected transformation of the YM field,
\begin{align}\label{4.37}
\delta_{\mathrm{Z}} \mathcal{A}^{\da\a} &= -{\frac{\om}{2}}\,\bep^{\da}_A \psi^{\a A} \,,
\end{align}
while the $G-$term compensates the $F-$term in \p{462} {due to the field equation} and we find
\begin{align}\notag
\delta_{\mathrm{CS + Z}+\Lambda} \ \bar \psi_{\da B} = \bar{\ep}^\db_B \ \tilde F_{\da\db}\,.
\end{align}

For the other prepotential $A^{++}$ {the $\bar Q$ transformation rules are the same as in the self-dual theory, \p{0242} and \p{4.28'}}, so we have 
\begin{align}\notag
 \delta A^{++} &= \bar{\ep}^\db_B\q^{+B} [\pa^-_\db A^{++} + A^+_\db - \cA^+_\db]\,.
\end{align}
Substituting the WZ gauges \p{23} and \p{25} {in this formula}, we find
\begin{align}\notag
 &\delta \phi_{AB} = 2 \bep^\db_{[A} \bar\psi_{B]\db}  \nt
 &\delta \psi^{A}_{\a} = -2\bep^\db_B \pa_{\a\db} \phi^{AB}\nt
 &\delta G^{\a\b}  = \frac12 \bep^\db_B \pa^{(\a}_\db \psi^{\b) B} \,.  \notag
  \end{align}
{Here we eliminated the auxiliary  fields $B^-$ and $\tau^{--}$  by means of the field equations \p{59}.} 

The component field transformations found here are in agreement with those in  \cite{Bullimore:2011kg}. In particular, the non-self-dual modification  concerns only the gauge field \p{4.37}.

\section{Quantization}\label{s6}

{In this section we  quantize the $\cN = 4$ SYM theory with the classical action \p{N4}. In order to define propagators for the gauge connections we need to fix a gauge. In Sect.~\ref{s6.1} we discuss the light-cone gauge. Then in Sect.~\ref{s5.2} 
we derive the propagators for the dynamical superfields $A^{++}$ and $A^+_{\da}$ in this gauge, and after that we summarize the
Feynman rules in Sect.~\ref{s6.3}. In Sect.~\ref{Full} we give some simple examples of LHC supergraph calculations.}

\subsection{Light-cone gauge}\label{s6.1}

We wish to implement the so-called ``axial" or ``CSW" gauge of  \cite{Boels:2006ir}, \cite{Cachazo:2004kj}. Translating to our notation, it is simply
\begin{equation}\label{CSW}
\xi^\da A^+_\da =0\; ,
\end{equation}
where $\xi^\da$ is an arbitrary but fixed commuting anti-chiral spinor.\footnote{In  \cite{Adamo:2011cb} the gauge is generalized by replacing $\xi_\da$ by a full `reference super-twistor' $Z_*=(\mu_\a, \xi_\da, \zeta^A)$. For our purposes the simpler form \p{CSW} suffices. Moreover, it seems unnatural to use  the fermion $\zeta$ in $Z_*$ to fix a gauge for the \emph{bosonic} semi-superfields $A^{++}, A^+_\da$. } We prefer the name ``light-cone" to the name ``axial" used in \cite{Boels:2006ir} and other twistor papers. Indeed, recall that the ordinary Yang-Mills field $\cA_{\a\da}(x)$ lives in the connection $A^+_\da$, see the first line in \p{59}. Then the condition \p{CSW} restricted to this particular component becomes $\xi_\da u^+_{\a}\cA^{\da\a}(x) =0$. Here we see the light-like vector $n_{\a\da} = \xi_\da u^+_{\a}$ typical for the light-cone gauge, rather than the non-null vector $n^\mu$ of the standard axial gauge \cite{Leibbrandt:1987qv}.

An important advantage of the gauge \p{CSW} is that the cubic interaction term in the Chern-Simons Lagrangian \p{CS} vanishes. Indeed, the general solution of \p{CSW} is $A^+_\da = \xi_\da A^+$, hence $A^{+\da} A^+_\da =0$. Thus, the connection $A^+_\da$ becomes  non-interacting but still propagating.  Another advantage of this type of gauge is the decoupling of the ghosts \cite{Leibbrandt:1987qv}.

One should make sure that the gauge \p{CSW} is possible. To answer this question, we make an Abelian gauge transformation and try to find a parameter $\Lambda$ which can eliminate the projection in \p{CSW}:
\begin{equation}\notag
\xi^\da A^+_\da =\xi_\da\pa^{+\da} \Lambda(x,\q^+,u)\; .
\end{equation}
The operator $\xi_\da\pa^{+\da}$ is invertible on the space of $\Lambda$, just like inverting a particular  projection of the momentum. So the gauge \p{CSW} is possible under appropriate boundary conditions.  

Let us now rewrite the gauge-fixing condition \p{CSW},
\begin{align}\label{417}
\xi^{\- \da} A^+_\da =0 
\end{align}
and give it a new interpretation. We consider the parameter $\xi^\da$ as belonging to a new set of LH variables, this time for the second (anti-chiral) factor of the Lorentz group $SU(2)_L\times SU(2)_R$. These new harmonics satisfy the usual $SU(2)$ conditions (recall \p{2})
\begin{align}\label{6.4}
\xi^{\+\da} \xi^{\-}_\da= 1\,, \qquad (\xi^{\+\da})^* = \xi^{\-}_{\da}
\end{align}
and are defined up to a $U(1) $ phase. To distinguish the harmonic charges of $u^\pm_\a$ and of $\xi^{\dt\pm}_\da$  we put a dot above the latter (like the anti-chiral spinor indices $\da$). The main difference between $\xi^{\dt\pm}_\da$ and $u^\pm_{\a}$ is that the former are fixed parameters while the latter are coordinates. With the help of the two sets of harmonics we can project any Lorentz tensor onto its \emph{light-cone components}. For example, the space-time coordinate $x^{\da\a}$ can be decomposed into
\begin{align}\label{6.5}
x^{\da\a} = - \xi^{\-\da} u^{+\a}   x^{\+ -} - \xi^{\+\da} u^{-\a}   x^{\- +} + \xi^{\-\da} u^{-\a}   x^{\+ +} + \xi^{\+\da} u^{+\a} x^{\- -} \,, 
\end{align}
where we see the \emph{covariant light-cone projections} 
\begin{align}\label{419}
x^{\dt\pm \pm} = \xi^{\dt\pm}_{\da} x^{\da\a} u^{\pm}_{ \a}
\end{align}
obtained by using the four light-like vectors $n^{\dt\pm \pm}_\mu =  \xi^{\dt\pm}_\da (\sigma_\mu)^{\da\a} u^{\pm}_\a$. If we make some standard choice of the harmonics, e.g., $\xi^{\+}_\da = (1,0)\,, \ \xi^{\-}_\da = (0,1)$, etc., the projections \p{419} become the usual light-cone projections. 

\subsection{Green's functions}\label{s5.2}

{Now we proceed to the derivation of the propagators for  the dynamical superfields $A^{++}$ and $A^+_{\da}$.}
To simplify the derivation, we shall adopt the approach of \cite{Boels:2006ir}, which is to treat the bilinear term in the second line of \p{3.23}, coming from the Zupnik Lagrangian \p{lint}, as a `bivalent vertex' (see also the comments in Sect.~\ref{Full}).  Then the relevant kinetic terms are contained in the first line of \p{3.23}, coming from the Chern-Simons Lagrangian \p{CS}.  Adding sources for the two dynamical superfields, we obtain
\begin{equation}\label{91}
L_{\rm kin} =  A^{++}\pa^{+\da}A^+_\da-
{1\over 2} A^{+\da}\pa^{++} A^+_\da - A^{++} J^{++} - A^{+\da} J^{+3}_\da \,.
 \end{equation}
Like the super-connections, the sources are chiral-analytic semi-superfields, $J(x,\q^+,u)$. Due to the gauge invariance \p{21}, the sources satisfy the conservation condition
\begin{align}\label{Jconserv}
\pa^{++} J^{++} + \pa^{+\da} J^{+3}_\da =0\,.
\end{align}  
The linearized field equations are
\begin{align}\label{EOM}
\delta/\delta A^{++}: \quad \pa^{+\da}A^+_\da= J^{++}\,, \qquad \delta/\delta A^{+\da}: \quad  \pa^{+}_{\da} A^{++} - \pa^{++}A^+_\da = J^{+3}_\da\,.
\end{align}
They are solved in terms of Green's functions:
\begin{align}
&A^{++}(1)   = \int_2 \vev{A^{++}(1) A^{++}(2)}J^{++}(2)  + \int_2 \vev{A^{++}(1) A^{+\da}(2)}J^{+3}_\da(2) \label{AJ1} \\ 
&A^{+}_\da(1)   = \int_2 \vev{A^{+}_\da(1) A^{++}(2)}J^{++}(2)  + \int_2 \vev{A^{+}_\da(1) A^{+\db}(2)}J^{+3}_\db(2)\,. \label{AJ2}
\end{align}
{Here we use the shorthand $A^{++}(k) = A^{++}(x_k,\theta^+_k,u_k)$ and omit the integration measure $d^4 x_2 d u_2 d^4 \q^+_2 $.}
The  Green's function equations are obtained by substituting \p{AJ1}, \p{AJ2} in \p{EOM}, shifting the derivatives from point 1 to 2, integrating by parts and using \p{Jconserv}. 

Below we show that a solution to the Green's function equations and  the gauge fixing condition \p{417}  is given by the set of propagators\footnote{Similar propagators, in momentum space and for the bosonic YM theory, appear in \cite{Jiang:2008xw}. } \footnote{The momentum space equivalents are presented in Appendix \ref{MomentProp}.}
\begin{align}
&\langle A^{+}_{\da}(1) A^{+\db}(2) \rangle=0\,; \label{prop3} \\
&\vev{A^{++}(x,\q^+,u_1) A^{++}(0,0,u_2)}= \frac{1}{\pi}\delta^2( x^{\+ +} )\, \delta (u_1, u_2)\, \delta^{(4)}(\q^+)\,; \label{prop1}\\
&\langle A^{+}_{\da}(x,\q^+,u_1) A^{++}(0,0,u_2) \rangle=-\langle A^{++}(1) A^{+}_{\da}(2) \rangle=\frac{1}{\pi}\frac{\xi^{\-}_{\da}} { x^{\- +}  }\,
\delta^2( x^{\+ +} )\, 
\delta(u_1,u_2) \,\delta^{(4)}(\q^+)\,,  \label{prop2} 
\end{align}
where we have used translation invariance to set $x_2=\q_2=0$. {We omit for simplicity the color inidices $\delta_{ab}$ of the propagators.} The bosonic delta functions are defined in App.~\ref{apa2}, \ref{apa3}.  Notice that we do not specify if the projections \p{419} are made with the harmonic $u_1$ or $u_2$, in view of the harmonic delta functions above.

Let us check that this is a solution of  \p{EOM}. The gauge fixing condition \p{417} is obviously satisfied by \p{prop2}. From \p{aa21}, \p{a20}, \p{a21} and \p{a22} we get
\begin{align}
\pa^{+\da} \langle A^{+}_{\da}(1) A^{++}(2) \rangle &=   
\delta^2( x^{\- +} )\, \delta^2( x^{\+ +} )\,
\delta(u_1,u_2) \,\delta^{(4)}(\q^+)
\nt
& =  \delta^4(x) \,\delta(u_1,u_2) \,\delta^{(4)}(\q^+) \,; \label{dif1}
\\
\pa^{++} \langle A^{+}_{\da}(1) A^{++}(2) \rangle &= 
\frac{1}{\pi}\ \xi^{\-}_{\da}\, \pa_{x^{\- -}}\delta^2( x^{\+ +} ) \delta(u_1,u_2) \,\delta^{(4)}(\q^+) 
+ \nt & +
\frac{1}{\pi} \frac{\xi^{\-}_{\da}} { x^{\- +}  }\,
\delta^2( x^{\+ +} )\, 
\pa^{++} \delta(u_1,u_2) \,\delta^{(4)}(\q^+)\,;  \label{dif2}\\
\pa^{+}_{\da} \, \vev{A^{++}(1) A^{++}(2)} &= 
\frac{1}{\pi}\ \xi^{\-}_{\da}\pa_{x^{\- -}}\delta^2( x^{\+ +} ) \, \delta (u_1, u_2)\, \delta^{(4)}(\q^+)\,.  \label{dif3}
\end{align}

We start with the first equation in \p{EOM} 
and substitute \p{AJ2}, \p{prop2} and  \p{prop3} in it. Taking into account  \p{dif1} we get
$$
\pa^{+\da}A^{+}_\da(1)   = \int_2 \pa^{+\da} \vev{A^{+}_\da(1) A^{++}(2)}J^{++}(2) = J^{++}(1)\,.
$$
The second  equation in \p{EOM} is more subtle.
We substitute \p{AJ1}, \p{AJ2} and \p{prop3} in it:
\begin{align}
\int_2 \Big[ \pa^{+}_{\da} \vev{A^{++}(1) A^{++}(2)}J^{++}(2)  +  \pa^{+}_{\da} \vev{A^{++}(1) A^{+\db}(2)}J^{+3}_\db(2) 
-  \pa^{++}\vev{A^{+}_\da(1) A^{++}(2)}J^{++}(2) \Big]\,. \label{IIeqGF}
\end{align}
We need to show that this expression equals $J^{+3}_{\da}$. 
In the second term  we use translation invariance to  
bring $\pa^+_\da$ from the first point of the propagator $\vev{A^{++}(1) A^{+\db}(2)}$ to the second point. Then we integrate $\pa^+_\da$ by parts onto the source and use its conservation  \p{Jconserv},
$$
\pa^{+\da} J^{+3}_\db = \pa^{+}_{\db} J^{+3\,\da} + \delta^{\da}_{\db} \left(\pa^{+\dg}J^{+3}_{\dg} \right) =
\pa^{+}_{\db} J^{+3\,\da} - \delta^{\da}_{\db} \,\pa^{++} J^{++}\,.
$$
Thus in view of \p{dif1} the second term in \p{IIeqGF} equals
\begin{align}
\int_2 \pa^{+}_{\db} \vev{A^{++}(1) A^{+\db}(2)}J^{+3\,\da}(2) - \int_2 \vev{A^{++}(1) A^{+\da}(2)}\,\pa^{++} J^{++}(2) = \nt =
J^{+3\,\da}(1) - \int_2 \vev{A^{++}(1) A^{+\da}(2)}\,\pa^{++} J^{++}(2)\,. \label{IIeqGFb}
\end{align}
We substitute \p{dif2} in the third term in \p{IIeqGF}, identify the first term of \p{dif2} with \p{dif3}, {and
use \p{4.8.6b}  to} integrate $\pa^{++}$ in the second term of \p{dif2} by parts onto the source:
\begin{align}
- \int_2 \Big[ \pa^{+}_{\da}\vev{A^{++}(1) A^{++}(2)} J^{++}(2) 
+  \vev{A^{+}_\da(1) A^{++}(2)}\,\pa^{++}J^{++}(2)\Big] \,.\label{IIeqGFc}
\end{align}
The first term in \p{IIeqGFc} cancels the first term in \p{IIeqGF} and 
the second term in \p{IIeqGFc} cancels the second term on the right-hand side of \p{IIeqGFb} due to the property $\vev{A^{+}_\da(1) A^{++}(2)} = - \vev{A^{++}(1) A^{+}_\da(2)}$.

This concludes the verification of the propagators \p{prop3}--\p{prop2} in the  gauge \p{417}.

\subsection{Feynman rules}\label{s6.3}

The action \p{N4}, the light-cone gauge condition \p{417} and the propagators \p{prop3}--\p{prop2} yield simple Feynman rules:

\begin{figure}
\begin{center}
\includegraphics[width = 12 cm]{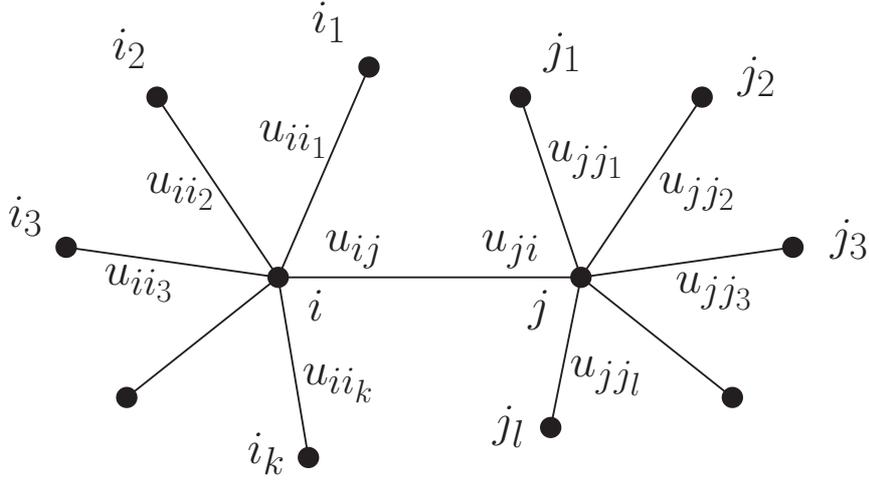}
\end{center}
\caption{Part of a supergraph. The two interaction vertices $i$ and $j$ of valences $k$ and $l$  are connected by a propagator $ij$. Two sets of LHs $u_{i j}, u_{i i_1}, \ldots, u_{i i_k}$ and $u_{j i}, u_{j j_1}, \ldots, u_{j j_l}$ are associated with the vertices. The harmonic delta functions in the propagator identify $u_{ij} = u_{ji}$. The vertices $i_1, i_2 , \ldots, i_k$ and $j_1 , j_2 ,\ldots, j_l$ can be either external or interaction.
}
\end{figure}

\begin{itemize}
\item 
To each internal (interaction) vertex $i$ connected with the vertices $j_1,j_2,\ldots, j_k$ we assign a set of LHs $u_{ij_1}, u_{ij_2}, \ldots, u_{ij_k}$.
\item An LH integral and factor are attached to the vertex, 
namely
\begin{align}\notag
 \om\,(-1)^k \int \frac{\prod_{l=1}^k du_{ij_l}}{(u_{ij_1}^+u_{ij_2}^+)(u_{ij_2}^+u_{ij_3}^+)\ldots (u_{ij_k}^+u_{ij_1}^+)}  \,.
\end{align}
\item An integration $\int d^4 x_i d^8 \theta_i $ is associated with each internal  vertex $i$. 
\item {The color structure at the vertex is ${\rm tr}\bigl( T^{a_{j_1}} \ldots T^{a_{j_k}} \bigr)$. The propagators are diagonal in the color space.}
\item 
{ The superfield $A^+_{\da}$ can appear only at external vertices, but not at interaction vertices.}
\item
We assign $\frac{1}{\pi}\delta^2(x_{ij}^{\+ +})\delta^4(\theta_{ij}^{+}) \delta(u_{ij},u_{ji})$
to each propagator $\vev{A^{++}(i) A^{++}(j)}$ connecting vertices $i$ and $j$. If vertex $i$ is external and  $A^+_\da(i)$ is present there, an additional factor $\xi^{\-}_\da/x^{\- +}_{ij}$ appears.
\item All the  Grassmann integrals are done with the help of $\delta^4(\theta_{ij}^{+})$ from the propagators.
\item
Using the harmonic delta functions $\delta(u_{ij},u_{ji})$ from the propagators, half of the harmonic integrations can be done immediately leading to the identification $u_{ij}=u_{ji}$. The other half of the harmonic integrals are lifted with the help of the bosonic  $\delta^2(x_{ij}^{\+ +})$ from the propagators. 
\end{itemize}

{These Feynman rules prove to be extremely useful in the calculation of correlation functions of gauge invariant operators.
The main reason is that the Born-level graphs do not contain interaction vertices, and so each relevant graph is a rational function. 
In \cite{Chicherin:2014uca} we used the twistor version of the Feynman rules to calculate the multipoint correlators of the chiral truncation of the stress-tensor supermultiplet. In \cite{twin} we use the LHC Feynman rules  for the study of the multipoint correlation functions of the full (non-chiral) stress-tensor supermultiplet.  We illustrate the Feynman rules in the next Section on the simple example of  the classical and quantum corrections to the propagators.}  

In conclusion, just a word about prescriptions. One of our propagators, eq.~\p{prop2}, contains a pole of the type $1/x^{\-+}$. After the harmonic integration at the vertices in a Feynman graph many more poles of the same type will appear (see Sect.~\ref{Full} for an example). Such poles are typical for the light-cone gauge. Indeed, we can write $x^{\-+} = (xn)$ where $n_{\a\da}= \xi^\-_\da u^+_\a$ is a light-like vector. Their treatment is known to be problematic and special care is needed when choosing the right prescription (see \cite{Leibbrandt:1987qv} for a review). Here we do not discuss this subtle issue because in this and the twin paper  \cite{twin} we are only interested in rational Born-level correlators. They do not involve non-trivial space-time integrals, where the prescription becomes relevant.

\subsection{Full propagators}\label{Full}

Above we found the set of propagators using the quadratic Lagrangian \p{91}.
We recall that there is one more quadratic term in the full Lagrangian coming from $L_{\rm Z}$ (see \p{3.23}). Following \cite{Boels:2006ir}, we chose to treat it as an interaction vertex. This is useful because the quadratic part of $L_{\rm CS}$ supplies rather simple propagators
which are very convenient in Feynman diagram calculations. 
The quadratic term in $L_{\rm Z}$ is rather different from the quadratic part of $L_{\rm CS}$
since it is non-local in the harmonics and it contains four extra Grassmann integrations.

Still, we can ask the question how  the extended quadratic Lagrangian modifies the form of  the propagators. Instead of repeating the entire derivation of the previous subsection, it is easier to 
calculate the corrections to the `bare' propagators \p{prop3}--\p{prop2}
inserting into them the bivalent vertices from $L_{\rm Z}$. 
Counting the Grassmann degrees of the relevant Feynman graphs we conclude that
each insertion of the quadratic vertex from $L_{\rm Z}$ lowers the Grassmann degree by 4 units. 
Since the `bare' propagators have Grassmann degree 4, then only a single insertion can give rise to a non-zero correction and the result does not depend on $\theta$. 
A straightforward calculation yields the corrections
\begin{align}
&\vev{A^{++}(x,\q^+,u_1) A^{++}(0,0,u_2)}_{\rm corr}= {-\frac{\om}{4 \pi^2}} \frac{(u_1^+ u_2^+)}{(u_1^- u_2^-)} \notag\\
&\langle A^{+}_{\da}(x,\q^+,u_1) A^{++}(0,0,u_2) \rangle_{\rm corr}= {\frac{\om}{4 \pi^2}} \xi^{\-}_{\da} \frac{(u_1^+ u_2^+)}{(x^{\-} u_{2}^-)} \notag \\
&\langle A^{+}_{\da}(x,\q^+,u_1) A^+_{\db}(0,0,u_2) \rangle_{\rm corr}= {\frac{\om}{4 \pi^2}}
\xi^{\-}_{\da} \xi^{\-}_{\db}\frac{(u_1^+ u_2^+)(u_1^- u_2^-)}{(x^{\-} u_{1}^-)(x^{\-} u_{2}^-)} \,, \label{corr3}
\end{align}
which are to be added to the bare propagators \p{prop3}--\p{prop2}. These full propagators
take into account the quadratic vertex from $L_{\rm Z}$. {If we use the full propagators instead of the bare ones, then}
all interaction vertices will be at least cubic  in the fields. Following the lines of  Sect.~\ref{s5.2}, we have checked  that the corrected propagators 
solve the analogs of the Green's function equations \p{EOM} corresponding to the 
extended quadratic Lagrangian.

As an illustration, let us calculate the last correction in \p{corr3} explicitly. The insertion of a bivalent vertex at point 0  depicted in Figure~\ref{figAA}
\begin{figure}
\begin{center}
\includegraphics[width = 7 cm]{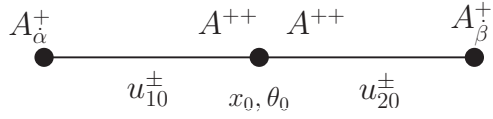}
\end{center}
\caption{Bivalent vertex correction to the bare propagator $\langle A^{+}_{\da}(1) A^+_{\db}(2) \rangle$.}
\label{figAA}
\end{figure}
contributes 
\begin{align}
\langle A^{+}_{\da}(1) A^+_{\db}(2) \rangle_{\rm corr} = & - \omega \int d^8\q_0 d^4 x_0 \int \frac{d u_{10} d u_{20}}{(u^+_{10} u^+_{20})^2} 
\vev{A^+_{\da}(x_1,\q^+_1,u_1) A^{++}(x_0,\q_0 u_{10}^+,u_{10})} \nt &\times
\vev{A^{++}(x_0,\q_0 u_{20}^+,u_{20})A^+_{\db}(x_2,\q^+_2,u_2)}\,. \notag
\end{align}
Substituting the explicit expressions for the propagators \p{prop2} we obtain 
\begin{align}
&\langle A^{+}_{\da}(1) A^+_{\db}(2) \rangle_{\rm corr}= {\frac{\om}{\pi^2}}
\int d^8\q_0 d^4 x_0 \int \frac{d u_{10} d u_{20}}{(u^+_{10} u^+_{20})^2} 
\frac{\xi^{\-}_{\da} }{(x_{10}^{\-} u^{+}_1)}\delta(u_1,u_{10})
\delta^4(\q_{10} u_1^+)\delta^2(x_{10}^{\+} u_1^+) \times \nt &
\frac{\xi^{\-}_{\db}}{(x_{20}^{\-} u^{+}_2)} \delta(u_2,u_{20})\delta^4(\q_{20} u_2^+) 
\delta^2(x_{20}^{\+} u_2^+) 
= {\frac{\om}{\pi^2}} (u_1^+ u_2^+)^2 \xi^{\-}_{\da}\xi^{\-}_{\db} 
\int d^4 x_0  \frac{\delta^2(x_{10}^{\+} u_1^+) \delta^2(x_{20}^{\+} u_2^+)}{(x_{10}^{\-} u^{+}_1)(x_{20}^{\-} u^{+}_2)}\,. \notag
\end{align}
On the support of the delta functions $(x_{10}^{\-} u^{+}_1) = - \frac{(x_{12}^{\-} u_2^-)}{(u_1^- u_2^-)}$ and
$(x_{20}^{\-} u^{+}_2) = -\frac{(x_{12}^{\-} u_1^-)}{(u_1^- u_2^-)}$. 
The four delta functions make the integration over $x_0$ trivial and we obtain \p{corr3}.

As another simple illustration of the Feynman rules from Sect.~\ref{s6.3} let us show that the first {and the second} quantum corrections 
(orders $O(g^2)$ and $O(g^4)$) to the two-point function $\vev{A^{++}(1) A^{++}(2)}$ vanish. There are two potentially contributing Feynman graphs {depicted in Figure~\ref{figg2}.}
\begin{figure}
\begin{center}
\includegraphics[width = 5 cm]{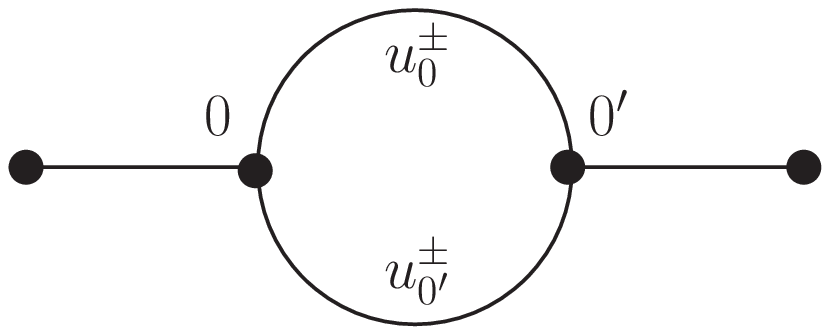} \qquad
\raisebox{23 pt}{\includegraphics[width = 5 cm]{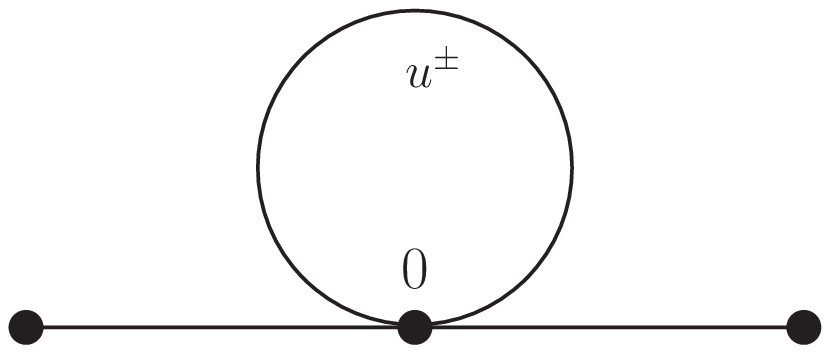}}
\end{center}
\caption{Graphs contributing to the two-point function $\vev{A^{++}(1) A^{++}(2)}$ at order $O(g^2)$.}
\label{figg2}
\end{figure}
The blobs denote the integration $\int d^4x d^8 \q$. Both  graphs vanish, indeeed. 
The loop in the first graph contains
\begin{align}\notag
\frac{1}{(u^+_0 u^+_{0'})^2}\delta^4(\q_{00'}u^+_{0}) \delta^4(\q_{00'}u^+_{0'}) \delta^2(x^{\+}_{00'} u^+_{0}) \delta^2(x^{\+}_{00'} u^+_{0'})  =
(\q_{00'})^8 (u^+_0 u^+_{0'})^2 \delta^2(x^{\+}_{00'} u^+_{0}) \delta^2(x^{\+}_{00'} u^+_{0'}) = 0
\end{align}
because on the support of the bosonic delta functions $u^+_0 \sim u^+_{0'}$. 
The second graph contains a self-closed line which is proportional to $\delta^4((\q_0-\q_0)u^+) = 0$.

{The LHC supergraphs of order $O(g^4)$ contributing to the two-point function are depicted in Figure~\ref{figg4}.}
\begin{figure}
\begin{center}
\begin{tabular}{ccc}
\includegraphics[width = 4 cm]{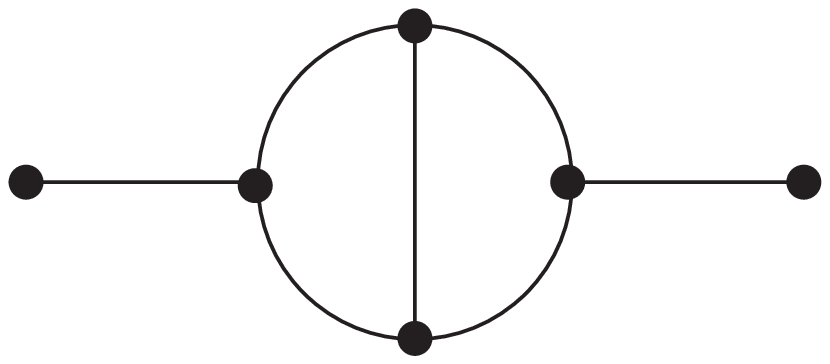} & \includegraphics[width = 4 cm]{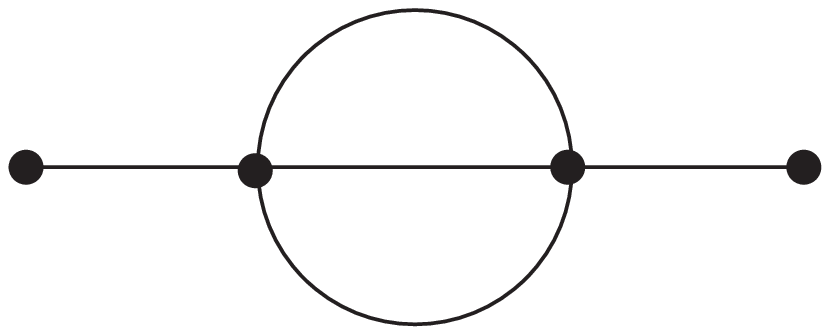} &\includegraphics[width = 4 cm]{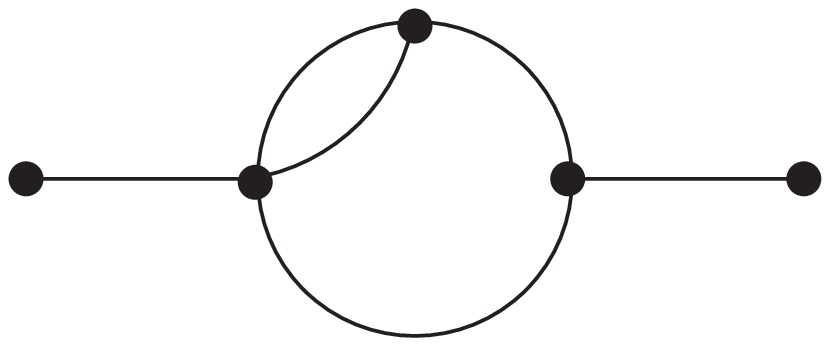} \\
\raisebox{-17 pt}{\includegraphics[width = 4 cm]{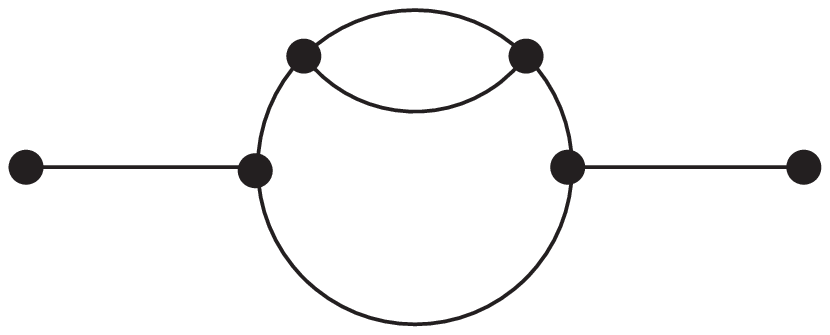}} & \includegraphics[width = 4 cm]{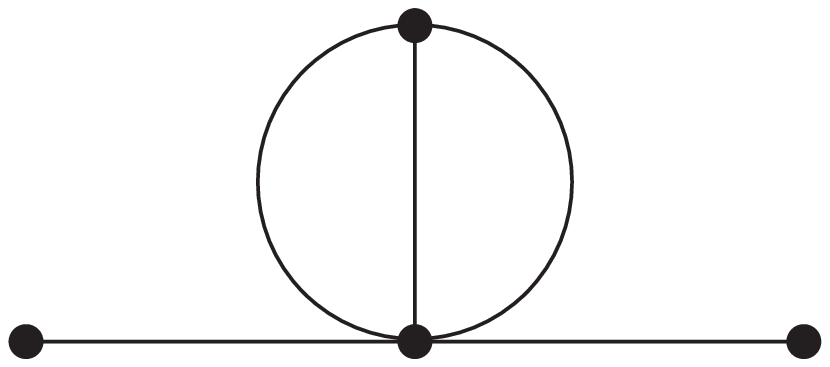} & \includegraphics[width = 4 cm]{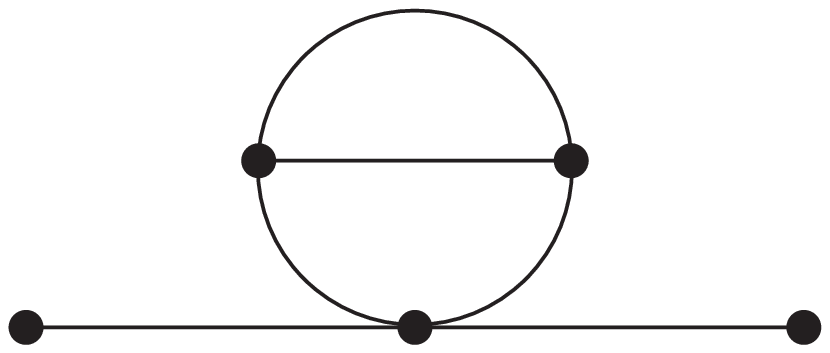}
\end{tabular}
\end{center}
\caption{Graphs contributing to the two-point function $\vev{A^{++}(1) A^{++}(2)}$ at order $O(g^4)$.}
\label{figg4}
\end{figure}
{All of them vanish. Indeed,  the formal counting of the Grassmann degree of the first graph gives $-4$, i.e. it has to be zero.
The remaining graphs contain the one-loop subgraph from Figure~\ref{figg2}. It vanishes as we have already explained above, so the whole graph containing it vanishes as well.}

{At higher loop orders there are graphs that do not vanish, at least not for obvious reasons. The above calculation is presented just for illustration purposes. Much more useful applications of the LHC supergraphs are worked out in \cite{twin} where we deal with correlation functions of composite operators.}

\section{Analogy with $\cN=2$ harmonic superspace}\label{n2}

$\cN=4$ SYM can be formulated in terms of $\cN=2$ superfields in harmonic superspace \cite{Galperin:1984av,Galperin:2001uw}. The main tool there are the harmonic variables on $SU(2)$, but for the R-symmetry group rather than half of the Lorentz group as in this paper. Like the formulation in LHC superspace, the $\cN=2$ formulation has only half of the $\cN=4$ supersymmetry manifest, but it is a different half: there one has two $Q$ and two $\bQ$ manifest supersymmetries, here we have  all four $Q$'s but no $\bQ$'s. Still, both the formulation and the way the non-manifest half of $\cN=4$ supersymmetry is realized are strikingly similar in the two harmonic superspaces. 

Here we give a very brief summary of the $\cN=2$ formulation, referring the reader to \cite{Galperin:2001uw} for more detail. The notation of \cite{Galperin:2001uw} is  modified in order to make the similarity with the LHC formulation more transparent. We make use of Grassmann (or G-)analytic superfields
\begin{align}\label{61}
 \Phi(z_{\rm an}^{a' a},\q^{+ a}, \bq^{+ a'}, u)\,, \qquad {\rm where} \ \q^{+ a} = u^+_{\a} \q^{\a a}\,, \quad  \bq^{+ a'} = u^+_{\a} \bq^{\a a'}\,.
\end{align}
Here $\a=1,2$ is an index of the R-symmetry group $SU(2)$ and $a,a'=1,2$ are indices of the Euclidean Lorentz group $SO(4) \sim SU(2)_L\times SU(2)_R$. The space-time variable $z^{a' a}$ (to be distinguished from $x^{\da\a}$ in \p{chan}) is defined in the analytic basis
\begin{align}\label{62}
z_{\rm an}^{a' a} = z^{a' a}  + \bq^{+ a'} \q^{-a} + \bq^{- a'} \q^{+a}\,.
\end{align}
From the point of view of $\cN=4$ supersymmetry these are `semi-superfields' because they do not depend on the other half of the Grassmann variables, $ \q^{\da a}, \bq^{\da a'}$ where $\da=1,2$ completes the $SU(4)$ index $A=(\a, \da)$. The G-analytic superfields \p{61} depend only on a quarter of the $\cN=4$ odd variables, hence the $\cN>2$ half of the supersymmetry will be realized non-linearly and on shell.

We make use of two basic superfields  
\begin{equation}\label{2.3'}
A^+_\da = A^+_\da (z_{\rm an},\q^{+ }, \bq^{+ }, u) \,, \qquad A^{++} = A^{++}(z_{\rm an},\q^{+ }, \bq^{+ }, u) \,.
\end{equation}
The first of them describes the $\cN=2$ hypermultiplet matter (HM), with an additional index $\da=1,2$ of the automorphism Pauli-G\"ursey group $SU(2)$. The second is the gauge connection (prepotential) for the supersymmetrized harmonic derivative
\begin{align}\label{B4}
D^{++}= \pa^{++} -2 \q^{+a} (\pa_{z_{\rm an}})_{aa'} \bq^{+a'}  \qquad \Rightarrow \qquad  \nabla^{++} = D^{++} +     A^{++}\,,
\end{align} 
with respect to a gauge group with  {analytic parameters} 
\begin{equation}\label{21'} \delta A^+_{\da} = \ [A^+_{\da}, \Lambda]\,,  \ \ \ \delta A^{++} = D^{++} \Lambda
+ [A^{++}, \Lambda]\,, \qquad \Lambda=\Lambda(z_{\rm an},\q^{+ }, \bq^{+ }, u) \,.
\end{equation}
Note that the HM $A^+_\da$ now transforms as a matter field, not as a gauge field as in \p{21}. The reason is the absence of a coordinate like $x^-_\da$ and the corresponding derivative $\pa^+_\da$, whose covariantization requires the gauge field in \p{3.3}.  Another difference is the space-time derivative term in the harmonic derivative $D^{++}$ due to the change of basis \p{62}.

\subsection{Action}

The $\cN=4$ SYM action for the superfields \p{2.3'} consists of two terms,
\begin{align}\label{N4'}
S_{\cN=4} = S_{\rm HM} + S_{\rm Z}  \,.
\end{align}
The first  involves only the HM  $A^+_\da$, the second  depends only on $A^{++}$ and is the original Zupnik's action  \cite{Zupnik:1987vm}. In the linearized (Abelian) theory the action \p{N4'} has the form
\begin{align}\label{2.7'}
 S_{\cN=4} &=  -{1\over 2}\int d^4z_{\rm an} du d^2\q^+ d^2\bq^+\;    A^{+\da}D^{++} A^+_\da\nt
& - \frac{\om}{2} \int d^4z d^4\q d^4\bq du_1 du_2\; {A^{++}(z_{1\, \rm an}, \q^+_1,\bq^+_1,u_1) A^{++}(z_{2\, \rm an}, \q^+_2,\bq^+_2,u_2) \over
(u^+_1u^+_2)^2} + O(g) \,, 
\end{align}
where $\q^+_i= \q^\a (u_i)^+_\a\,, \ \bq^+_i= \bq^\a (u_i)^+_\a$ and $z_{i\, \rm an} = z  + \bq^{+ }_i \q^{-}_i + \bq^{- }_i \q^{+}_i$, with $i=1,2$. Note that the first term involves a G-analytic superspace integral, while the second has a full $\cN=2$ superspace integral. 

In the interacting (non-Abelian) theory the action \p{2.7'} becomes 
\begin{align}\label{2.7''}
 S_{\cN=4} &=  \int d^4z_{\rm an} du d^2\q^+ d^2\bq^+\;    \tr \left(-{1\over 2}A^{+\da}D^{++} A^+_\da + A^{++} A^{+\da}  A^+_\da \right) \nt
& +  \om \int d^4z d^4\q d^4\bq  \; \log\det \left( \frac1{D^{++}} \nabla^{++}\right) \,.
\end{align}
Here the HM action has an additional cubic term, which is the standard gauge-matter coupling. Compared to the Chern-Simons action \p{CS}, the only difference is the absence of the mixed kinetic term $A^{++}\pa^{+\da}A^+_\da$. This is again due to the absence of a coordinate like $x^-_\da$ and the corresponding derivative $\pa^+_\da$ in the $\cN=2$ formalism. The non-polynomial original Zupnik's term is almost identical with the LHC action \p{sint}, \p{logdet}. The only difference is the Grassmann shift of the coordinates $z_{i\, \rm an}$ in \p{2.7'}.

\subsection{Equations of motion and component field content}

The linearized field equations following from \p{2.7'} are
\begin{align}\label{2.8'}
&\frac{\delta   S_{\cN=4}}{\delta A^{+\da}}: \qquad D^{++} A^+_\da = 0 \\
&\frac{\delta   S_{\cN=4}}{\delta A^{++}}: \qquad (D^+)^4 A^{--} = 0 \label{2.9'}
\end{align}
where $A^{--}$ is the non-analytic connection for the harmonic derivative $D^{--}$. It is given by the non-polynomial and non-local (in the harmonic space) expression
\begin{align}\label{2.10'}
A^{--}(z,\q,\bq,u) = \int du_2   {A^{++}(z_{2\, \rm an}, \q^+_2,\bq^+_2,u_2) \over
(u^+ u^+_2)^2} + O(g)\,.
\end{align}
The role of the supersymmetric spinor derivatives $(D^+)^4=(D^+_a)^2 (\bar D^+_{a'})^2$ in  \p{2.9'} is to project $A^{--}$ onto the G-analytic subspace. These equations of motion are to be compared with the (linearized) LHC ones, \p{3.6} and \p{3.14'}, respectively. The main difference is that the field equations \p{2.8'}, \p{2.9'} are decoupled, once again due to the absence of $\pa^+_\da$. 

The component field content of the gauge prepotential is revealed in the WZ gauge
\begin{align}\label{2.11'}
A^{++}= &(\q^+)^{2}\phi(z) + (\bq^+)^{2}\bar\phi(z) + \bq^{+a} \q^{+a'} \cA_{aa'}(z) \nt
&+ (\bq^+)^{2} \q^{+a} \psi^{-}_a(z) 
+  (\q^+)^{2} \bq^{+a'} \bar\psi^-_{a'}(z) + 3(\q^+)^{2}(\bq^+)^{2} u^{-a} u^{-\b}G_{\a\b}(z)\,.
\end{align}
Here we see the full content of the $\cN=2$ \emph{off-shell} vector multiplet, including the triplet of \emph{auxiliary fields } $G_{\a\b}$. Inserting \p{2.11'} in the second field equation \p{2.9'}, we obtain the free equations of motion for the component fields. In particular, the auxiliary field satisfies the homogeneous equation $G_{\a\b}=0\,$,
in contrast with the inhomogeneous  equation $\om G=F$ following from the action \p{3.42}. This time the YM equation is obtained from \p{2.9'} directly in the second-order form $\pa_z^\mu F_{\mu\nu}(z)=0\,$.

The first field equation \p{2.8'} determines the content of the \emph{on-shell} $\cN=2$ HM:
\begin{align}\label{2.14'}
A^+_\da = f^\a_\da(z) u^+_\a +   \q^{+a} \chi_{a \da}(z) + \bq^{+a'} \bar\chi_{a' \da}(z) +2 \q^{+a} \bq^{+a'} \pa_{aa'}  f^\a_\da(z) u^-_\a \,,
\end{align} 
where the physical fields (the four scalars $f^\a_\da$ and the two Dirac fermions $\chi_\da, \bar\chi_\da$)  satisfy their free equations of motion $
\Box_z f^\a_\da = \pa^{a' a}_z    \chi_{a \da} =0\,$.

We observe that the $\cN=4$ vector multiplet is split in two halves. The prepotential $A^{++}$ contains 2 of the 6 physical scalars, the gauge field and half of the gluinos (the triplet of auxiliary fields $G$ vanishes on shell). The HM $A^+_\da$ contains the remaining 4 scalars and the other half of the gluinos.

\subsection{Realization of the hidden $\cN>2$ supersymmetry}

The two extra
supersymmetries are realized in terms of the gauge and matter
superfields as follows (see \cite{Galperin:2001uw} and also \cite{Bossard:2015dva} for a $d=6$ harmonic superspace version):
\begin{equation}\label{2.16'}
\delta A^{++} = (\epsilon^{\da a}\theta^+_a +
\bar\epsilon_{a'}^\da \bar\theta^{+ a'})\, A^+_\da\,, \qquad \delta
A^+_\da = \om (D^+)^4 [(\epsilon^a_\da \theta^-_a +
\bar\epsilon_{a' \da} \bar\theta^{-a'}) A^{--}]\,,
\end{equation}
where $\epsilon^a_\da$ and $\bar\epsilon^{a' \da} \equiv
\overline{(\epsilon^a_\da)}$ are the relevant Grassmann parameters. The variation of $A^+_\da$ is non-linear and non-local in the harmonic space because of the presence of $A^{--}$ in it (see \p{448}). The supersymmetry algebra closes modulo field equations and compensating gauge transformations, as usual for on-shell supersymmetry.

Note the similarity with the realization of the hidden $\bQ$ supersymmetry in \p{4.27}, \p{4.28'}. The absence of space-time derivative terms in \p{2.16'} is due to the fact that   now $\pa_x A=0$. 

\subsection{Interpretation as eight-dimensional SYM}

We have shown the remarkable similarity of the two formulations, up to a few details. A possible interpretation is that both are obtained from a master SYM theory in eight-dimensional Euclidean superspace enhanced with a harmonic two-sphere,\footnote{We are grateful to Evgeny Ivanov for a discussion on this point. } 
\begin{align}\notag
R^8_{x,z} \times S^2_u  \,,  
\end{align}
having 8 odd coordinates $(\q^{\a a},\bq^{\a a'})$. The full $d=8$ superspace has twice as many odd coordinates, the missing ones being   $(\tilde\q^{\da a},\tilde\bq^{\da a'})$. Here the $d=8$ Lorentz group $SO(8)$ is broken down to $SO(4)\times SO(4) \sim [SU(2)]^4$, and one of the $SU(2)$ factors is `harmonized'.
On this superspace we consider the analytic superfields $A^{++}(x,z,u; \q^+,\bq^+)$ and $A^+_\da(x,z,u;\q^+,\bq^+)$ depending only on the harmonic projected odd variables $\q^{+a}=\q^{\a a} u^+_\a$, $\bq^{+a'}=\bq^{\a a'} u^+_\a$. They are gauge fields for the harmonic derivative $D^{++}$  defined in \p{B4} and for the space-time derivative $u^{+\a}(\pa_x)_{\a\da}$, respectively. The linearized action is a hybrid of  \p{CS} and \p{2.7'}:
\begin{align}\label{B18}
 S_{d=8} &=  \int d^4x d^4z_{\rm an} du d^2\q^+ d^2\bq^+\;  \left(A^{++} \pa^{+\da}_x A^+_\da   -
{1\over 2} A^{+\da}D^{++} A^+_\da\right)\nt
& - \frac{\om}{2}  \int d^4x d^4z  d^4\q d^4\bq\, du_1 du_2\; {A^{++}(x,z_{1\, \rm an}, \q^+_1,\bq^+_1,u_1) A^{++}(x,z_{2\, \rm an}, \q^+_2,\bq^+_2,u_2) \over
(u^+_1u^+_2)^2}  \,, 
\end{align}
where $\q^+_i= \q^\a (u_i)^+_\a\,, \ \bq^+_i= \bq^\a (u_i)^+_\a$ and $z_{i\, \rm an} = z  + \bq^{+ }_i \q^{-}_i + \bq^{- }_i \q^{+}_i$, with $i=1,2$. The field content of the theory is seen in the WZ gauge for $A^{++}$ and after solving {the analog of the kinematical} constraint \p{3.53} on $A^+_\da$,  $D^{++} A^+_{\da} = (\pa_x)^+_{\da} A^{++}$ (bosons only { and Abelian approximation})
\begin{align}\label{2.11''}
A^{++}&= (\q^+)^{2}\phi(x,z) + (\bq^+)^{2}\bar\phi(x,z)  -2 \q^{+a} \bq^{+a'} \cA_{aa'}(x,z)  + 3(\q^+)^{2}(\bq^+)^{2} u^{-\a} u^{-\b}G_{\a\b}(x,z)\nt
A^+_\da &= u^{+\a}\cA_{\a\da}(x,z)  +  (\q^+)^{2} u^{-\a}(\pa_x)_{\a\da}\phi + (\bq^+)^{2}u^{-\a}(\pa_x)_{\a\da} \bar\phi   -2  \q^{+a} \bq^{+a'} u^{-\a} \cF_{aa'\, \a\da} \nt & \qquad + (\q^+)^{2}(\bq^+)^{2} u^{-\a} u^{-\b}u^{-\gamma}\pa_{\a\da} G_{\b\gamma} \ .
\end{align} 
Here $\cA_{aa'}, \cA_{\a\da}$ are the two halves of the $d=8$ gauge field and $\phi,\bar\phi$ are the two scalars of $d=8$ SYM. The field $G_{\a\b}$ is a Lagrange multiplier  which becomes identified on shell with part of the $d=8$  field strength, $\om G_{\a\b}=\cF_{\a\b}= (\pa_x)^\da_{(\a} \cA_{\b)\da}$.  Another part, $\cF_{aa'\, \a\da}=(\pa_x)_{\a\da} \cA_{aa'}   - (\pa_z)_{aa'}  \cA_{\a\da}$, appears on shell as a component of $A^+_\da$ in \p{2.11''}.   The last part $\cF_{aa'\, bb'}=(\pa_z)_{aa'} \cA_{bb'}   - (\pa_z)_{aa'}  \cA_{bb'}$ is contained in the second line in  \p{B18}. The action \p{B18} exhibits only  half of $d=8$ supersymmetry, the other half is realized non-linearly. 

This master $d=8$ theory gives rise to two $d=4$ reductions. In the first we suppress the dependence on the bosonic space-time coordinates $z^{a'a}$ and combine the odd coordinates into $\q^{\a A}=(\q^{\a a},\bq^{\a a'})$ carrying an $SU(4)$ R-symmetry index. The gauge field $\cA_{aa'}$ is unified with the two scalars $\phi,\bar\phi$ to form the sextuplet of scalars $\phi^{AB}$.  Thus, we recover the LHC formulation of $\cN=4$ SYM from this paper. 
In the second case, we suppress the dependence on $x^{\da \a}$ instead. The gauge field $\cA_{\a\da}$ becomes the four scalars of the $\cN=2$ matter HM, while $\phi,\bar\phi$ are part of the $\cN=2$ gauge multiplet. In the second scenario the Lorentz index $\a$ becomes the $SU(2)$ R-symmetry index, while $\da$ is identified with the Pauli-G\"ursey automorphism of the HM. The gluinos are split into halves, chiral in $A^{++}$ and anti-chiral in $A^+_\da$ in the first scenario, or  $SU(2)\times SU(2)\times U(1)$ halves in the second. 

The Chern-Simons Lagrangian looses its first term  in the second scenario due to the absence of $x$, thus becoming the HM Lagrangian.  The space-time variable $x$ in the first case is supposed to be in the chiral basis, but because of the absence of the antichiral $(\tilde\q^{\da a},\tilde\bq^{\da a'})$ we cannot tell the difference. In the second scenario the space-time variable $z$ is in the G-analytic basis \p{62}, and we can see it because both $\q^{\a a}$ and $\bq^{\a a'}$ are present.

\section{Conclusions}

In this paper we have presented a formulation of $\cN=4$ SYM in LHC superspace. It is based on the use of auxiliary variables (Lorentz harmonics) which allow us to have the chiral half of the supersymmetry manifest, together with the full R-symmetry $SU(4)$. The other half  is realized in a rather non-trivial way on the LHC dynamical semi-superfields. In the twin paper \cite{twin} we will use it to construct the full non-chiral stress-energy tensor supermultiplet in terms of LHC semi-superfields. This result, together with the Feynman rules derived in the present paper will allow us to compute the correlation functions of the full stress-tensor multiplet  at Born level. 

One issue that we mentioned in the beginning of the Introduction has not been addressed in this work. It is the  notoriously difficult `$\cN=4$ barrier' for an off-shell formulation of SYM theory with manifest supersymmetry. In this paper we discussed two approaches where only half of the $\cN=4$ supersymmetry is manifest -- the chiral half in LHC superspace or the $\cN=2$ half in the original harmonic superspace approach of \cite{Galperin:1984av}. It is also possible to have three quarters of the supersymmetry off shell in the $\cN=3$  harmonic superspace approach of \cite{Galperin:1984bu}, but $\cN=4$ has so far resisted all attempts. Could LHs or twistors help break this barrier? A natural idea seems to be to use LHs on both halves of the Lorentz group $SO(4) \sim SU(2)_L\times SU(2)_R$. This approach, known also as `ambitwistor', has been advocated in \cite{Witten:1978xx,Devchand:1993rt,Mason:2005kn}, but it seems that it again stops at the step $\cN=3$. Hopefully, the future will tell us what else we need.

\section*{Acknowledgements}

 We profited from numerous discussions with T. Adamo, I. Bandos, B. Eden, P. Heslop, E. Ivanov,  L. Mason and D. Skinner.  We are particularly grateful to G. Korchemsky for making comments on the draft. We acknowledge partial support by the French National Agency for Research (ANR) under contract StrongInt (BLANC-SIMI-4-2011). The work of D.C. has been supported by the ``Investissements d'avenir, Labex ENIGMASS'' and partially supported by the RFBR grant 14-01-00341.

\section*{Appendices} 

\appendix

\section{Lorentz harmonics and harmonic distributions}\label{Hd}

In this Appendix we give a summary of the main formulas of the harmonic analysis on $SU(2)$. More details can be found in \cite{Galperin:1985bj,Galperin:2001uw}.

\subsection{Harmonic coset and invariant integral}

In this paper we are dealing with harmonics $u^\pm_\a$ on one half of the Euclidean Lorentz group $SO(4) \sim SU(2) \times SU(2)$. 
To give an explicit parametrization of the coset it is convenient to use stereographic coordinates:
\begin{equation}
\parallel u\parallel\   = \left(\begin{array}{cc} u^+_1 & u^-_1 \\
u^+_2 & u^-_2 \end{array} \right) = {1 \over \sqrt{1+t\bar t}}
\left(\begin{array}{cc} e^{i\psi} & -\bar t e^{-i\psi} \\
te^{i\psi} & e^{-i\psi}\end{array} \right), \qquad 0\leq \psi < 2\pi\,.
\label{4.1.3}\end{equation} The phase $\psi$ is the coordinate on the subgroup
${U}(1)$. The harmonic functions \p{3} are by definition homogeneous in $e^{i\psi}$:
\begin{equation}
f^{(q)}(t,\bar t, \psi) = e^{iq\psi} f^{(q)}(t,\bar t)\,,
\label{4.2.1}\end{equation} thus they are equivalent to functions on the coset
${SU}(2)/{U}(1)\sim S^2$ parametrized by the complex variable $t$.

In the parametrization \p{4.1.3} the invariant integral on
${SU}(2)$ has the form
\begin{equation}
\int du\;  f^{(q)}(u) \equiv {i\over 4\pi^2} \int^{2\pi}_0 d\psi \int {dt\wedge
d\bar t\over (1+t\bar t)^2} f^{(q)}(t,\bar t,\psi)\,.
\label{4.2.2}\end{equation}  
The invariant measure is obtained by multiplying the three independent Cartan forms (up to normalization),  $du= (u^{+\a} du^+_\a) (u^{-\b} du^-_\b)(u^{-\gamma} du^+_\gamma)$. Obviously, if $q\neq 0$ the $\psi$ integral in \p{4.2.2} vanishes, so we
derive our first integration rule:
\begin{equation}
\int du\;  f^{(q)}(u) = 0 \qquad \mbox{if} \ q\neq 0\,.
\label{4.2.3}\end{equation} The second rule
\begin{equation}
\int du\; 1 = 1 \label{4.3.1}\end{equation} is just the
normalization condition for the integral in \p{4.2.2}. Finally, our third rule
is
\begin{equation}
\int du\; u^+_{(\a_1}\ldots u^+_{\a_n}u^-_{\b_1}\ldots u^-_{\b_n)} = 0
\qquad \mbox{for} \ n\geq 1\; . \label{4.3.2}\end{equation}
 It follows from the fact that the
harmonic variables $u^\pm_\a$ transform under the fundamental representation of 
${SU}(2)$, while the measure is ${SU}(2)$ invariant; thus the
left-hand side in \p{4.3.2} must be an invariant, constant and totally
symmetric tensor of ${SU}(2)$, which does not exist. {The three rules are summarized in eq.~\p{6}.}

An important property of the harmonic integral \p{4.2.2} is the vanishing
of the integral of a total derivative {(recall \p{2.10})}:
\begin{equation}
\int du\;  \pa^{++}f^{(-2)}(u)  =0\,.
\label{4.4.4}\end{equation} 
It follows from the facts that the integral
projects out the singlet part of the integrand whereas the charged function
$f^{(-2)}$ in \p{4.4.4} does not contain an ${SU}(2)$ singlet. A more
direct proof can be given using the parametrization \p{4.1.3} of the harmonic
variables, the invariant measure \p{4.2.2} and the expression for $\pa^{++}$ in
this parametrization
\begin{equation}
\pa^{++} = e^{2i\psi} \left[-(1+t\bar t){\partial\over \partial
\bar t} + {it\over 2} {\partial\over \partial\psi}\right]\,.
\label{4.5.1}\end{equation}
Substituting \p{4.2.2}, \p{4.5.1} and \p{4.2.1} into \p{4.4.4},
one finds
\begin{eqnarray}
\int du\;  \pa^{++}f^{(-2)}(u) &=& {i\over 4\pi^2} \int^{2\pi}_0 d\psi \int
{dt\wedge d\bar t\over (1+t\bar t)^2}\left[-(1+t\bar t){\partial f\over
\partial\bar t } + tf\right] \nonumber\\ &=& {1\over 2\pi i}\int dt\wedge
d\bar t\; {\partial\over
\partial\bar t} \left({f\over 1+t\bar t}\right) =0.
\end{eqnarray}
The last integral vanishes since the function $f(t,\bar t)$
satisfies suitable boundary condition, being globally defined on $S^2$.

\subsection{Harmonic distributions} \label{apa2}

The first singular harmonic distribution that we need is the  delta function. It is defined
by the property
\begin{equation}
\int dv\; \delta^{(q,-q)}(u,v)\; f^{(q)}(v) = f^{(q)}(u)\,,
\label{4.7.4}\end{equation} where $f^{(q)}(u)$ is a test function.
In the parametrization \p{4.1.3} the harmonic delta function has the form
\begin{equation}
\delta^{(q,-q)}(u_1,u_2) = \pi e^{iq(\psi_1-\psi_2)}(1+t_1\bar t_1)^2
\delta^2(t_1-t_2)\,, \label{4.8.1}
\end{equation} 
where $\delta^2(t) \equiv \delta(t,\bar t)$ is the delta function on the complex plane. The harmonic delta functions can be differentiated in a natural
way, e.g.,
\begin{equation}
\pa^{++}_2\delta^{(q,-q)}(u_1,u_2) =
-\pa^{++}_1\delta^{(q-2,2-q)}(u_1,u_2)\,. \label{4.8.6b}
\end{equation}

Another important distribution is 
\begin{equation}
{1\over (u^+_1u^+_2)} = - {1\over (u^+_2 u^+_1)}\,, \label{4.8.7}\end{equation} where
\begin{equation}
(u^+_1u^+_2) \equiv u^{+\a}_1u^+_{2\a} = e^{i(\psi_1+\psi_2)} {t_2-t_1\over
\sqrt{(1+t_1\bar t_1)(1+t_2\bar t_2)}}\,. \label{4.8.8}\end{equation} 
Acting on it, the harmonic derivative $\pa^{++}$ produces a delta function:
\begin{eqnarray}
\pa^{++}_1 {1\over (u^+_1u^+_2)} &=& e^{i(\psi_1-\psi_2)}(1+t_1\bar t_1)^{3/2}
(1+t_2\bar t_2)^{1/2} {\partial\over \partial\bar t_1}{1\over t_1-t_2}
\nonumber\\
&=& \pi e^{i(\psi_1-\psi_2)}(1+t_1\bar t_1)^2\delta^2(t_1-t_2) \nonumber\\ &=&
\delta^{(1,-1)}(u_1,u_2)\,. \label{4.9.1}\end{eqnarray} 
Here we assume that the
singular distribution $t^{-1}$ is defined so that the relation
\begin{equation}
{\partial\over \partial\bar t}\;\frac1{t} = \pi \delta^2(t)
\label{4.9.2}\end{equation} holds. In a similar manner, one can define the distribution $1/(u^+_1u^+_2)^2$ with the property
\begin{equation}
\pa^{++}_1{1\over (u^+_1u^+_2)^2} =  \pa^{--}_1
\delta^{(2,-2)} (u_1,u_2)\,. \label{4.10.2}\end{equation}

The chargeless harmonic delta function can be written as a complex delta function  with argument $z\equiv (u^+_1u^+_2)$:
\begin{align}\label{a.18}
\delta^{(0,0)}(u_1,u_2) = \pi  \delta^2\left((t_1-t_2)(1+t_1\bar t_1)^{-2}\right) = \pi \delta^2\left((u^+_1u^+_2)\right)\,,
\end{align}
using the fact that $\delta^2(t) \equiv \delta(t,\bar t)$ is invariant under phase rotations of the complex argument. Note also the relation $\delta^{(q,-q)}(u_1,u_2)  = (u^+_1 u^-_2)^q \delta^{(0,0)}(u_1,u_2) $.

\subsection{Mixed harmonic/space-time distributions}  \label{apa3}

In Sect.~\ref{s6} we encounter the singular distribution $1/x^{\- +}$, i.e. the inverse of one of the light-cone projections \p{419}. The latter can be treated as pairs of complex conjugate variables,
\begin{align}\label{cv}
z \equiv x^{\+ +}\,, \quad   \bar z \equiv x^{\- -} \qquad {\rm and} \qquad \zeta \equiv x^{\- +}\,, \quad \bar\zeta \equiv x^{\+ -}\,.
\end{align}
Here we assume that $x$ is a hermitian matrix while the left- and right-handed LHs $u^\pm$ and $\xi^{\dt\pm}$ have the conjugation properties \p{2} and \p{6.4}, respectively.  In terms of the new variables \p{cv} the space-time derivative becomes (see \p{6.5})
\begin{align}\notag
\pa_{\a\da} = \xi^{\+}_\da u^{+}_{\a}   \pa_{z}  + \xi^{\-}_{\da} u^{-}_{\a}   \pa_{\bar z} + \xi^{\+}_{\da} u^{-}_{\a}  \pa_{\bar\zeta} + \xi^{\-}_{\da} u^{+}_{\a}   \pa_{\zeta}\,,   
\end{align}
or projecting with $u^{+\a}$,
\begin{align}\label{aa21}
\pa^+_\da =  \xi^{\-}_{\da}   \pa_{\bar z} + \xi^{\+}_{\da}   \pa_{\bar\zeta}\,.
\end{align}
From this we find 
\begin{align}\label{a20}
\pa^+_\da\frac{1}{x^{\- +}} =   \Big(\xi^{\-}_{\da}   \pa_{\bar z} + \xi^{\+}_{\da}   \pa_{\bar\zeta}\Big) \frac1{\zeta} = \pi\xi^{\+}_{\da}  \delta^2(x^{\- +}) \,.
\end{align}
Here the delta function is to be understood in the complex sense, $\delta^2( x^{\- +}) \equiv \delta( \zeta, \bar\zeta)$. 

In the light-cone basis the harmonic derivative $\pa^{++}$ becomes
\begin{align}\label{a21}
\pa^{++} = \pa^{++}_u +   x^{\- +} \pa_{\bar z} +   x^{\+ +} \pa_{\bar\zeta}\,,
\end{align} 
so that $\pa^{++} x^{\dt\pm -} = x^{\dt\pm +}$. With this we find
\begin{align}\label{a22}
\pa^{++}\frac{1}{x^{\- +}} =   x^{\+ +}\pa_{\bar\zeta} \frac{1}{\zeta} = \pi x^{\+ +} \delta^2(x^{\- +}) \,.
\end{align}

Finally, the harmonic integral $\int du\, \delta^2( x^{\+ +})$ is to be understood as follows. The support of the complex delta function $\delta^2( x^{\+ +}) = \delta(x^{\+ +}, x^{\- -})$ is given by
\begin{align}\notag
x^{\+ +} =0 \ \rightarrow \   x^{\+\a} = a u^{+\a}\,, \qquad x^{\- -} =0 \ \rightarrow \   x^{\-\a} = b u^{-\a}\,.
\end{align}
Remembering the defining properties \p{2}, we find $a^* = b$ and $|a|^2 = x^2$ since $x^{\+\a}  x^{\-}_\a = x^2 \xi^{\+\da}  \xi^{\-}_\da =x^2$. We can then choose $a=b= \sqrt{x^2}$ up to an insignificant $U(1)$ phase (the freedom in the definition of the $SU(2)$ harmonics). This yields the identifications
\begin{align}\label{5.39}
x^{\+ \a} =  \sqrt{x^2} \,  u^{+ \a} \;\;,\;\;  x^{\-\a}=\sqrt{x^2}u^{-\a} 
\end{align}
and
$x^{\- +} =  -\sqrt{x^2}$.
So, the harmonic integration produces a Jacobian factor,  
\begin{align}\label{6.9}
\int du\, \delta^2( x^{\+ +})= \frac1{x^2} \int d u\, \delta^2\left(\frac{x^{\+}u^{+}}{\sqrt{x^2}} \right) = \frac1{\pi x^2} \int d u\, \delta^{(0,0)}(v,u) =\frac1{\pi x^2} \,,
\end{align}
where we have used \p{a.18}, treating $v^+ \equiv x^{\+}/ \sqrt{x^2}$ as another harmonic variable.


\section{Propagators in momentum space} \label{MomentProp}

In Section \ref{s6} we quantized the theory in the light-cone gauge \p{417} and found the set of propagators \p{prop3} -- \p{prop2}
which solve the Green's function eqsuations \p{EOM}. A similar calculation in momentum space yields the following set of propagators,
\begin{align}
&\langle A^{+}_{\da}(p,\q^+,u_1) A^{+\db}(-p,0,u_2) \rangle=0\,, \\
&\vev{A^{++}(p,\q^+,u_1) A^{++}(-p,0,u_2)}= 4\pi \delta^2( p^{\- +} )\, \delta (u_1, u_2)\, \delta^{4}(\q^+)\,, \label{prop1m}\\
&\langle A^{+}_{\da}(p,\q^+,u_1) A^{++}(-p,0,u_2) \rangle=
2i \xi^{\-}_{\da} / p^{\- +}  \,
\delta(u_1,u_2) \,\delta^{4}(\q^+)\,.  \label{prop2m} 
\end{align}

Here we show that the two sets of propagators are related by a Fourier transform.
The most non-trivial Fourier transform is that of the propagator $\vev{A^{++}A^+_{\da}}$, see \p{prop2}, \p{prop2m}.
We need to show that
\begin{align} \label{FTprop2}
\int d^4 x\, e^{-i p \cdot x} \frac{1}{\pi}\frac{1}{x^{\- +}} \delta^2(x^{\+ +}) = \frac{2i}{p^{\- +}}\,,
\end{align}   
where the harmonic light-cone projections are defined in eq.~\p{419}. According to eq.~\p{6.5} the scalar product can be decomposed as follows
\begin{align}
2(p \cdot x) = p^{\+ +} x^{\- -} + p^{\- -} x^{\+ +} - p^{\+ -} x^{\- +} - p^{\- +} x^{\+ -}\,.
\end{align}
In Euclidean space we can treat the light-cone projections as pairs of complex conjugate variables, see eq.~\p{cv}.
So the 4D measure $d^4 x$ splits into a pair of 2D measures $d^2 z \equiv d z d \bar z$ and $d^2 \zeta \equiv d \zeta d \bar\zeta$ over the 
complex plane, i.e. $d^4 x = d^2 z d^2\zeta$. The integration $d^2 z$ is trivial due to $\delta^2(x^{\+ +}) 
= \delta^2(z) \equiv \delta(z,\bar z)$. 
The remaining integral is done with the help of the following formula from \cite{Gelfand} 
for the Fourier transform of a class of homogeneous distributions on the complex plane,
\begin{align} \label{GelfandFT}
\int \frac{d^2 \zeta}{\pi}\, \frac{e^{i y \zeta + i \bar y \bar \zeta}}{\zeta^{1+\a} \bar \zeta^{1+\bar\a}} 
= i^{|\a-\bar\a|} \frac{\Gamma\left( \frac{|\a-\bar\a|-\a-\bar\a}{2}\right)}
{\Gamma\left(\frac{|\a-\bar\a|+\a+\bar\a + 2}{2}\right)}y^{\a} \bar y^{\bar\a}\,,
\end{align} 
where $\a$ and $\bar\a$ are arbitrary complex numbers (not complex conjugate in general) such that $\a-\bar\a \in \mathbb{Z}$.
The light-cone projections of the momentum are complex conjugate variables, $p^{\+ -} \equiv y$ and $p^{\- +} \equiv \bar y$.
In the case of interest we have $\a = 0$, $\bar\a = -1$, so we obtain the desired result \p{FTprop2},
\begin{align}
\int \frac{d^2 \zeta}{\pi}\, e^{i y \zeta + i \bar y \bar \zeta}\ \frac{1}{\zeta} = i \frac{1}{\bar y} \,. \notag
\end{align}
Similarly, one derives the Fourier transform of $\vev{A^{++}A^{++}}$, eqs. \p{prop1}, \p{prop1m}.
This time the integration $d^2 \zeta$ gives
\begin{align}
\int \frac{d^2 \zeta}{\pi}\, e^{i y \zeta + i \bar y \bar \zeta} = \pi \delta^2(y) \,. \notag
\end{align}
Let us note that the last formula can be obtained as the limit of \p{GelfandFT} at $\alpha,\bar\alpha \to -1$.  



\section{Relationship with the supertwistor approach} \label{Dict}

The LHC formulation of $\cN =4$ SYM is very similar to the formulation in terms of fields living on supertwistor space. In this Appendix we explain the similarities as well as the differences between the two approaches (see also \cite{Lovelace:2010ev} for an earlier version of such a comparison). The twistor formulation we wish to compare with is that of Refs.~\cite{Mason:2005zm,Boels:2007qn} for bosonic Yang-Mills theory and of Refs.~\cite{Boels:2006ir,Adamo:2011cb} for the $\cN=4$ supersymmetric case.

\subsection{Lightning review of supertwistors}

The main advantage of the (super)twistor variables is the linear realization of the (super)conformal symmetry $PSU(2,2|4)$ on them.

Let us start with the bosonic twistor variables. One considers the complexified conformal symmetry group $SL(4,\mathbb{C})$.
The non-projective twistor space $\mathbb{T}$ is its fundamental representation space.
The complexification enables us to 
employ the embedding formalism, in which complexified compactified Minkowski space is realized as a light-cone in the complex projective space $\mathbb{CP}^5$ with homogenous coordinates $X^{IJ} \sim c X^{IJ}$, $I,J=1,\dots,4$,
\begin{align}
X\cdot X \equiv X_{IJ} X^{IJ} =0\,,  \label{quadric}
\end{align}
where $X_{IJ} =\frac12 \epsilon_{IJKL} X^{KL}$ and $X^{IJ} = - X^{JI}$. The complex coordinates $x_{\alpha\dot\alpha}$ of  Minkowski space
define a particular parametrization of $X^{IJ}$,
\begin{align}\label{X-x}
X^{IJ}(x) =  \left(\begin{array}{cc}  \epsilon_{\alpha\beta} & -i x_\alpha^{\dot\beta}  \\ i x_\beta^{\dot\alpha} & -\frac12 x^2 \epsilon^{\dot\alpha\dot\beta} \end{array}\right) .
\end{align}
Conformal transformations acting nonlinearly on $x_{\alpha\dot\alpha}$ 
correspond to linear  $SL(4,\mathbb{C})$ transformations of $X^{IJ}$.

Bosonic  twistor space $\mathbb{PT}$ is the complex projective space $\mathbb{CP}^3$ whose homogeneous
coordinates $Z^I\sim c Z^I$, $I=1,\dots,4$, belong to $\mathbb{T}$, i.e. they transform in the fundamental representation of 
$SL(4,\mathbb{C})$. A space-time point  $X^{IJ}(x)$ corresponds to a line  in twistor space given by the incidence relation
\begin{align}\label{inc}
X_{IJ} Z^J = 0\,.
\end{align}
For a given point $X^{IJ}$ this relation defines a line in  twistor space since eq.~\eqref{quadric} is the condition that the matrix $X_{IJ}$ has rank two. Choosing two arbitrary points on this line, $Z_1^J$ and $Z_2^J$, we
can reconstruct $X^{IJ}$ as
\begin{align}\label{X-Z}
X^{IJ} = Z_1^I Z_2^J - Z_1^J Z_2^I  = \epsilon^{ab} Z_a^I Z_b^J \, .
\end{align}
Combining eqs.~\p{X-x} and \p{X-Z} we obtain that each point in complexified Minkowski space-time $x_{\alpha\dot\alpha}$ is mapped into
a line $Z^I(\sigma)$ in projective twistor space defined by the equation $X_{IJ}(x) Z^J(\sigma) = 0$. It has the explicit solution 
\begin{align}
Z^I(\sigma) = Z_1^I \sigma^1 + Z_2^I \sigma^2 \equiv Z_a^I \sigma^a \label{sigma-param}
\end{align}
with $\sigma^a=(\sigma^1,\sigma^2)$ being homogeneous coordinates on the line, $\sigma^a \sim c\, \sigma^a$. 
The $\sigma$-coordinates are local in the sense that the twistor line is invariant under $\rm GL(2)$ reparametrizations, 
$\sigma^a \to m^{a}_{b} \sigma^{b}$, $m \in \rm GL(2)$.
Using the parametrization \p{X-x} we
can rewrite the relation between a point in Minkowski space and a line in twistor space as
\begin{align}\label{linebos}
Z^I = (\pi_\alpha, i x^{\dot\alpha\beta}\pi_\beta)
\end{align} 
with $\pi_\alpha$ replacing the homogeneous coordinates $\sigma^a$ on the line in twistor space.
So, the line in twistor space is a $\mathbb{CP}^{1}$ fiber.
The square of the distance between two points $x_{i}$ and $x_j$ in Minkowski space is proportional to
\begin{align}\notag
\frac12 X_i\cdot X_j  = \frac14 \epsilon_{IJKL} X_i^{IJ} X_j^{KL} = \frac14 \epsilon_{IJKL}  \epsilon^{ab} Z_{i,a}^I Z_{i,b}^J
\epsilon^{cd} Z_{j,c}^K Z_{j,d}^L \equiv \langle Z_{i,1} Z_{i,2} Z_{j,1} Z_{j,2}\rangle\,,
\end{align}
where $Z_{i,a}$ and $Z_{j,a}$, $a=1,2$, are two pairs of points belonging to two lines with moduli $X_i$ and $X_j$, respectively.
If two lines intersect, we can choose $Z_{i,2}^I=Z_{j,1}^I$ leading to $x_{ij}^2=0$. 
In order to establish a precise relation between the distance in Minkowski space and the determinant of four twistors 
we choose the parametrization \p{linebos} of the twistor lines
\begin{align}\label{skpr}
\langle \pi_{i,1} \pi_{i,2} \rangle \langle \pi_{j,1} \pi_{j,2} \rangle \, x_{ij}^2 = \langle Z_{i,1} Z_{i,2} Z_{j,1} Z_{j,2}\rangle\,.
\end{align}

To deal with $\mathcal N=4$ \emph{chiral} supersymmetry, we extend the projective twistor space to include four Grassmann 
coordinates
\begin{align}\label{ZZ}
\mathcal Z  = (Z^I,\chi^A)
\end{align}
(with $I,A=1,\dots,4$), subject to the equivalence relation $\mathcal Z  \sim c \mathcal Z $. The odd twistor 
coordinates $\chi^A$ satisfy an incidence relation. In the parametrization \p{X-x}
it has the form $\chi^A = \theta^{A\beta} \pi_\beta$. 
So, the relation between a point in chiral Minkowski (super)space-time $(x^{\dot\alpha\alpha},\theta^{A\alpha})$
and a line in projective twistor superspace is  
\begin{align}\label{line}
\mathcal{Z} = (\pi_\alpha, i x^{\dot\alpha\beta}\pi_\beta, \theta^{A\beta} \pi_\beta)\,.
\end{align} 
The $\cN=4$ superconformal transformations correspond to  linear $SL(4|4,\mathbb{C})$ transformations of the supertwistors $\mathcal Z$.

Along with the holomorphic coordinates one considers their Euclidean complex conjugates,
$\hat \pi_{\a} = (-(\pi_2)^*, (\pi_1)^*)$ and $\hat{\hat{\pi}}_{\a} = - \pi_{\a}$.
The (complex) Minkowski coordinates $x_{\a\da}$ are real with respect to this conjugation (involution).
Instead of working directly with derivatives, in the twistor formalism one introduces the exterior derivative $\bar\pa$ and the basis of 
$(0,1)$-differential forms $\bar e^{0}$, $\bar e^{\da}$,
\begin{align} \label{01form}
\bar e^{0} = \frac{\hat{\pi}^{\a} d \hat{\pi}_{\a}}{(\pi \cdot \hat\pi)^2} \quad, \qquad
\bar e^{\da} = \frac{\hat{\pi}_{\a} d x^{\da\a}}{(\pi \cdot \hat \pi)}\,. 
\end{align}
In this basis the exterior derivative is decomposed as $\bar\pa = \bar e^0 \bar \pa_{0} + \bar e^{\da} \bar \pa_{\da}$
where the basis of vector fields $\bar \pa_{0}$, $\bar \pa_{\da}$ on twistor space is dual to the basis of $(0,1)$-forms,
\begin{align}
\bar \pa_{0} = (\pi \cdot \hat\pi) \pi_{\a} \pa/\pa \hat\pi_{\a}  \quad, \qquad
\bar \pa_{\da} = \pi^{\a} \pa / \pa x^{\da\a}\,. 
\end{align}
The natural volume form on a twistor line (or a $\mathbb{CP}^1$ fiber) is $D \pi = \pi^{\a} d \pi_{\a}$,
and the volume form on the projective twistor space $\mathbb{PT}$ is
$\mathcal D ^{3} Z = \frac1{4!} \epsilon_{IJKL} Z^I d Z^J d Z^K d Z^L$. The latter is an exterior product of $(1,0)$-forms 
$\mathcal D ^{3} Z = (\pi \cdot \hat{\pi})^4 e^0 \wedge e^{\da} \wedge e_{\da}$ which are the Euclidean complex conjugates 
of the $(0,1)$-forms \p{01form}.
The volume form has an obvious extension to the 
supertwistor space $\mathcal D ^{3|4} \mathcal Z =  \mathcal D ^{3} Z  d^4 \chi$.

 \subsection{Dictionary twistors-harmonics}

Now it is easy so see that  twistor superspace is very similar to Lorentz harmonic chiral superspace, Table \ref{twvshar}.
The  harmonics $u^+_\a$ are replaced by the holomorphic spinor coordinates $\pi_{\a}$ on $\mathbb{CP}^1$ fibers.
The harmonic derivative $\pa^{++}$ corresponds to the twistor derivative $\bar\pa_0$.
The local $U(1)$ charge of the harmonic functions corresponds to the degree of homogeneity of the functions on the (super)twistor space (called `holomorphic weight').
The L-analyticity implies that the corresponding twistor form depends on the 
holomorphic projection $\chi^A = \pi_{\a} \q^{\a A}$ of the Grassmann variable $\q$ but is independent of $\hat\pi_{\a}\q^{\a A}$. 
So, $\pi$ and $\hat\pi$ are the analogs of the harmonics $u^+$ and $u^-$, respectively. 
Ordinary space-time fields are extracted from the differential forms by an integral  Penrose transform. It corresponds to picking out the first term in the harmonic expansion \p{3}. So, in the LHC approach the notion of Penrose transform is replaced by the simpler notion of Fourier expansion on the LH two-sphere.
The harmonic measure $d u \sim (u^{+\a} du^+_\a) (u^{-\b} du^-_\b)$ carries zero $U(1)$ charge, 
while in the twistor approach one also introduces the projective measure
$D \pi = \langle \pi d \pi \rangle$ (the equivalent of $u^{+\a} du^+_\a$) with holomorphic weight $2$. 

The main conceptual difference between the two approaches is in the prominent role played by the harmonic derivative $\pa^{--}$ in the LHC approach. Together with $\pa^{++}$ and the $U(1)$ charge generator $\pa^0$ they form the algebra of $SU(2)_L$ realized on the charges $\pm$ of the harmonics (see \p{2.7}). These notions are absent in the  holomorphic twistor description, which makes the construction of gauge connections and curvatures less transparent (see the comment at the end of Sect.~\ref{twi2}). 

\begin{table}
\begin{center}
\begin{tabular}{c|c}
Lorentz harmonic chiral superspace & Twistor superspace \\[0.2 cm] \hline \\[-0.4 cm]
$u^{+}_{\a}$ & $\pi_{\a}$ \\[0.2 cm]
$u^-_{\a}$ & $\hat{\pi}_{\a}$ \\[0.2 cm] 
$u^{+\a} u^-_{\a} = 1$ & $ \pi^{\a} \hat{\pi}_{\a} \neq 0$ \\[0.2 cm] 
harmonic variables & $\mathbb{CP}^1$ fibers \\[0.2 cm]
$\q^{+A} = u^+_{\a} \q^{\a A} $ & $\chi^A = \theta^{A\beta} \pi_\beta$ \\ [0.2cm]
space-times coordinate $x_{\a\da}$ & moduli $X^{IJ}$ of a line in twistor space \\ [0.2cm]
LHC superspace & complex projective space $\mathbb{CP}^{3|4}$ \\ [0.2cm]
LHC superspace coordinates $u^{\pm}_{\a}$, $x_{\a\da}$, $\q^{+A}$ 
     & supertwistor $\mathcal{Z}= (\pi_\alpha, i x^{\dot\alpha\beta}\pi_\beta, \chi^A )$ \\ [0.2cm]
L-analytic superfields $A(x,\q^+,u)$ &	twistor fields $f(\mathcal{Z})$ \\ [0.2cm]
harmonic expansion & Penrose transform \\ [0.2 cm]
$U(1)$ charge & holomorphic weight \\ [0.2 cm] 
harmonic derivative $\pa^{++}$ & $(0,1)$-vector on $\mathbb{CP}^1$ fiber $\bar\pa_0$ \\ [0.2 cm]
space-time derivative $\pa^{+}_{\da}$ &  $(0,1)$-vector on twistor space $\bar\pa_{\da}$ \\ [0.2 cm]
harm. measure $d u \sim (u^{+\a} du^+_\a) (u^{-\b} du^-_\b)$ & volume form of $\mathbb{CP}^1$ fiber 
$D \pi = \langle \pi d \pi \rangle$ \\ [0.2 cm]
LHC superspace measure $d^4 x du d^4 \q^+ $ 
   & $\mathcal D ^{3|4} \mathcal Z = \frac1{4!} \epsilon_{IJKL} Z^I d Z^J d Z^K d Z^L  d^4 \chi$ \\ [0.2 cm]
$SU(2)_L$ algebra of harmonic derivatives & ??? \\ [0.2 cm]
harmonic derivative $\pa^{--}$ & ??? 
\end{tabular}
\end{center}
\caption{Comparison of LHC superspace with  twistor superspace} \label{twvshar}
\end{table}


\subsection{$\cN = 4$ SYM on twistor superspace} \label{twi2}

\begin{table}\vskip5mm
\begin{center}
\begin{tabular}{c|c}
Lorentz harmonics chiral superspace & Twistor superspace \\[0.2 cm] \hline \\[-0.4 cm]
L-analytic gauge connections $A^{++}$, $A^+_{\da}$ & twistor superfield $\cA = \bar e^{0} \cA_0 + \bar e^{\da} \cA_{\da} $ \\[0.2 cm] 
Chern-Simons term $L_{\rm CS}(x,u,\q^+)$ & Chern-Simons term $L_{\rm CS}(\mathcal{Z})$ \\[0.2 cm] 
Zupnik's action $L_{\rm Z}(x,\q)$ & $L_{\rm int}(x,\q)$ \\[0.2 cm] 
Wess-Zumino gauge & space-time gauge  \\[0.2 cm] 
chiral gauge connection $A^{--}(x,u,\q)$ & ??? \\[0.2 cm] 
chiral supercurvature $W_{AB}(x,u,\q)$ & ???
\end{tabular}
\end{center}
\caption{Comparison of the LHC and twistor superspace formulations
of $\cN = 4$ SYM} \label{twvshar2}
\end{table}
  
The fields of $\mathcal N=4$ SYM theory are described on projective twistor space $\mathbb{PT}$ by a superfield $\mathcal{A}$ which is a 
$(0,1)$-form with values in the Lie algebra of the gauge group $SU(N_c)$.  Expanding $\mathcal{A}$ in the fermionic coordinates $\chi^A$ one obtains
\begin{align}\notag
\mathcal A(Z,\bar Z,\chi) &{}= a(Z,\bar Z) + \chi^A \tilde \gamma_A(Z,\bar Z) + \frac12 \chi^A\chi^B \phi_{AB}(Z,\bar Z)
\\
&{}+ \frac1{3!} \epsilon_{ABCD}\chi^A\chi^B\chi^C \gamma^{D}(Z,\bar Z)
 +\frac1{4!} \epsilon_{ABCD}\chi^A\chi^B\chi^C\chi^D  b(Z,\bar Z)  \, .
\end{align}  
The coefficients accompanying $\chi^n$ are $(0,1)$-differential forms on twistor space, homogeneous of degree $n$ that are
related to the various component fields of $\mathcal N=4$ SYM in the space-time gauge: $b$ and $a$ give
rise to the self-dual and anti self-dual part of the YM curvature, $\tilde \gamma_A$ and $\gamma^{D}$ are mapped
into the (anti)-gluino fields and $\phi_{AB}$ are the scalar fields. 
The twistor action of $\mathcal N=4$ SYM takes the form
\begin{align} \label{S-tw}
 &{} S[\mathcal A] = \int_{\mathbb{CP}^{3|4}} \mathcal D ^{3|4} \mathcal Z \wedge \tr\left(\frac12 \mathcal A \,\bar{\partial}\mathcal A  
 + \frac13 \mathcal A^3 \right) +   \int d^{4} x \,d^8 \theta \, L_{\rm int}(x,\theta)\,.
\end{align}  
In the first term in eq. \p{S-tw} the $(0,3)$-form constructed out of the twistor field $\mathcal{A}$ and the exterior derivative $\bar\pa$
is integrated against the $(3,0)$-volume form $\mathcal D ^{3|4} \mathcal Z$ over the projective twistor space.
It is the holomorphic Chern-Simons action, which describes the self-dual part of the $\mathcal{N}=4$ action.
The second term on the right-hand side of eq. \p{S-tw} is the non self-dual part of the action, which describes the interactions
\begin{align} \label{tw-log-det}
L_{\rm int}(x,\theta)=g^2  \left[ \ln \det (\bar{\partial}_0 + \mathcal A_0 ) -\ln \det  \bar{\partial}_0 \right].
\end{align}
It involves the logarithm of the chiral determinant of the Cauchy-Riemann operator $\bar\pa + \cA$ 
restricted to the line in twistor space ($\mathbb{CP}^1$ fiber) with moduli $(x,\q)$, and then integrated over the
moduli space of the lines. This separation of the action $S[\mathcal A]$ into a sum of two terms corresponds 
to the expansion of the $\mathcal{N}=4$ theory around the self-dual sector.

The logarithm of the determinant in eq. \p{tw-log-det} can be written in a more explicit form
as an expansion in powers of the superfields
\begin{align}\notag\label{Sint}
 L_{\rm int} (x,\theta) &{}= g^2  \sum_{n\ge 2} \frac{(-1)^n}{n}  
\tr\left[\bar{\partial}^{-1}_0\mathcal A_0\dots \bar{\partial}^{-1}_0\mathcal A_0 \right]
 \\
&{} =g^2   \sum_{n\ge 2} \frac{(-1)^n }{ n}   \int\limits_{(\mathbb{CP}^1)^n}  
\frac{ \tr\left[\mathcal A(\mathcal Z(\sigma_1))\wedge D \sigma_1 \dots \mathcal A(\mathcal Z(\sigma_n))\wedge D \sigma_n \right] }{ \vev{\sigma_1\sigma_2}\dots
 \vev{\sigma_n\sigma_1}}\ .
\end{align}
It involves copies $\mathcal Z(\sigma_i)$ of the same twistor line with  moduli $(x,\q)$.
They are parametrized by the local coordinates $\sigma_i^a$, eq. \p{sigma-param}, and 
the projective measures $D\sigma_i = \vev{\sigma_i d\sigma_i}$.
The $(0,1)$-forms are restricted to $\mathbb{CP}^1$ fibers, so only the $\cA_0$ component of the $(0,1)$-form $\cA$ contributes.  
An analogous expression exists in terms of the local coordinates $\pi_{i\,\a}$, eq. \p{line}.

The $(0,1)$-differential form $\mathcal A$ 
is equivalent to the pair of L-analytic gauge connections $A^{++}$ and $A^{+}_{\da}$ in the LHC
formulation, Table \ref{twvshar2}. The two terms in the twistor action, eq. \p{S-tw}, correspond to the splitting of the action in 
the LHC superspace formulation, eq. \p{N4}. The nonpolynomial Lagrangian $L_{\rm int}$, eq. \p{Sint}, is the twistor analog of the Zupnik Lagrangian $L_{\rm Z}$, eq. \p{lint}.

The main conceptual difference between the twistor and harmonic approaches
is the absence of a twistor analog of the LH gauge connection $A^{--}$ for the harmonic derivative $\pa^{--}$ 
and of the associated  $SU(2)_L$ algebraic structure, eq. \p{2.7}. 
The notion of $A^{--}$ is crucial for the construction of the remaining gauge connections, 
supercurvatures, eq. \p{3.12}, and composite operators.

The harmonic derivative $\pa^{--}$ and the associated analytic gauge frame and  gauge connection $A^{--}$, eq. ~\p{448}, in the LHC formulation makes the supercurvature geometry very transparent. The remaining gauge connections and curvatures are constructed simply by commuting covariant derivatives. In the twistor approach one replaces these notions by an indirect construction in terms of a gauge bridge $H(x,\q,u)$ (`holomorphic frame' in the twistor language). It is defined by the equation $H^{-1}(\pa^{++} + A^{++}) H=0$ and thus trivializes the gauge connection for $\pa^{++}$. Rotating every superfield by $H$, $\Phi^\Lambda = H\Phi^\tau H^{-1}$, one  can switch between the real (with LH-independent parameter $\tau(x,\q)$) and analytic (with $\q^-$-independent parameter $\Lambda(x,\q^+,u)$) gauge frames. The drawback of this approach is that the bridge $H$ cannot be expressed unambiguously in terms of the prepotential $A^{++}$. In contrast, in the LHC approach every geometric object (gauge connection or curvature) is given directly and manifestly in terms of the prepotentials. Thus, one can only compare expressions for gauge invariant objects in the two approaches, like the Lagrangian or composite operators.

Another important difference is the lack of a discussion of the $\bar Q$ supersymmetry transformations in the twistor literature. In Sect.~\ref{s4} we have shown that the gauge connection $A^{--}$ plays a crucial role in theses transformations. Once again, the absence of the notion of $A^{--}$ in the twistor approach may explain the difficulties in realizing $\bar Q$ supersymmetry.

 \subsection{Gauge fixing}
 
Next, we compare the gauge-fixing conditions. Just before (3.28) in Ref.~\cite{Boels:2006ir} we read $\eta^A \bar\pa_A \lrcorner \cA=0$ where $\eta^A$ is an arbitrary but fixed commuting {chiral} spinor (not to be confused with the antichiral $\pi_{A'}$). We translate this condition as \p{CSW}. The spinor $\xi^\da$ has no $U(1)$ charge and hence we cannot mix $\xi^\da A^+_\da$ with $A^{++}$, unless we put in front of it a negative-charged parameter $\mu^- = \mu^{\a} u^-_{\a}$. We can thus consider the following generalized gauge:
\begin{align}\label{Z*}
\xi^\da A^+_\da + \mu^{\a} u^-_{\a}A^{++}=0\,.
\end{align}
 The pair $Z_*=(\xi^\da, \mu_{\a})$ forms a four-component twistor. The gauge (3.9)  in Ref.~\cite{Adamo:2011cb} makes use of such a `reference twistor', but one should be allowed to set $\mu^{\a}=0$  (see the special choice ${\cal Z}_*=(i^A,0,0)$ just above eq.~(3.9)  in Ref.~\cite{Adamo:2011cb}). Indeed, we can write $\mu^{\a} = \xi_\da x_*^{\da\a}$ and then we should be able to set $x_*=0$ by a translation in Minkowski space. However, it is hard for us to find an analog of the fermionic part of the so-called `reference supertwistor' of the axial gauge given in Ref.~\cite{Adamo:2011cb}. Its role is to maintain the appearance of superconformal symmetry in the gauge, but we know that the very presence of the fixed reference supertwistor ${\cal Z}_*$ breaks all symmetries. They are only restored once  the gauge-fixing parameter has been eliminated from the sum of all Feynman graphs. This is not so simple, as discussed in \cite{Chicherin:2014uca} and \cite{twin}.

Finally, we find it not very easy to understand page 14 in Ref.~\cite{Adamo:2011cb} with the derivation of the propagators in the presence of  ${\cal Z}_*$. It starts by writing the propagator equation for the truncated (gauge-fixed) CS action. This equation has just a delta function as its right-hand side, while   we would expect that the gauge fixing yields additional contact terms (compare, e.g., with the propagator equation in the usual axial gauge). Then the authors give the solution (3.11). It is not immediately obvious, but nevertheless possible to extract our propagators $\vev{A^{++} A^{++}}$ and  $\vev{A^{+}_\da A^{++}}$  from it.\footnote{We thank Lionel Mason for the explanation.}  Last but not least, the equation after (3.11) states that this propagator satisfies not quite the required equation, but produces ``essentially vanishing error terms". It appears in fact that the Green's function equation at the top of page 14 is not correct, but the solution (3.11) is. The authors  of Ref.~\cite{Adamo:2011cb} remark that choosing ${\cal Z}_*=(\xi,0,0)$ would lead to `error terms' of the type $\delta(\pi_{\a}) \delta(\mu_\da + \xi_\da) =0$ since  $\pi_{\a}\neq 0$ by definition. This confirms once more that the most natural gauge-fixing parameter is $\xi_\da$ and not the full reference supertwistor ${\cal Z}_*$.


\end{document}